\documentclass[letterpaper,twocolumn,10pt]{article}

\usepackage{zhanggroup}
\usepackage{titlesec}
\usepackage[absolute]{textpos}
\usepackage{cite}
\usepackage{amsmath,amssymb,amsfonts}
\usepackage{algorithmic}
\usepackage{graphicx}
\usepackage{textcomp}
\usepackage{xcolor}
\usepackage{mathrsfs}
\usepackage{amsfonts}  
\usepackage{amsthm}
\usepackage{booktabs}
\usepackage{multirow}
\usepackage{url}
\usepackage{caption,subcaption}
\usepackage{amsmath}
\usepackage{dsfont}
\usepackage{mathtools}
\usepackage{cite}
\usepackage[capitalize]{cleveref}
\usepackage{xfrac}
\usepackage[linesnumbered,ruled,vlined]{algorithm2e}
\newcommand{\mypara}[1]{\noindent{\bf {#1}.}}
\newcommand{\etal}{\textit{et al.}\xspace}
\definecolor{ultramarine}{RGB}{76, 153, 0}
\newcommand{\commentcolor}[1]{\textcolor{ultramarine}{#1}}

\newcommand{\customTableFont}{\fontsize{8pt}{9pt}\selectfont}
    
\begin{document}

\begin{textblock}{15}(3.2,1)

To Appear in the 45th IEEE Symposium on Security and Privacy, May 20-23, 2024.
\end{textblock}

\date{}

\title{\Large \bf Test-Time Poisoning Attacks Against Test-Time Adaptation Models}

\author{
{\rm Tianshuo Cong\textsuperscript{1}}\ \ \
{\rm Xinlei He\textsuperscript{2}}\ \ \
{\rm Yun Shen\textsuperscript{3}}\ \ \
{\rm Yang Zhang\textsuperscript{2}}\ \ \
\\
\\
\textsuperscript{1}\textit{Tsinghua University}\ \ \
\textsuperscript{2}\textit{CISPA Helmholtz Center for Information Security}\ \ \ 
\textsuperscript{3}\textit{NetApp}\ \ \
}

\maketitle

\begin{abstract}

Deploying machine learning (ML) models in the wild is challenging as it suffers from distribution shifts, where the model trained on an original domain cannot generalize well to unforeseen diverse transfer domains.
To address this challenge, several test-time adaptation (TTA) methods have been proposed to improve the generalization ability of the target pre-trained models under test data to cope with the shifted distribution.
The success of TTA can be credited to the continuous fine-tuning of the target model according to the distributional hint from the test samples during test time.
Despite being powerful, it also opens a new attack surface, i.e., test-time poisoning attacks, which are substantially different from previous poisoning attacks that occur during the training time of ML models (i.e., adversaries cannot intervene in the training process).
In this paper, we perform the first test-time poisoning attack against four mainstream TTA methods, including TTT, DUA, TENT, and RPL.
Concretely, we generate poisoned samples based on the surrogate models and feed them to the target TTA models.
Experimental results show that the TTA methods are generally vulnerable to test-time poisoning attacks.
For instance, the adversary can feed as few as 10 poisoned samples to degrade the performance of the target model from 76.20\% to 41.83\%.
Our results demonstrate that TTA algorithms lacking a rigorous security assessment are unsuitable for deployment in real-life scenarios.
As such, we advocate for the integration of defenses against test-time poisoning attacks into the design of TTA methods.\footnote{Our code is available at \url{https://github.com/tianshuocong/TePA}.}

\end{abstract}

\section{Introduction}

\begin{figure}[t]
\centering
\includegraphics[width=5cm]{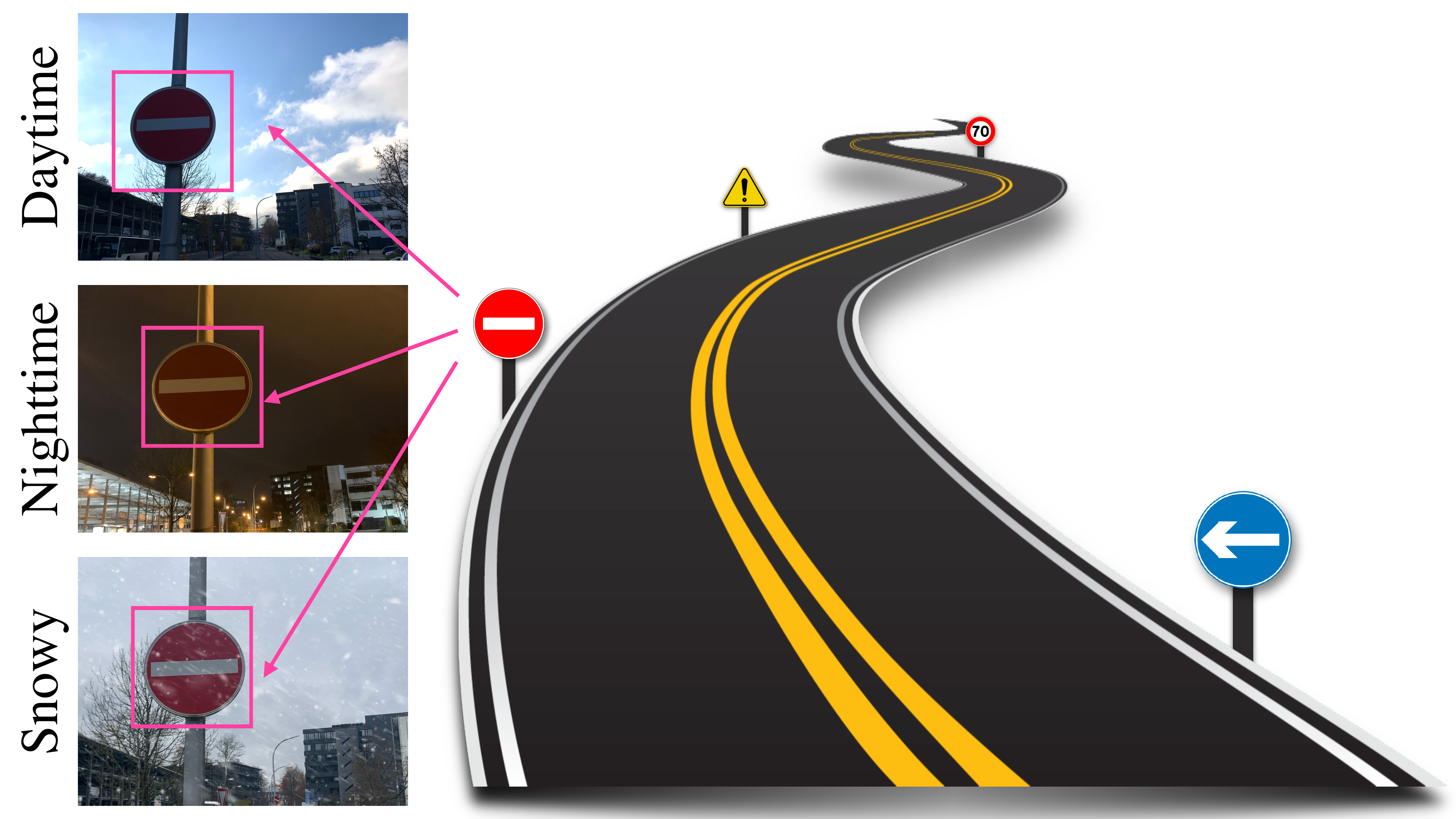}
\caption{Targeting the challenge of ``distribution shifts,'' test-time adaptation (TTA) methods can aid in the identification of traffic signs across diverse weather conditions.}
\label{fig:self-driving-scenario}
\end{figure}

In recent years, machine learning (ML) has achieved remarkable performance\cite{HZRS16}.
Nevertheless, deploying these ML models in the real world poses a significant challenge, as distribution shifts may occur when the training and test datasets come from different distributions~\cite{KSMXZBHYPGLDSGEHBLKPLFL21,ZDZ23}.
Take image classification as an example; the data used to train the ML models are often carefully curated, e.g., selecting the object images with a clean background and cropping the center area of the objects.
However, those models must deal with the test images coming from a different distribution in the real world (See \autoref{fig:self-driving-scenario}), which usually leads to degraded model performance~\cite{SKLGXSKHYMBDSGLSHLFL22}.
Prior approaches to enhancing the ML model's generalization under distribution shifts have focused on the training process to prompt the target model to learn more distribution types in advance, such as leveraging a large number of labeled data~\cite{THSD17}, novel data augmentation~\cite{HMCZGL20}, etc.
However, test data usually comes from an unseen distribution.
Consequently, target models with fixed parameters trained on the original domain will no longer be applicable to the diverse transfer domains, leading to an increasing interest in the dynamic adaptation of ML models during inference.

Test-time adaptation (TTA) is an emerging technique to tackle distribution shifts and has been leveraged in several real-world security-sensitive scenarios, such as autonomous driving~\cite{MMPB22}, medical diagnosis~\cite{KECK21}, etc.
In TTA settings, the test data from the transfer domain is delivered as a data stream and the target model is updated online.
In other words, the target model only has access to the current test data instead of the whole test dataset.
This is particularly relevant in latency-sensitive scenarios, such as autonomous driving, which necessitate immediate prediction of arrival data.
To address these realistic constraints, various TTA methods~\cite{SWLMEH20,MMPB22,WSLOD21,RSPEGBBB22} have been proposed to enhance the performance of prediction by fine-tuning the model's parameters based on the current test data before making predictions. 

Though proven successful in improving the generalization of ML models, TTA paradigms may introduce a new attack surface for adversaries to tamper with the parameters of a target model at test time by fine-tuning it with potential malicious samples. 
This can directly impact the predictions for benign samples.
To explore this possibility, in this work, we propose the first untargeted {\bf \emph{test-time poisoning attacks (TePAs)}} against TTA models, i.e., an adversary aims to degrade a TTA model's performance at test time.
Our approach is drastically different from previous poisoning attacks that are executed during the model's training process, i.e., \emph{training-time poisoning attacks (TrPAs)}~\cite{CT22,SHNSSDG18,FGCGCG21}.
Compared to TrPAs, TePAs face the following non-trivial challenges:
(i) TrPAs require modification access to the target model's training dataset, while TePAs do not poison the training dataset nor control the training process of the target model.
(ii) For TrPAs, poisoned samples are mixed with clean training samples where they can be learned in multiple epochs by the model and become more memorable. 
However, considering effectiveness and efficiency, TTA methods usually update the model using \emph{one} epoch based on each arrival of test data hence a different setting for TePAs.
(iii) In TePAs, poisoned and benign samples are in the same pipeline, and the model is in a state of dynamic adjustment. 
Therefore, the poisoning effectiveness is also affected by the benign samples.
(iv) Since TePAs are test-time attacks, the adversary must take the query budget into account to maintain the attack's stealthiness.
(v) To avoid the target models ``forgetting'' the original task, TTA methods usually only update part parameters of the model.
However, for TrPAs, the poisoned samples are used to update the whole model parameters.
In summary, these differences make TePAs harder to succeed than TrPAs.

\mypara{Our Work}
In this paper, we take the first step toward understanding the TePAs against TTA models and the plausible countermeasures.
Our study aims to demonstrate that the current TTA methods are prone to TePAs.
Considering their use in safety-critical applications where a deterioration in their efficacy could result in severe consequences, exposing the model modification right to the adversaries is irresponsible, and taking into account TePAs during TTA methods design becomes crucial.
Surprisingly, to the best of our knowledge, no prior research has investigated the vulnerability of TTA models with respect to TePAs.

We first systematically define the threat model of TePAs against TTA models.
The goal of the adversary is to launch an indiscriminate poisoning attack against the target model, resulting in its performance degradation.
The adversary's ability is limited to the query access to the target model, meaning that they are unable to access important details (such as the loss value, gradients, and parameters) nor the outputs (such as posterior or predicted label) of the target model.
Additionally, we assume that the adversary has background knowledge of the distribution of the target model's training dataset.
This knowledge allows the adversary to construct a surrogate model with a similar distribution dataset.
They can later generate poisoned samples based on the surrogate model and feed them to the target model.

To better demonstrate the vulnerability of the TTA techniques to TePAs, we consider four prominent TTA methods in our paper, including \underline{T}est-\underline{t}ime \underline{t}raining ({\bf TTT})~\cite{SWLMEH20}, \underline{D}ynamic \underline{u}nsupervised \underline{a}daptation ({\bf DUA})~\cite{MMPB22},  \underline{T}est \underline{ent}ropy minimization ({\bf TENT})~\cite{WSLOD21}, and \underline{R}obust \underline{p}seudo-\underline{l}abeling ({\bf RPL})~\cite{RSPEGBBB22}.
We launch TePAs against the above TTA methods.
Specifically, we propose a poisoned sample generation framework, \texttt{PoiGen}, which creates poisoned samples based on a surrogate model and transfer-based adversarial attacks.
Our experimental results indicate that only 10 poisoned samples or a small poisoning ratio of 0.1 can cause a 34.37\% drop or a 6.13\% drop in the target TTT-model's performance, respectively. 
To mitigate TePAs, we investigate several defense mechanisms such as adversarial training~\cite{MMSTV18}, bit-depth reduction~\cite{XEQ18}, etc. 
However, our experiments show that these defenses are not effective against TePAs, which prompts the need for more effective defense mechanisms.

In summary, we make the following contributions:

\begin{itemize}
\item We propose the first test-time poisoning attacks against four TTA methods, including TTT, DUA, TENT, and RPL.
\item Empirical evaluations show that our attacks are effective in degrading the target model's performance even with limited poisoned samples and small fractions of poisoned data.
\item To mitigate the attacks, we investigate four defense mechanisms and find that none of them are effective to defend against the proposed TePAs.
\end{itemize}

\section{Background}

\subsection{Preliminaries}

\begin{figure}[t]
\centering
\subfloat[Inference w/o TTA. \label{fig:overview_tta_a}]{%
\includegraphics[width=0.36\linewidth]{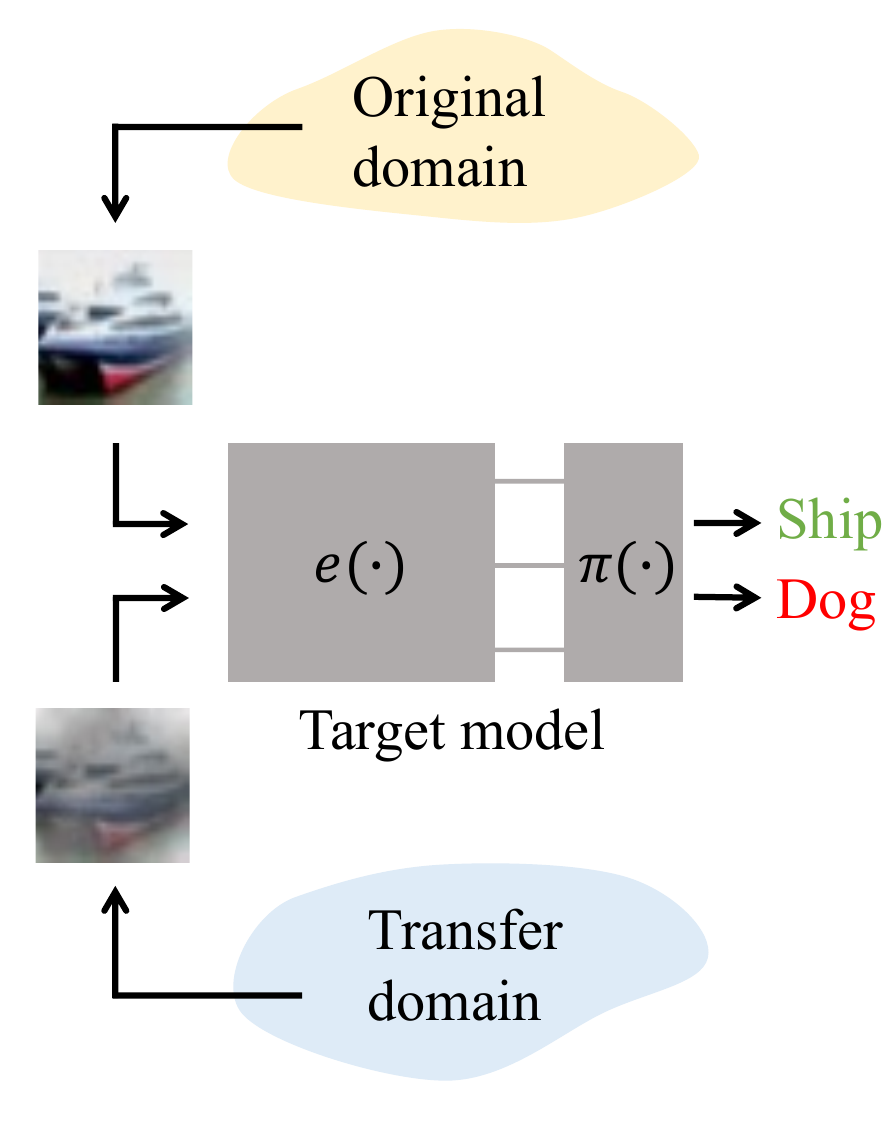}}
\hspace{1mm}
\subfloat[Inference w/ TTA. \label{fig:overview_tta_b}]{%
\includegraphics[width=0.6\linewidth]{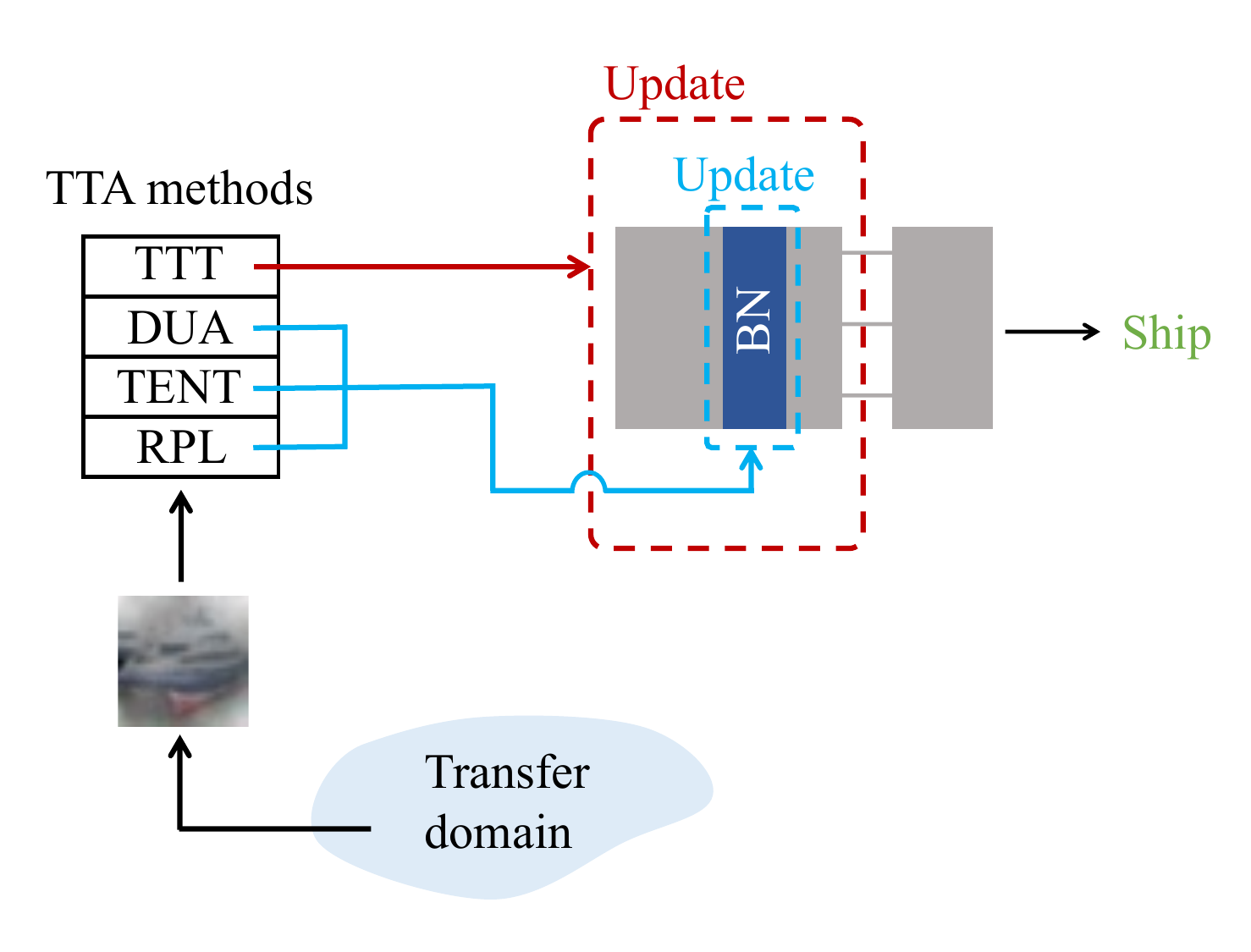}}
\caption{Overview of TTA methods. 
(a) A target model with fixed parameters cannot cope with distribution shifts. 
(b) TTA methods can improve the target model's performance by adjusting the target model's parameters.} 
\label{fig:overview_tta} 
\end{figure}

\mypara{Notations}
We use $f: x\in [0,1]^D \rightarrow \mathbb{R}^C$ to denote a $C$-class classification model, where $x$ is the input (such as an image) and $D$ is the input size. 
$f(x)=[f_1,...,f_C]$ is the output logits vector.
$p(x)=[p_1,...,p_C]=\sigma(f(x))$ is the confidence vector where $\sigma(\cdot)$ is the softmax function and $p_j=p(j|x)$ is the prediction probability on the $j$-th class.
Then the final prediction can be calculated by
$z=\mathop{\arg\max}_{j=1,...,C}p_j.$

\mypara{Overall Goal of TTA}
An illustration of TTA methods is shown in \autoref{fig:overview_tta}.
A target model $f$ which is trained on the original domain $\mathcal{D}_{ori}=\{x\sim P(x)\}$ does not generalize well to a different transfer domain $\mathcal{D}_{trans}=\{x~\sim Q(x)\}$ ($Q(x)\neq P(x)$) due to distribution shifts~\cite{KSMXZBHYPGLDSGEHBLKPLFL21}.
TTA methods aim to improve the performance of the target model $f$ by updating its parameters at test time to cope with such distribution shifts.
In essence, $f$ can be split into a feature extractor $e(\cdot)$ and a linear classifier $\pi(\cdot)$, i.e., $f(x) = \pi(e(x))$ where $h=e(x)\in \mathbb{R}^d$ is the feature vector.
Different TTA methods update different parts of the target model to attain the above goal.
For instance, TTT updates the feature extractor $e(\cdot)$ at test time. DUA, TENT, and RPL update the parameters in the batch normalization (BN) layers\cite{IS15}.
More details can be found in \autoref{background_ttt}.

\begin{figure*}[t]
\centering
\includegraphics[width=13cm]{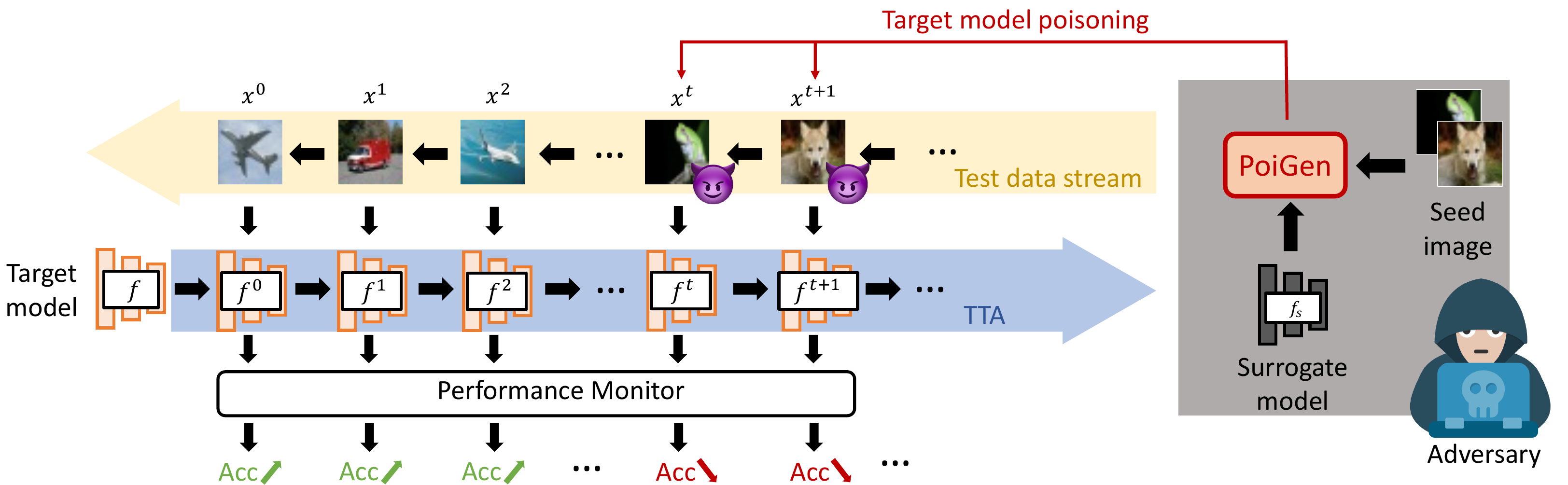}
\caption{Workflow of TePA. 
The adversary uses \texttt{PoiGen} to generate poisoned samples which will be fed into the test data stream (the yellow indicating arrow). 
The target model $f$ will be updated via TTA methods to $f^t$ (the blue indicating arrow) according to the arrived test data. 
When meeting benign samples, the performance of $f^t$ (Acc) will be improved. 
However, the poisoned samples could degrade the prediction ability of $f^t$.}
\label{fig:workflow_attack}
\end{figure*}

\mypara{Test-Time Behavior of TTA}
It is important to note that the target model does not have access to the whole test samples which are from the transfer domain.
Note that under TTA assumptions, the test samples come in sequential order, i.e., $x^0 \gets \cdots x^t \gets \cdots$ where $x^t$ denotes the test data at timestamp $t$.
The target model $f$ will be adapted by $x^t$ to $f^t$  using TTA methods (See \autoref{fig:workflow_attack}).
That is, it can only process the current arrived test data instead.
For TTT and DUA, the test data come in a ``point-by-point'' manner, i.e., $x^t$ stands for a single image.
For TENT and RPL, the test data come in a ``batch-by-batch'' manner, i.e., $x^t=\{x_i^t\}_{i=1}^B$ stands for a batch of test samples where $B$ is the batch size.
The technical details are outlined below.

\subsection{TTA Methods}
\label{background_ttt}

\mypara{TTT~\cite{SWLMEH20}}
As a classical TTA method, TTT has been widely used in the real world~\cite {FLIMSK21,BGB21,KECK21}. 
Briefly, TTT updates the feature extractor based on a self-supervised learning (SSL) task at test time, in turn, adapting to the distribution shifts on-the-fly.
The overview of TTT is shown in \autoref{fig:workflow_TTT}.

During training time, TTT requires a Y-structured model as the target model.
For instance, the training process of TTT can be considered as a multi-task learning problem~\cite{CDBCT19}, which jointly learns from two tasks (i.e., a main task and an auxiliary task).
The main task is a classification task, and the cross-entropy loss is applied (denoted as $\mathcal{L}_m$).
The auxiliary task is an SSL task, i.e., rotation prediction~\cite{DGZ21}.
Concretely, it rotates each training image into $0$, $90$, $180$, and $270$ degrees first and then predicts the rotation angle (i.e., a $4$-class classification problem).
We denote the SSL task loss function as $\mathcal{L}_s$.
As shown in \autoref{fig:workflow_TTT}, the TTT model has a Y-structure architecture with a shared feature extractor $e(x;\theta_e)$ and two branches $\pi_m(x;\theta_m)$ and $\pi_s(x;\theta_s)$, where $\pi_m$ is used for the main task and $\pi_s$ is used for the auxiliary task.
Given a training sample, TTT first feeds it into $e(x;\theta_e)$ to obtain its feature vector $h$.
Then, $h$ is fed into $\pi_m$ and $\pi_s$ to calculate the $\mathcal{L}_m$ and $\mathcal{L}_s$, respectively.
The total loss function for training the Y-structured target model $f(x;\theta)$ can be thus defined as 
\begin{equation}
\label{ttt_train_loss}
\min_{e, \pi_s, \pi_m}\frac{1}{N}\sum_{i=1}^N(\mathcal{L}_m(x_i,y_i;e, \pi_m)+\mathcal{L}_s(x_i; e,\pi_s)),
\end{equation}
where $\{(x_i,y_i), i \in N\}$ is the training data, and $\theta^*=(e^*, \pi_s^*, \pi_m^*)$ are the optimized parameters of \autoref{ttt_train_loss}.

During inference, TTT adapts the model based on the test data first and then makes a prediction using the updated model.
Concretely, TTT updates $e(x;\theta_e)$ and $\pi_s(x;\theta_s)$ based on the SSL task $\mathcal{L}_s$,
and $\pi_m(x;\theta_m)$ is fixed throughout.
At $t=0$, the model's initial state is $\theta^*$.
Given $x^0$, TTT first fine-tunes its feature extractor and auxiliary branch by minimizing $\mathcal{L}_s$ as
\begin{equation}
\label{ttt_standard_loss}
e^0,\pi_s^0 = \min_{e^*,\pi_s^*}\mathcal{L}_s(x^0; e^*, \pi_s^*).
\end{equation}
Once getting the optimized $\theta^0 = (e^0, \pi_s^0, \pi_m^*)$, TTT then makes a prediction with the updated parameters as $z^0=\pi_m^*(e^0(x^0))$.
Since TTT updates the model in an online manner, the model first is initialized with $(e^{t-1}, \pi_s^{t-1}, \pi_m^*)$ at time $t$ ($t>0$), and then uses the updated parameters $\theta^t=(e^{t}, \pi_s^t, \pi_m^*)$ to make a prediction.\footnote{The inference process we introduce here is the TTT-online version. Besides, the TTT-offline version always initializes the model with $\theta^*$ when meeting each test data. We focus on the online version, whose performance has been proven that is much better than the offline version~\cite{SWLMEH20}.}

\mypara{DUA~\cite{MMPB22}}
DUA is a newly proposed TTA method.
Compared to TTT, DUA is more lightweight because it requires no back-propagation process and only updates $<1\%$ parameters of the target model.
Specifically, DUA aims to update the normalization statistics of the BN layers in an unsupervised manner and fix all remaining parameters of the target model.
Here we first introduce the BN layers and then explain the detailed updating rule of DUA.

Batch normalization (BN) layers are widely used components in modern deep neural networks. 
They are applied to stabilize the training process by reducing internal covariate shift~\cite{IS15}.
In particular, once the training process is finished, the output of BN layers can be formulated as
\begin{equation}
\label{eq:BN}
{\rm BN}(x;\mu_{ori},\sigma_{ori}^2,\gamma_{ori},\beta_{ori}) = \frac{x-\mu_{ori}}{\sqrt{\sigma_{ori}^2+\varepsilon}}\cdot \gamma_{ori} + \beta_{ori},
\end{equation}
where $\mu_{ori}=\mathbb{E}[\mathcal{D}_{ori}]$ and $\sigma_{ori}^2={\rm Var}[\mathcal{D}_{ori}]$ are normalization statistics of the original domain, $\gamma_{ori}$ and $\beta_{ori}$ are the affine transformation parameters learned via back propagation during training process.
These parameters are all fixed at test time in the traditional inference paradigm.
However, recent work has found that recalculating normalization statistics in the transfer domain (e.g., test-time normalization (TTN)~\cite{SREBBB20}) can improve the robustness of the target model.
Therefore, DUA continues to update the normalization statistics in a momentum-updating manner.

An illustration of DUA is shown in \autoref{fig:workflow_DUA}.
The main intuition of DUA is to adapt the target model by aligning the activation distribution between the original domain and the transfer domain.
The adaptation rule of DUA is as follows: Given a test sample $x^t$, DUA first expands it to a small batch $x^t=\{x_i^t\}_{i=1}^{B_{dua}}$ through data augmentation including random horizontal flipping, random cropping, and rotation, where $x_i^t$ is an augmented version of $x^t$.
Then, DUA updates the values of the normalization statistics using \autoref{eq:dua_normalization_stats}.
\begin{equation}
\label{eq:dua_normalization_stats}
\begin{aligned}
\hat \mu_t &= (1-(\rho_t+\xi))\cdot \hat \mu_{t-1} +(\rho_t+\xi)\cdot \mu_t, \\
\hat \sigma_t^2 &= (1-(\rho_t+\xi))\cdot \hat \sigma_{t-1}^2 +(\rho_t+\xi)\cdot \sigma_{t}^2,
\end{aligned}
\end{equation}
where $\mu_t=\mathbb{E}[x^t]$, $\sigma_t^2={\rm Var}[x^t]$ are the current running normalization statistics.
We use $\hat{\mu}_t$, $\hat{\sigma}_t$ to denote the updated statistics where $\hat{\mu}_0=\mu_{ori}$, $\hat{\sigma}_0^2=\sigma_{ori}^2$.
In addition, $\rho_t$ is a decaying momentum term defined in \autoref{eq:dua_decaying}.
\begin{equation}
\label{eq:dua_decaying}
\rho_t = \rho_{t-1} \cdot w, ~\rho_0=0.1.
\end{equation}
There are two hyperparameters in DUA: 
$w \in (0,1)$ controls the decay of $\rho$, and $\xi \in (0, \rho_0)$ defines the lower bound of the momentum.

\mypara{TENT~\cite{WSLOD21}}
Compared to TTT, TENT does not require an auxiliary task but regards prediction confidence as a self-supervision signal.
Similar to DUA, TENT also only adjusts the parameters in the BN layers, and all other parameters of the target model are frozen.
However, besides updating the normalization statistics $\mu$ and $\sigma^2$, TENT updates the affine parameters, $\gamma$ and $\beta$, as well.

\autoref{fig:workflow_TENT} shows an illustration of TENT.
The intuition behind TENT is straightforward. 
Regularizing entropy during training can assist domain adaptation~\cite{GB04}, TENT demonstrates that minimizing entropy during inference can further improve the model's adaptability.
Concretely, given a batch of test samples $x^t=\{x_i^t\}_{i=1}^{B_{tent}}$, TENT updates $\gamma$ and $\beta$ by minimizing the Shannon entropy~\cite{S01} as
\begin{equation}
\begin{aligned}
\gamma_t &\leftarrow \gamma_{t-1} - \partial \mathcal{L}_{tent}(x^t)/\partial \gamma_{t-1},\\
\beta_t & \leftarrow \beta_{t-1} - \partial \mathcal{L}_{tent}(x^t)/\partial \beta_{t-1},
\end{aligned}
\end{equation}
where $(\gamma_0,\beta_0)$ = $(\gamma_{ori},\beta_{ori})$, and $\mathcal{L}_{tent}(\cdot)$ is defined in \autoref{eq:tent_loss}. 
\begin{equation}
\label{eq:tent_loss}
\mathcal{L}_{tent}(f(x^t))=-\frac{1}{{B_{tent}}}\sum_{i=1}^{{B_{tent}}}\sum_{j=1}^Cp(j|x_i^t)\log p(j|x_i^t).
\end{equation}
Specifically, TENT combines entropy minimization with test-time normalization~\cite{SREBBB20}.
It replaces the normalization statistics of the training data with the current statistics as $\mu_t=\mathbb{E}[x^t]$, $\sigma_t^2={\rm Var}[x^t]$.
Then, it uses the updated parameters $\{\mu_t$, $\sigma_t^2$, $\gamma_{t-1}$, $\beta_{t-1}\}$ to make a prediction on $x^{t}$. 
Note that TENT uses one forward process for efficiency, in turn, $\gamma_{t}$, $\beta_{t}$ will be used for predicting $x^{t+1}$.

\mypara{RPL~\cite{RSPEGBBB22}}
As \autoref{fig:workflow_TENT} shows, RPL improves upon TENT by updating the affine parameters based on the prediction confidence, which is treated as the self-supervision label.
However, the entropy-based loss functions are sensitive to label noise~\cite{ZBHRV17,ZS18}.
Therefore, RPL uses generalized cross entropy (GCE) to adapt the target model on the transfer domain.

Concretely, given a batch of test data $x^t=\{x_i^t\}_{i=1}^{B_{rpl}}$, RPL updates the affine parameters using \autoref{eq:RPL_update}.
\begin{equation}
\label{eq:RPL_update}
\begin{aligned}
\gamma_t &\leftarrow \gamma_{t-1} - \partial \mathcal{L}_{rpl}(f(x^t))/\partial \gamma_{t-1},\\
\beta_t & \leftarrow \beta_{t-1} - \partial \mathcal{L}_{rpl}(f(x^t))/\partial \beta_{t-1},
\end{aligned}
\end{equation}
where $\mathcal{L}_{rpl}$ is formulated by \autoref{eq:gce}.
\begin{equation}
\label{eq:gce}
\mathcal{L}_{rpl}(f(x^t))=\frac{1}{B_{rpl}}\sum_{i=1}^{B_{rpl}}q^{-1}(1-p(\Psi|x_i^t)^q).
\end{equation}
Here $\Psi=\mathop{\arg\max}_{j=1,...,C}p(j|x_i^t)$, and $q\in (0,1]$ is a hyperparameter.
From~\autoref{eq:gce} we can observe that $\lim_{q \rightarrow 0}\mathcal{L}_{rpl}(\cdot)$ is the cross entropy loss (which has implicit weighting
scheme~\cite{ZS18}) and $\lim_{q \rightarrow 1}\mathcal{L}_{rpl}(\cdot)$ is the MAE loss (which is noise-robustness~\cite{GKS17}).

\subsection{Poisoning Attacks}
\label{sec:poisoning_attacks_background}

\mypara{Overview}
Poisoning attacks are one of the most dangerous threats to the ML models~\cite{YWLC17,CT22}.
These attacks assume that the adversary can inject poisoned samples into the ML model's training dataset.
The assumption is reasonable as the training datasets of ML models are usually collected from the Internet and it is hard to detect the poisoned samples manually given the size of the dataset.
In poisoning attacks, the adversary's goal is to degrade model performance on a validation dataset $\mathcal{D}_{val}$ through some malicious modifications $\mathcal{A}$ to the training data $\mathcal{D}_{train}$ as
\begin{equation}
\begin{aligned}
\label{poisonattacks}
&\max_{\mathcal{A}}\mathcal{L}(\mathcal{D}_{val}; \theta^*),\\
&~{\rm where}~\theta^*=\underset{\theta}{\mathrm{argmin}} \mathcal{L}(\mathcal{A}(\mathcal{D}_{train});\theta).
\end{aligned}
\end{equation}

\noindent After being trained on the poisoned dataset $\mathcal{A}(\mathcal{D}_{train})$, the model's performance degrades at test time~\cite{PYDSZ212}.

\mypara{Goal}
Poisoning attacks can be broadly grouped into two categories - \emph{untargeted poisoning attacks}~\cite{YWLC17,MBDPWLR17} and \emph{targeted poisoning attacks}~\cite{BNL12,SHNSSDG18}.
The goal of untargeted poisoning attacks is to decline the overall performance of the target model.
The goal of targeted poisoning attacks is to force the target model to perform abnormally on a specific input class.
Backdoor attacks~\cite{PZGXJCW20} are a special case of targeted poisoning attacks where the poisoned target models only misclassify samples containing specific triggers. 

\mypara{Note}
Our work is substantially different from previous poisoning attacks.
We conduct the poisoning attack during the \emph{inference process} while previous work only conducts the poisoning attacks in the \emph{training process}.
Note that we focus on the \emph{untargeted} poisoning attacks in this paper.

\subsection{Adversarial Attacks}

\mypara{Overview}
Adversarial attacks aim to find a perturbed example $x^{adv}$ around $x$ which can be misclassified by the model.
Such $x^{adv}$ is called an adversarial example.
Find such adversarial examples can be formulated as the following constrained optimization problem:
\begin{equation}
\begin{aligned}
x^{adv}&=\arg\max_{x'} \mathcal{L}(x',y;\theta), \\
&s.t. ~||x'-x||_p \leq \epsilon,
\end{aligned}
\end{equation}
where $y$ is the ground-truth label, $||\cdot||_p$ is the $\ell_p$-norm, and $\mathcal{L}(\cdot)$ is usually the cross-entropy loss.
Fast Gradient Sign Method (FGSM)\cite{GSS15} is a widely used method to find adversarial examples, it can be formulated by 
\begin{equation}
\label{eq:fgsm}
x^{adv} = x + \epsilon \cdot {\rm sign}(\nabla_x \mathcal{L}(f(x), y)).
\end{equation}

\mypara{DIM~\cite{XZZBWRY19}}
As \autoref{eq:fgsm} shows, FGSM needs white-box access to the model to find adversarial examples.
However, the adversaries may only have black-box access.
Therefore, transfer-based adversarial attacks are proposed to generate adversarial examples against a surrogate model which can also misclassify the remote target model~\cite{DLPSZHL18,DFYPSXZ19,XZZBWRY19}.
Among them, Diverse Input-FGSM (DIM)~\cite{XZZBWRY19} is the state-of-the-art attack method.
In brief, DIM applies random resizing with probability $p$ to the input $x$ to alleviate the overfitting of the adversarial examples on the surrogate model to improve the transferability (i.e., $T(x,p)$ in Algorithm~\ref{algTePAs}).
In our paper, we integrate DIM to generate our poisoned samples.
Note that any advanced transfer-based attacks can be integrated into our algorithm.

\section{Threat Model}

\mypara{Adversary's Goal}
We assume that the target models (i.e., the models which the adversaries aim to attack) make predictions following the online TTA paradigm.
For example, if the target model uses TTT to adjust the parameters, then we denote the target model as TTT-model.
The adversary's goal is to degrade the target model's performance by nudging the model in a ``wrong direction'' by feeding poisoned samples at test time.
Meanwhile, the benign samples
uploaded by legitimate users and the poisoned samples fed by the adversaries are in the same pipeline, which means multiple users concurrently use and update
the parameters of the target model.
We use a fixed evaluation dataset to monitor the changes in model performance.

\mypara{Adversary's Knowledge}
We assume that the adversary has three pieces of knowledge:
(i) They know which TTA method the target model uses. 
This assumption is realistic since TTA methods should be publicly available so that they can be rigorously vetted for security before deployment like cryptanalysis. 
In addition, systems may eventually converge towards certain SOTA public TTA methods.
(ii) The adversary knows the API where the legitimate users upload the benign samples, hence they can upload the poisoned samples to covertly poison the target model.
(iii) They may collect a surrogate dataset that comes from a similar distribution of the target model's training dataset.
Notably, different from the previous poisoning attacks, the adversaries do not know the architecture or training hyperparameters of the target model.
(iv) They are unable to tamper with the training data or intervene in the target model's training process.
(v) They also do not have access to the model parameters of the target model at any time.
(vi) They cannot control the order of the poisoned samples reaching the target model, e.g., the target TTA model may have been updated by an unknown number of test samples.

\mypara{Adversary's Capability}
The surrogate dataset enables the adversary to train a surrogate model, which can then be utilized to generate poisoned samples.
However, it should be noted that the adversary cannot obtain information about the gradient of the loss from the target model and can only resort to transfer-based adversarial attacks, as demonstrated in previous works such as~\cite{XZZBWRY19,MFWJZLZLBW22}. 
That is, they can only feed these poisoned samples to the target TTA-online model.
Moreover, since the test data come to the target TTA models ``point-by-point'' or ``batch-by-batch'', the adversaries can set up the poisoned samples in advance to mix them with the benign samples.

\mypara{Attack Challenge}
TePAs lead to the following non-trivial challenges.
Previous poisoning attacks assume that the target model is trained on fixed poisoned training data or its training process is controlled by the adversary.
However, none of the assumptions are valid in the case of TTA.
First, the adversary cannot poison the training data and does not control the training process of the target TTA model.
They only have query access to the target TTA model.
Secondly, a TTA model updates its parameters for each query sample once deployed.
Even if the adversaries possess knowledge of the training data and process, such as hyperparameters like training epochs and batch size, they cannot assume that the target model is newly trained. 
Finally, the adversaries must take the budget (i.e., the number of poisoned samples) into consideration to stay stealthy and avoid detection.
Nevertheless, we show in our evaluation (see \autoref{sec::evaluation}) that our attacks are effective with a limited amount or limited fraction of poisoned samples.

\section{Attack Methodology}

\subsection{Attack Overview}
\label{attack_overview}

In general, TePAs consist of three steps -  surrogate model training, poisoned sample generation, and target model poisoning.
The overall workflow of TePAs is illustrated in \autoref{fig:workflow_attack}.

\begin{itemize}
\item  \textbf{Surrogate Model Training.}
The goal of this step is to construct a surrogate model $f_s$ with the surrogate dataset as a stepping stone to launch the attack.
It is essential to note that the adversary operates under the assumption that the target model's architecture is unknown and the distribution of the shadow dataset resembles that of the target model's training dataset.
Moreover, the surrogate model's training process is independent of the target model and does not need any supervision information from the target model, such as query results. 
\item \textbf{Poisoned Sample Generation.}
In this step, we introduce \texttt{PoiGen}, a poisoned sample generation framework.
The details of \texttt{PoiGen} are summarized in Algorithm~\ref{algTePAs}.
The goal of \texttt{PoiGen} is to create a poisoned sample $x'$ from a clean seed image $x_{in}$ that aims to decrease the inference performance of the target model.
Depending on the target TTA method $\mathcal{A}$, \texttt{PoiGen} uses different generation strategies (e.g., different loss function $\mathcal{L}_{poi}$) to generate poisoned samples with stronger transfer properties.
We stress that the poisoned sample generation process does not interact with the target model, which enhances the stealthiness of our attack.
Also, \texttt{PoiGen} allows the attacker to plug in different advanced transfer-based adversarial attack algorithms.
\item \textbf{Target Model Poisoning.}
The goal of this step is to employ different poisoning strategies to deliver the poisoned samples to the target model.
In this step, the adversary must take various factors, such as the budget (i.e., the number of poisoned samples) and the order (i.e., how the poisoned samples and the clean samples are mixed at inference time) into consideration to stay stealthy and avoid detection.
\end{itemize}

In conclusion, the core process of TePAs is \texttt{PoiGen} (poisoned sample generation).
To attack different TTA models, \texttt{PoiGen} chooses different loss functions $\mathcal{L}_{poi}$ and attack strategies.
We outline how \texttt{PoiGen} generates poisoned samples for four different TTA models in the rest of this section.
Note that since poisoning strategies are tightly coupled with performance evaluation, we defer the description of poisoning strategies in \autoref{section:exp_poisoning_strategy}.

\subsection{TePA Against TTT}
\label{sec:TePAs_method_ttt}

Recall that in the inference process, TTT fine-tunes the feature extractor $e(\cdot)$ and the SSL task branch head $\pi_s(\cdot)$ through the rotation prediction loss $\mathcal{L}_s$.
Our intuition is that if $e(\cdot)$ and $\pi_s(\cdot)$ learn the wrong information about the rotation from the test samples together, the feature extractor will be guided in incorrect directions, causing the model to lose the information learned from the training data. 
Previous work~\cite{DGZ21} also shows that the rotation prediction accuracy is strongly linked to the classification accuracy of the primary task.
Therefore, the adversary can generate poisoned samples according to the auxiliary loss $\mathcal{L}_s$ by adversarial attacks.
Specifically, the generated noise should maximize $\mathcal{L}_s$ in each angle.
Inspired by Universal Adversarial Perturbations~\cite{MFFF17}, given one original sample, the adversary may find a universal perturbation for all its rotations.

\begin{algorithm}[t]
\caption{\texttt{PoiGen}}
\label{algTePAs}
\SetKwInput{KwInput}{Input}               
\SetKwInput{KwOutput}{Output}              
\DontPrintSemicolon
\KwInput{Seed image $x_{in}$, surrogate model $f_s$, the target TTA method $\mathcal{A}$, loss function $\mathcal{L}_{poi}$, the perturbation budget $\epsilon$, updating step $\alpha$;}
\KwOutput{Poisoned sample $x'$;}
\SetKwFunction{FMain}{PoiGen}
\SetKwFunction{FMainn}{TePAs}
\SetKwFunction{FDIM}{DIM}

\SetKwProg{Fn}{Def}{:}{}
\Fn{\FDIM{$x$, $y$, $f$, $L$, $\epsilon$}}{
$g=0$; \;
$\mu=1$; \;
$p=0.5$; \;
\For{$j=1$ {\rm to} $I_{adv}$}{
$x = T(x,p)$;\;
\If{y {\rm is not None}}{
$g =\mu \cdot g + \frac{\nabla_{x_{in}} L(f(x),y)}{||\nabla_{x_{in}} L(f(x), y)||_1} $; \;
}
\Else{
$g =\mu \cdot g + \frac{\nabla_{x_{in}} L(f(x))}{||\nabla_{x_{in}} L(f(x))||_1} $; \;
}
$x^{adv} = x_{in} + \alpha \cdot {\rm sign}(g)$;\;
$\delta = {\rm Clip}(x^{adv}-x_{in}; -\epsilon, +\epsilon)$; \;
$x = {\rm Clip} (x_{in}+\delta;0,1)$;\;
}
\KwRet $x$.\;
}
\;
\SetKwProg{Fn}{Main function}{:}{\KwRet}
\Fn{\FMain($\mathcal{A}$, $x_{in}$, $f_s$, $\mathcal{L}_{poi}$, $\epsilon$)}{
\If{$\mathcal{A}$ {\rm is TTT}}
{
$x'=x_{in}$;\;
\For {$i = 1$ {\rm to} $I_{iter}$}
{
\For {y' = $1$ {\rm to} $4$}
{
$x_{rot} = {\rm rot90}(x', y')$; \;
$x' = \texttt{DIM} (x_{rot}, y', f_s, \mathcal{L}_{poi}, \epsilon)$;\;
}
}
}
\ElseIf{$\mathcal{A}$ {\rm is TENT or RPL}}{
$x' = \texttt{DIM} (x_{in}, y={\rm None}, f_s, \mathcal{L}_{poi}, \epsilon)$;\;
}
\ElseIf{$\mathcal{A}$ {\rm is DUA}}{
$x' = x + \epsilon \cdot \mathcal{N}(\mu,\sigma^2)$ (See \autoref{eq:adv_dua});\;
}
\KwRet $x'$.\;
}
\end{algorithm}

Specifically, when attacking TTT-models (Line 17-22), \texttt{PoiGen} first sets $\mathcal{L}_{poi}$ as $\mathcal{L}_{s}$ and generates adversarial perturbation for each rotation $x_{rot}$.
Here ${\rm rot90}(x,j)$ stands for rotating the image $x$ by $90 \times j$ degrees (Line 21).
Given an image rotated by 0 degrees and its corresponding rotation label $y'$ is 1, \texttt{PoiGen} first computes the loss $\mathcal{L}_{poi}$ and backpropagate the gradient that maximizes the loss to the original image (Line 8).
Based on this gradient, \texttt{PoiGen} obtains the generated noise $\delta$ (Line 12), which is added to the clean image $x_{in}$, and produces a new poisoned sample $x'$ (Line 13).
This sample is then rotated by 90 degrees (with a corresponding label of 2), and a new perturbation is generated to fool the model on rotation prediction.
The same procedure is followed for rotations of 180 and 270 degrees.
After that, four perturbations are added to the image for four rotations.
To make the generated perturbation more robust, \texttt{PoiGen} introduces a hyperparameter $I_{iter}$ to repeat the whole process.
After generating perturbation, we consider the adversarial examples as poisoned examples  $x'_{ttt}$ against the TTT-models.

\subsection{TePA Against DUA}
\label{sec:TePAs_method_dua}

Recall that DUA uses a momentum updating method to fine-tune the statistical parameters in the BN layers.
However, the statistical parameters calculated from $x^{t+1}$ may differ significantly from $x^{t}$, and this difference may disrupt the adaptation process of the model parameters.
Meanwhile, this disruption is persistent and continues to affect the downstream test data due to the momentum design.
We show that a test sample with Gaussian noise can disrupt the updating process of the statistical parameters.
Thus, the poisoned sample for DUA is:
\begin{equation}
\label{eq:adv_dua}
x'_{dua}=x+\epsilon_{dua} \cdot \mathcal{N}(\mu_{dua},\sigma^2_{dua}),
\end{equation}
where $\mu_{dua}$ and $\sigma^2_{dua}$ control the perturbation distribution and $\epsilon_{dua}$ controls noise intensity.

\mypara{Note}
We can observe that, compared to TePAs against TTT, the generation process of $x'_{dua}$ does not require a surrogate model.
Recall that the adaptation process of DUA does not rely on any SSL tasks, nor is it based on a loss function to adjust the target model.
Therefore, the adversary does not need a surrogate model to launch the gradient-based adversarial attack for generating poisoned samples, which makes the poisoning attacks cheaper and easier.

\subsection{TePA Against TENT \& RPL}
\label{sec:TePAs_method_tent}

\mypara{TePA Against TENT}
Recall that TENT minimizes the entropy of the prediction to adapt the affine parameters of the BN layers, and RPL uses GCE loss instead.
We aim to generate such following perturbation to compel the target model to learn ``wrong information'' from our poisoned samples:
\begin{equation}
\Delta = \mathop{\arg \max}_{\delta} \mathcal{H}(f_s(x+\delta)), {\rm where}~||\delta||_\infty \leq \epsilon_{tent}.
\end{equation}
Here $\mathcal{H}(y)=-\sum_c p_c\log p_c$ is the Shannon entropy.
Since TENT uses prediction logits as the self-supervision signal, we aim to generate such adversarial examples to make the entropy of the logits much larger than normal.
Therefore, \texttt{PoiGen} first sets $\mathcal{L}_{poi}$ as $\mathcal{H}$.
Meanwhile, compared to TePAs against TTT-models, we do not need a label to attack TENT-models, so \texttt{PoiGen} sets $y=$None and uses DIM to maximize $\mathcal{H}$ (Line 10).
Therefore, as shown in Line 24 of Algorithm~\ref{algTePAs}, the final poisoned samples against TENT-models can be formulated as
\begin{equation}
\label{eq:advtent}
x'_{tent}=\texttt{DIM} (x_{in}, y={\rm None}, f_s, \mathcal{H}, \epsilon_{tent}).
\end{equation}

\mypara{TePA Against RPL}
We note that the adversarial examples generated by~\autoref{eq:advtent} can also be used as poisoned samples to poison RPL-models.
This is because as entropy $\mathcal{H}$ increases, $p(\Psi|x)$ decreases, which causes $\mathcal{L}_{rpl}$ to increase as well.
Therefore, we set 
\begin{equation}
x'_{rpl} = x'_{tent}.
\end{equation}

\section{Evaluation}
\label{sec::evaluation}

\begin{figure*}[t]
\centering
\subfloat[TTT \label{fig:utility_tta_a}]{%
\includegraphics[width=0.2\linewidth]{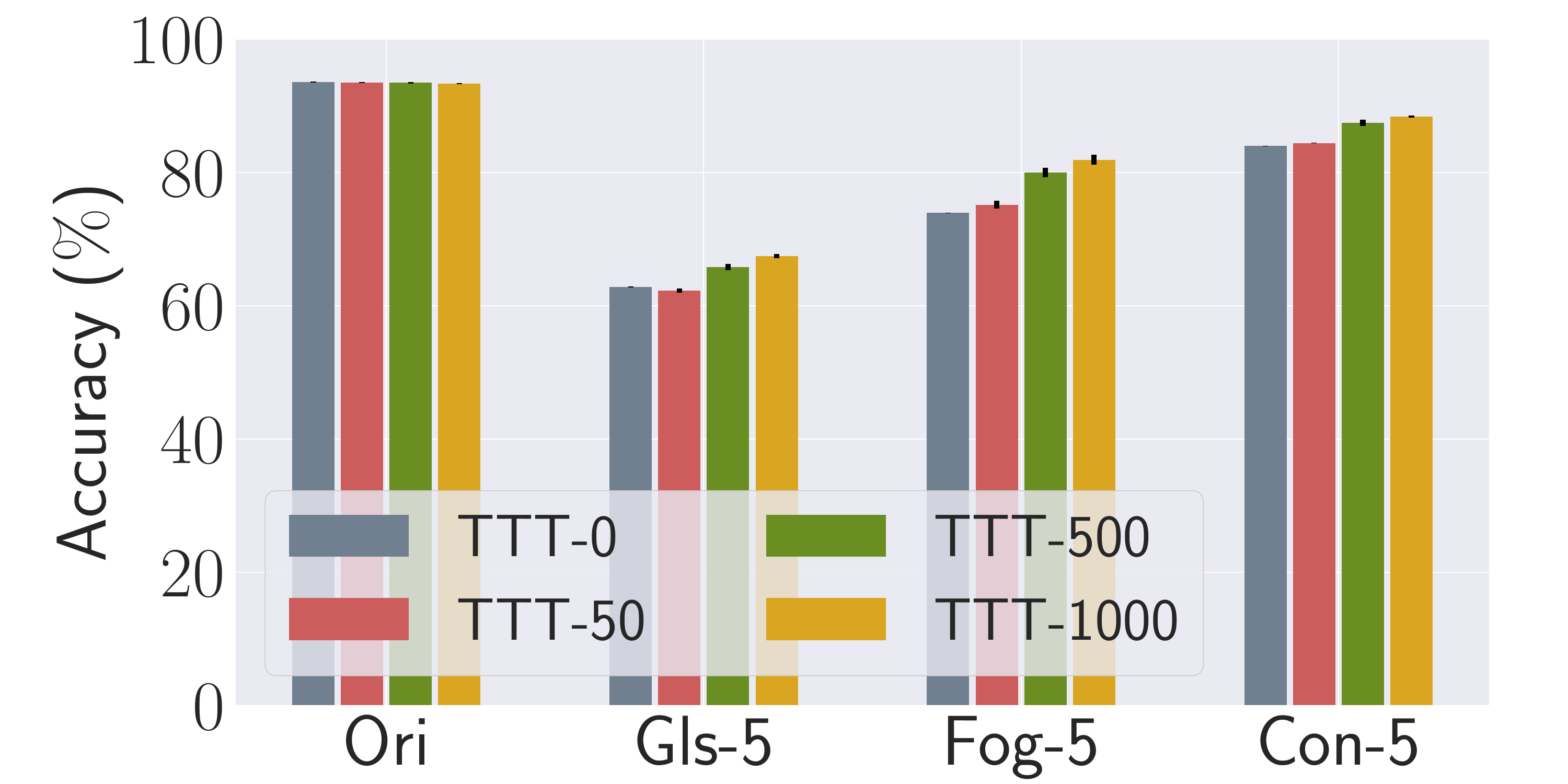}}
\hspace{3mm}
\subfloat[DUA \label{fig:utility_tta_b}]{%
\includegraphics[width=0.2\linewidth]{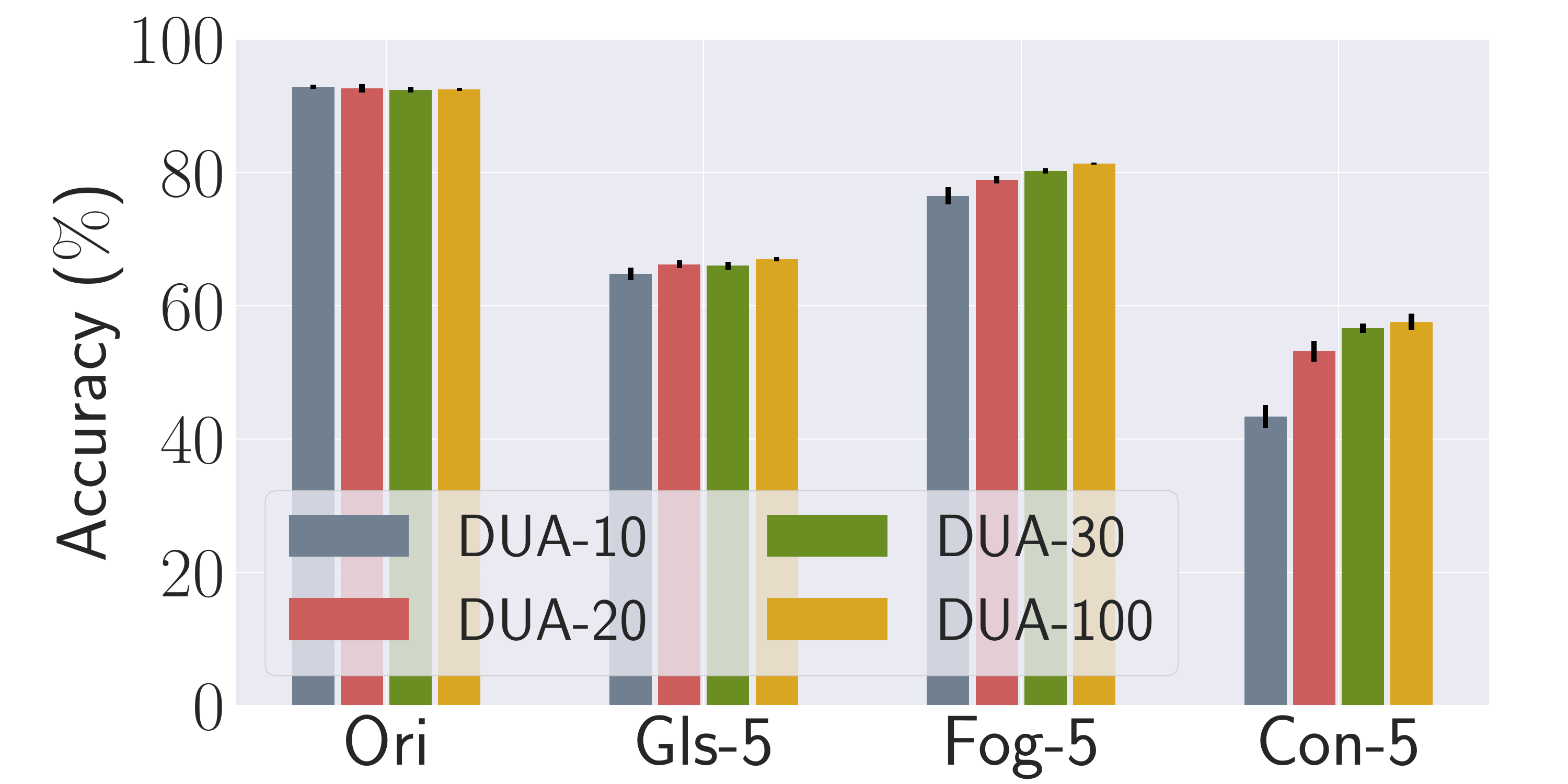}}
\hspace{3mm}
\subfloat[TENT\label{fig:utility_tta_c}]{%
\includegraphics[width=0.2\linewidth]{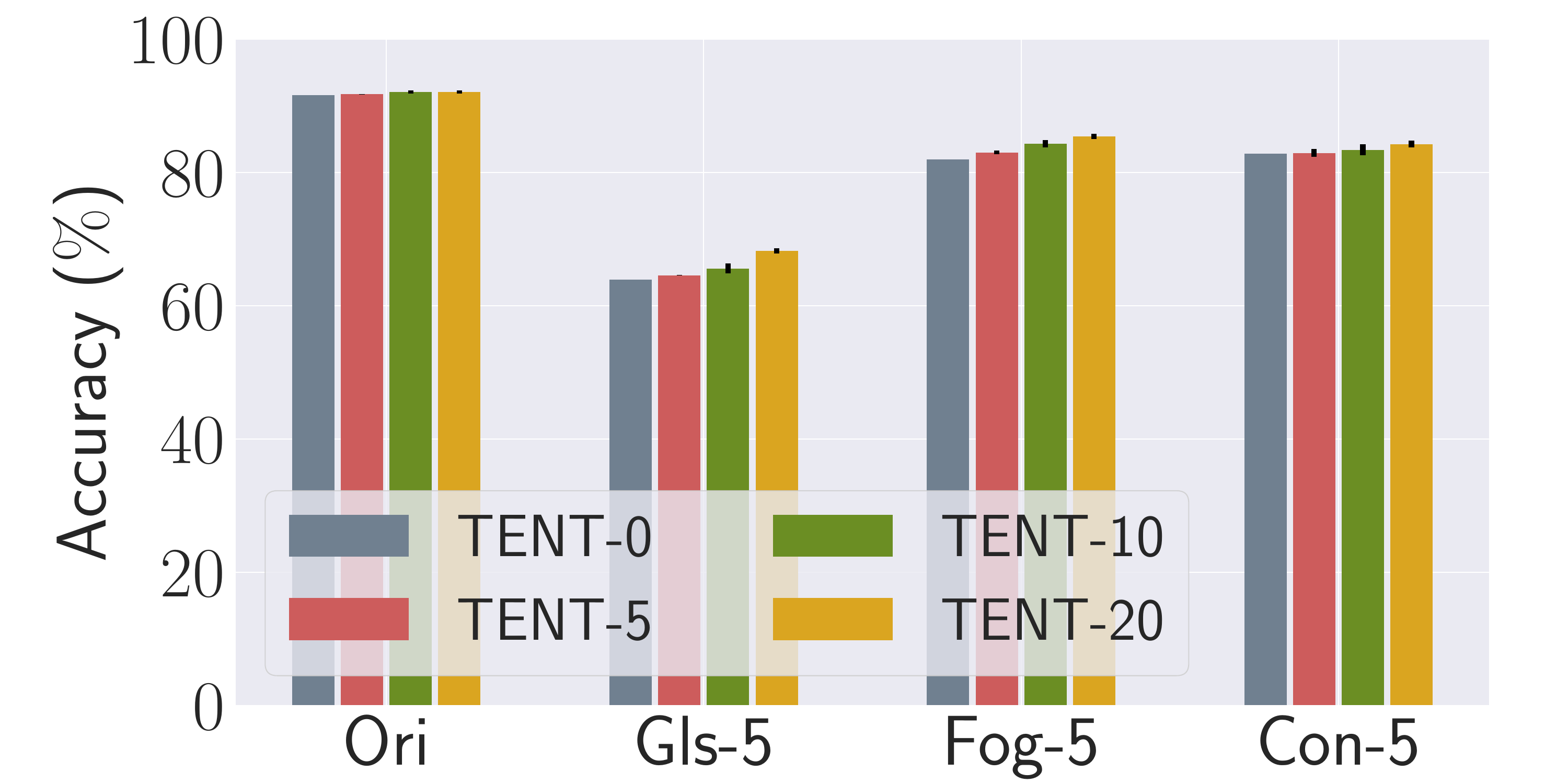}}
\hspace{3mm}
\subfloat[RPL \label{fig:utility_tta_d}]{%
\includegraphics[width=0.2 \linewidth]{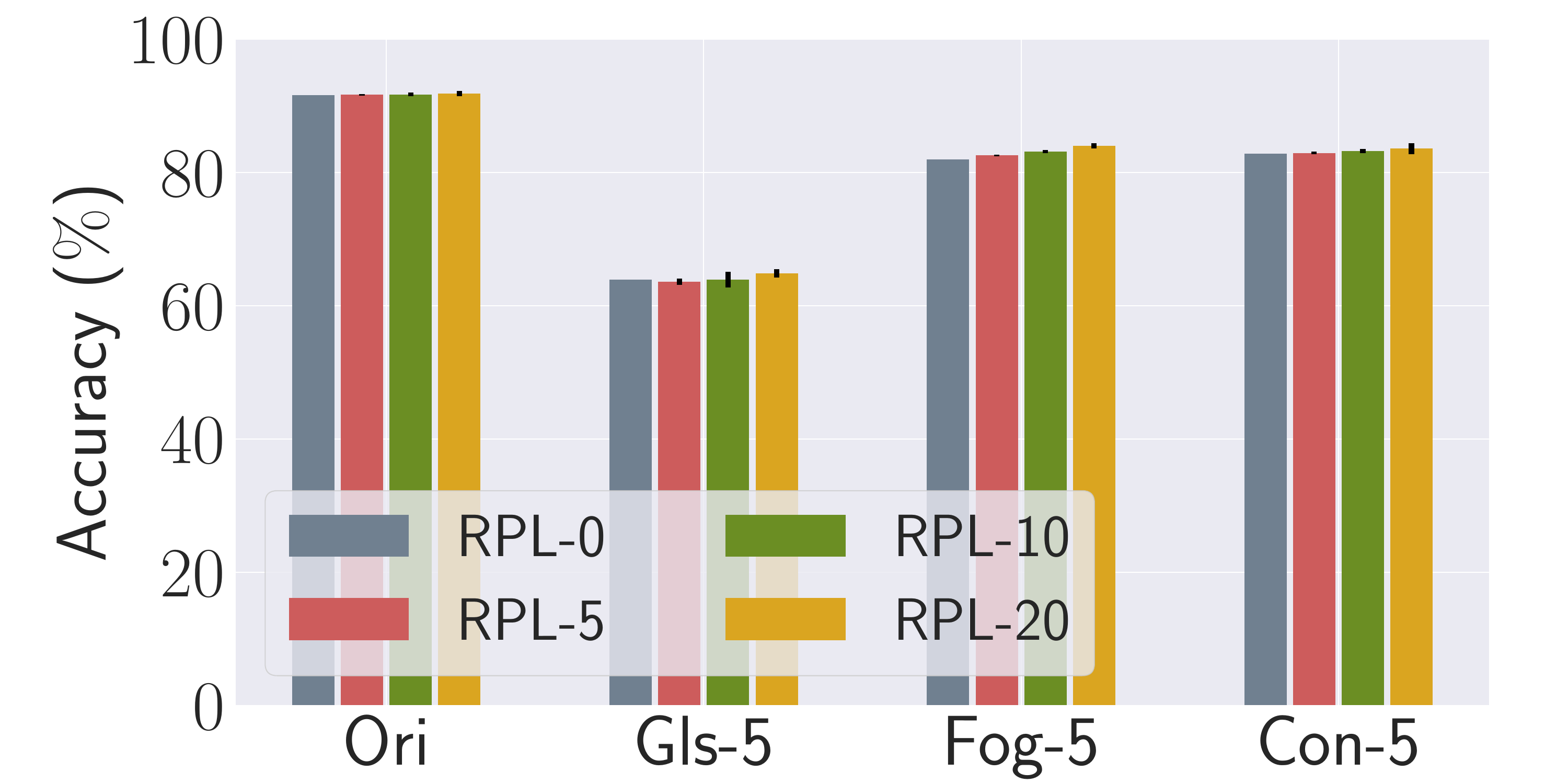}}
\caption{Utility of TTA methods. 
The target model is ResNet-18 trained on CIFAR-10. 
The x-axis represents different evaluation datasets. 
The y-axis represents the prediction accuracy.}
\label{fig:utility_tta} 
\end{figure*}

\subsection{Experimental Setup}

\mypara{Datasets}
We use 5 datasets to conduct our experiments, including CIFAR-10~\cite{CIFAR}, CIFAR-100\cite{CIFAR}, CIFAR-10-C\cite{CIFAR10-C}, CIFAR-100C\cite{CIFAR100-C}, and CINIC-10\cite{DCAS18}.
CIFAR-10/100-C are the corrupted datasets of CIFAR-10/100, which contain 5 different levels of corruption, in which level-5 is the highest corruption severity.
Specifically, we choose four corruptions to evaluate the target model's performance: Ori, Gls-5, Fog-5, and Con-5.
``Ori'' means the original dataset (i.e., the images from CIFAR-10/100), ``Gls-5'' stands for ``Glass blur'' with corruption severity level 5, and ``Fog'' (``Con'') means the ``Fog'' (``Contrast'') corruption.
To train the target models, we choose CIFAR-10 and CIFAR-100 as the training datasets $\mathcal{D}_{t}$.
Meanwhile, the CINIC-10 dataset contains images that are from CIFAR-10 and ImageNet~\cite{DDSLLF09}.
We use the images from ImageNet as the surrogate dataset $\mathcal{D}_{s}$ to train the surrogate model, which makes our poisoning attacks more realistic since $\mathcal{D}_{t} \cap \mathcal{D}_{s} = \phi$. 
The detailed descriptions of the above 5 datasets are shown in \autoref{sec:appendix_dataset}.

\mypara{Target Model}
We use ResNet-18 and ResNet-50 as the architectures of the target models.
We use C10-Res18 (C100-Res18) to denote the ResNet-18 model trained on CIFAR-10 (CIFAR-100).
Finally, we could get 4 target models: C10-Res18/50 and C100-Res18/50, which will be used as the target models for DUA, TENT, and RPL.
We train the above ResNets using public implementations.\footnote{\url{https://github.com/huyvnphan/PyTorch\_CIFAR-10}.}
Meanwhile, recall that the training process of TTT needs two learning tasks.
Therefore, we transform the ResNets into Y-structure.
For instance, we first choose the splitting point in ResNets, the parameters after the splitting point will be copied to form two identical branches: One is used for the main task and the other is for the auxiliary task.
We choose the end of the 4th resblock  (3rd resblock) in ResNets as the splitting point when $\mathcal{D}_{t}$ is CIFAR-10 (CIFAR-100).
Consequently, we get 4 target models as TTT-models: C10-Res18/50@Y4 and C100-Res18/50@Y3, where Y3/Y4 means the splitting point is the end of the 3rd/4th resblock.
We run the official training implementation\footnote{\url{https://github.com/yueatsprograms/ttt\_cifar\_release}.} to train the target TTT-models. 
Meanwhile, we replace BN with Group Normalization (GN)~\cite{WH18} in the ResNets following~\cite{SWLMEH20} for better performance.

\mypara{Surrogate Model}
As mentioned above, we use the images from ImageNet (resized to $32\times 32 \times 3$) as our surrogate dataset to train surrogate models.
When the target models are TTT-models, we choose Res18@Y3 as the architecture of the surrogate model.
Otherwise, when the target models are TENT-/RPL-models, we choose VGG-11 as the surrogate model, which is trained by a different public implementation\footnote{\url{https://github.com/kuangliu/pytorch-cifar}.} than training the target models.

\mypara{Hyperparameters of TTA Methods}
TTT uses an SGD optimizer to update the parameters of the target models for 1 epoch, the learning rate is 0.001.
For DUA, we set $\omega=0.94$, $\xi=0.005$, and $B_{dua}=64$.
For TENT and RPL, the batch size for the coming test samples is $B_{tent}=B_{rpl}=200$, and they both use an SGD optimizer with a momentum factor of 0.9 to update the affine parameters for 1 epoch.
Meanwhile, we set $q=0.8$ for RPL.

\mypara{Evaluation Metric}
To monitor the prediction ability of the target model $f^t$ promptly, we use an evaluation dataset $\mathcal{D}_e$ which contains 1,000 evaluation samples to evaluate the model's performance. 
The top-1 prediction accuracy (denoted as Acc) is the performance indicator.
We follow the evaluation methods from the official implementation of the TTA methods.
For instance, TTT,\footnote{\url{https://github.com/yueatsprograms/ttt\_cifar\_release}.} TENT,\footnote{\url{https://github.com/DequanWang/tent}.} and RPL\footnote{\url{https://github.com/bethgelab/robustness}.} all adapt the models once there comes new evaluation data.
Therefore, we input evaluation data to adjust the model first and then make predictions.
After each prediction, we reset the model to $f^t$ for the next prediction.
However, DUA adjusts the model in an online manner with a few unlabeled samples and then freezes the model to make predictions.\footnote{\url{https://github.com/jmiemirza/DUA}.}
Therefore, we also freeze the model when we evaluate it on our 1,000 evaluation samples.
Note that the evaluation samples in $\mathcal{D}_e$ suffer from the same corruption, and we will use different corrupted $\mathcal{D}_e$ to evaluate the target model's performance, i.e., Ori, Gls-5, Fog-5, and Con-5.

\mypara{Hyperparameters of TePAs}
We set the perturbation budget $\epsilon=\sfrac{32}{255}$ ($\ell_\infty$-norm) for default.
For TePA against TTT-models, we set $I_{iter}=3$ and $I_{adv}=20$. 
Meanwhile, we use a staged updated step strategy.
For instance, $\alpha=\sfrac{4}{255}$ when $I_{adv}\in [0,10)$, and $\alpha=\sfrac{2}{255}$ when $I_{adv}\in [10,15)$, otherwise,  $\alpha=\sfrac{1}{255}$.
For TePA against DUA-models, we set $\mu_{dua}=0.0$ and $\sigma^2_{dua}=0.8$.
For TePA against TENT-models and RPL-models, we set $I_{adv}=200$ and $\alpha=\sfrac{1}{255}$.

\subsection{Utility of Frozen Target Model}
\label{sec::TTT_ut_source}

We first evaluate the prediction ability of the frozen target models.
We use four kinds of evaluation datasets - Ori, Gls-5, Fog-5, and Con-5 - to evaluate the utility of our eight target models.
The results are shown in \autoref{tab:soure_model_utility}.
We observe the following two phenomenons:
(i) Deep neural networks (DNNs) cannot be robust enough on distribution shifts.
Take C10-Res18 as an example; when there is no corruption on evaluation samples, the Acc is $93.00\%$.
However, the model's performance drops to $58.10\%$ ($64.80\%$) when $\mathcal{D}_e$ is Gls-5 (Fog-5).
(ii) Y-structured DNNs are more robust than naive DNNs\cite{HD19}.
For example, the Acc of C10-Res18 on Con-5 is $19.20\%$, while the Acc is  $83.60\%$ for C10-Res18@Y4.
However, Y-structured DNNs are still not robust enough on all corrupted $\mathcal{D}_e$, e.g., the Acc of C10-Res18@Y4 is $61.90\%$ on Gls-5, which is $\sim 30\%$ lower than that on Ori.
Therefore, we need TTA methods to further improve the model's performance on distribution shifts.
Note that the results in \autoref{tab:soure_model_utility} are used as the baseline when discussing the enhancement capabilities of TTA methods and the attack performance of TePAs.

\begin{table}[ht]
\caption{The utility of the frozen target model ($\%$).}
\label{tab:soure_model_utility}
\centering
\customTableFont
\begin{tabular}{llcccc}
\toprule
\multirow{2}{*}{Dataset} &\multirow{2}{*}{Target Model} & \multicolumn{4}{c}{Acc} \\ \cmidrule(r){3-6}
~ & ~ & Ori & Gls-5 & Fog-5 & Con-5 \\ \midrule
\multirow{4}{*}{CIFAR-10} & C10-Res18@Y4  & 93.70 & 61.90 & 71.40 & 83.60   \\
~ & C10-Res50@Y4  & 92.80 & 56.60 & 68.00 & 78.50 \\
~&C10-Res18     & 93.00 & 58.10 & 64.80 & 19.20\\
~ &C10-Res50     & 94.20 & 62.60 & 70.80 & 24.90 \\
\midrule
\multirow{4}{*}{CIFAR-100} &C100-Res18@Y3 & 71.40 & 20.90 & 41.40 & 48.70   \\
~ &C100-Res50@Y3 & 65.20 & 24.70 & 31.40 & 30.80 \\
~ &C100-Res18    & 73.50 & 24.60 & 32.60 & 11.50 \\
~ &C100-Res50    & 76.20 & 25.50 & 38.30 & 12.30 \\ 
\bottomrule
\end{tabular}
\end{table}

\begin{figure*}[!t]
\centering
\subfloat[C10-Res18@Y4 \label{fig:attack2num_ttt_a}]{%
\includegraphics[width=0.2\linewidth]{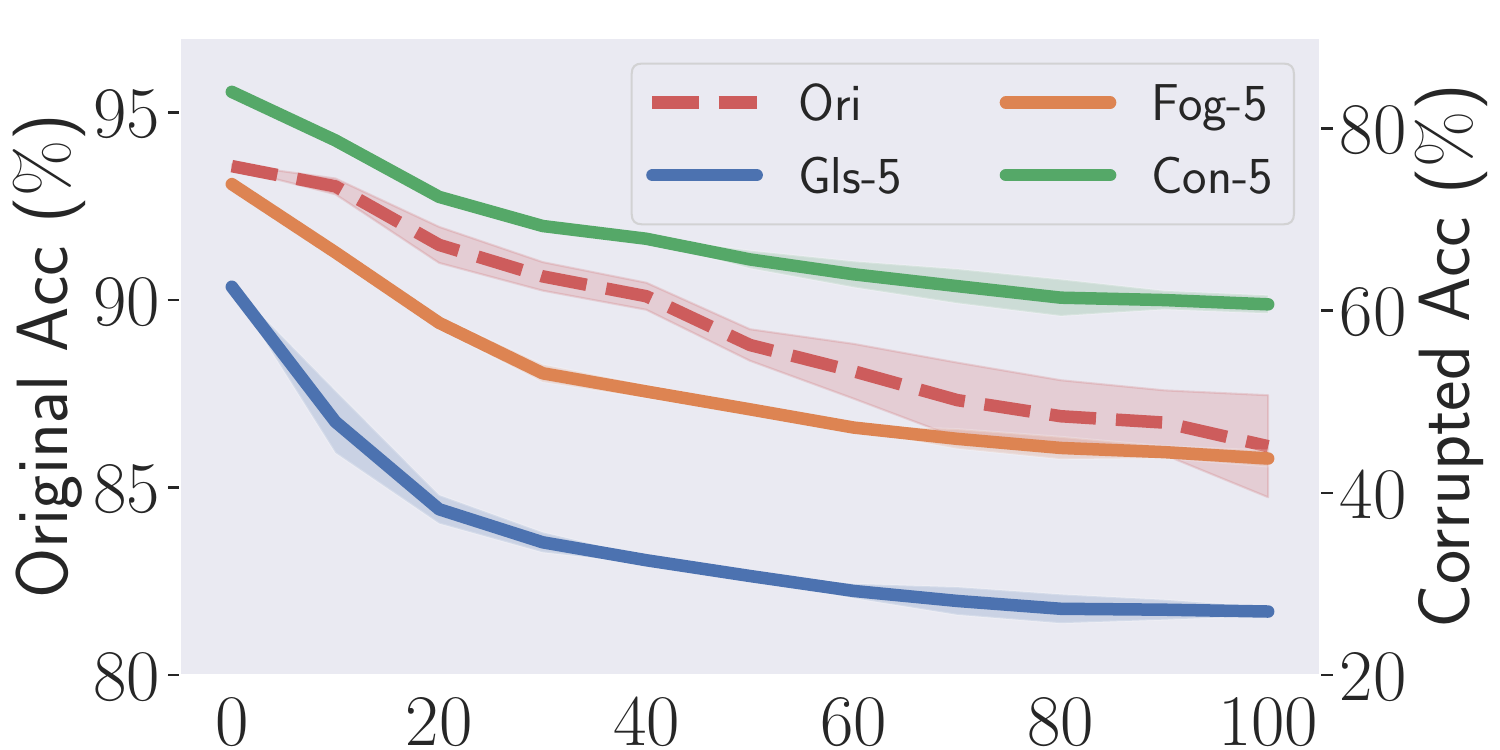}}
\hspace{3mm}
\subfloat[C10-Res50@Y4 \label{fig:attack2num_ttt_b}]{%
\includegraphics[width=0.2\linewidth]{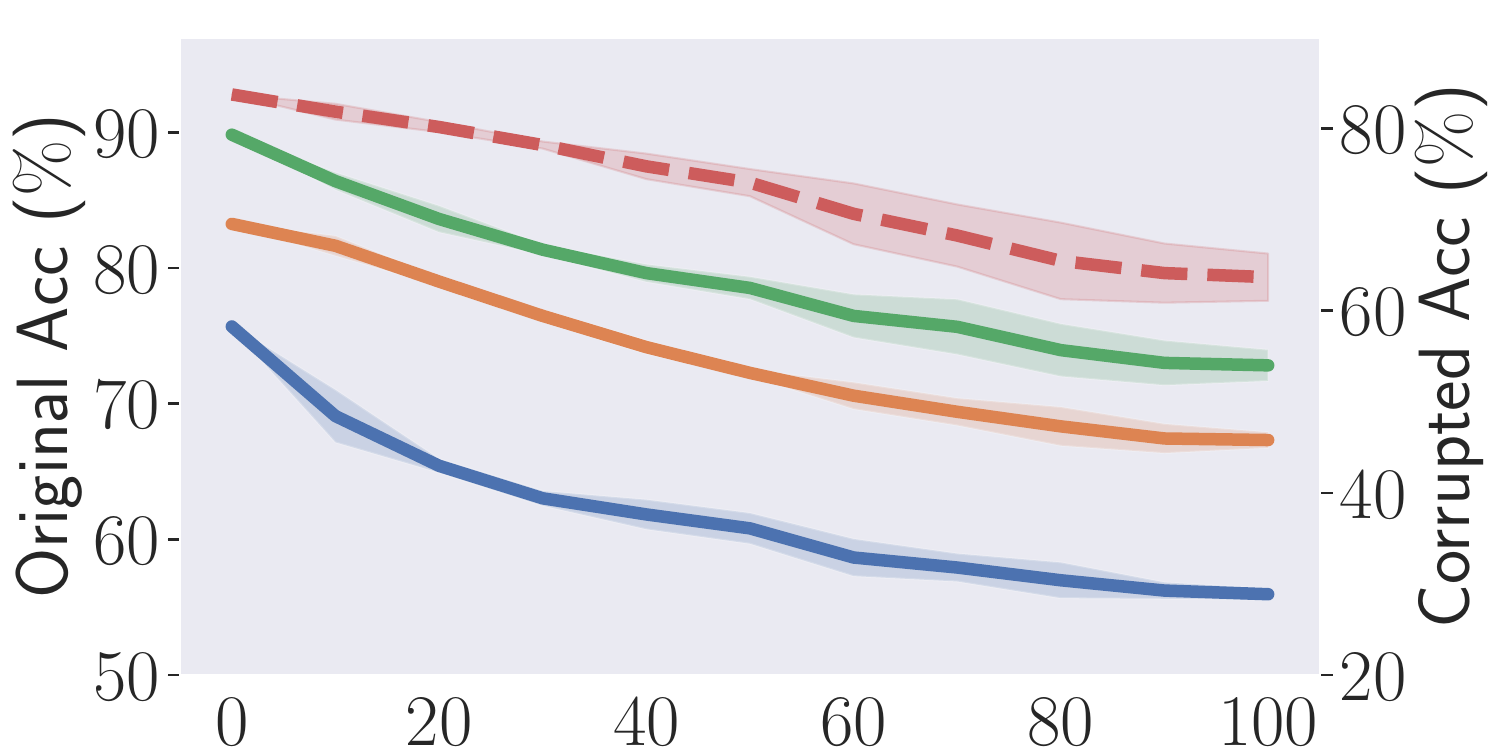}}
\hspace{3mm}
\subfloat[C100-Res18@Y3\label{fig:attack2num_ttt_c}]{%
\includegraphics[width=0.2\linewidth]{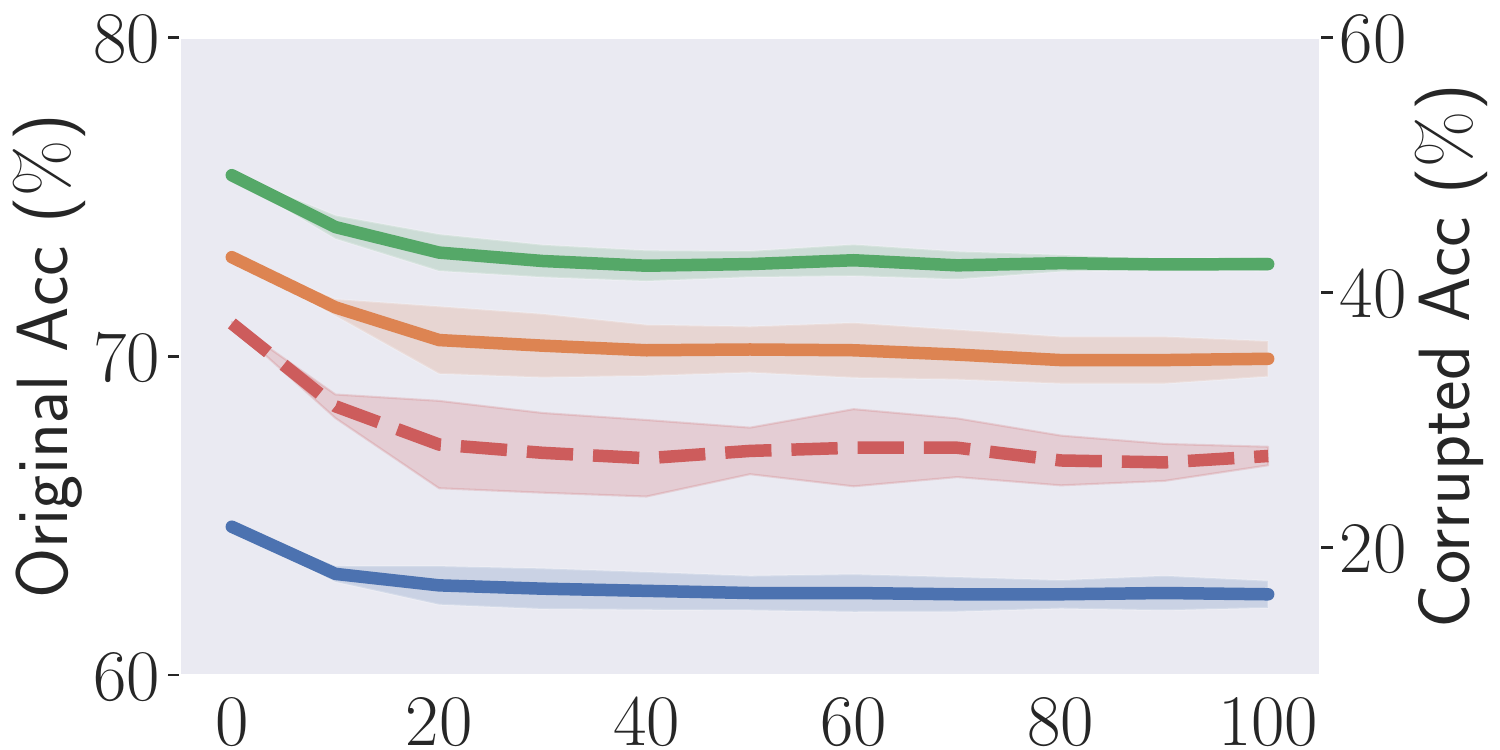}}
\hspace{3mm}
\subfloat[C100-Res50@Y3 \label{fig:attack2num_ttt_d}]{%
\includegraphics[width=0.2\linewidth]{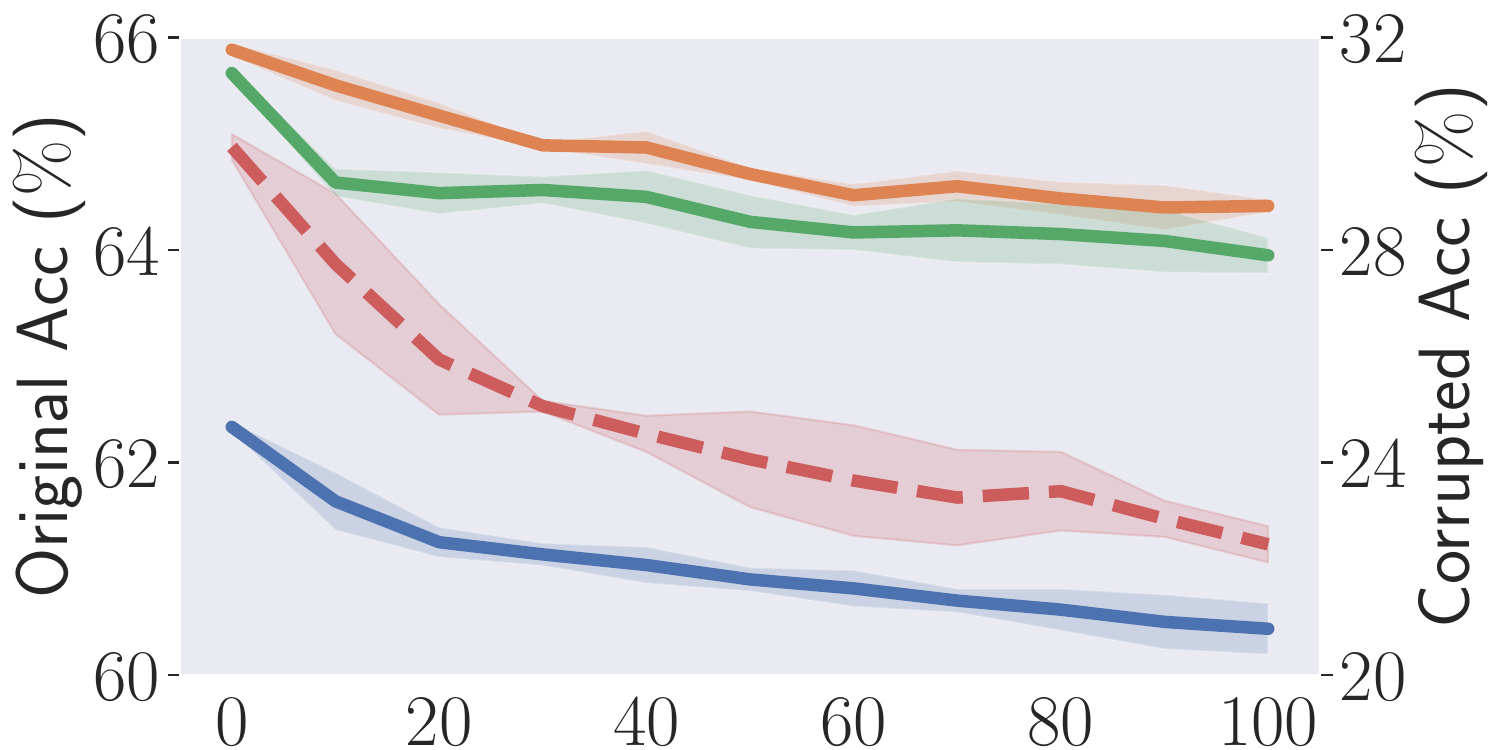}}
\caption{TePAs Against TTT-models. 
The left y-axis and the right y-axis represent the prediction accuracy on the original and corrupted evaluation datasets, respectively. 
The x-axis represents the number of poisoned samples.}
\label{fig:attack2num_ttt} 
\end{figure*}

\begin{figure*}[!t]
\centering
\subfloat[C10-Res18 \label{fig:attack2num_dua_a}]{%
\includegraphics[width=0.2\linewidth]{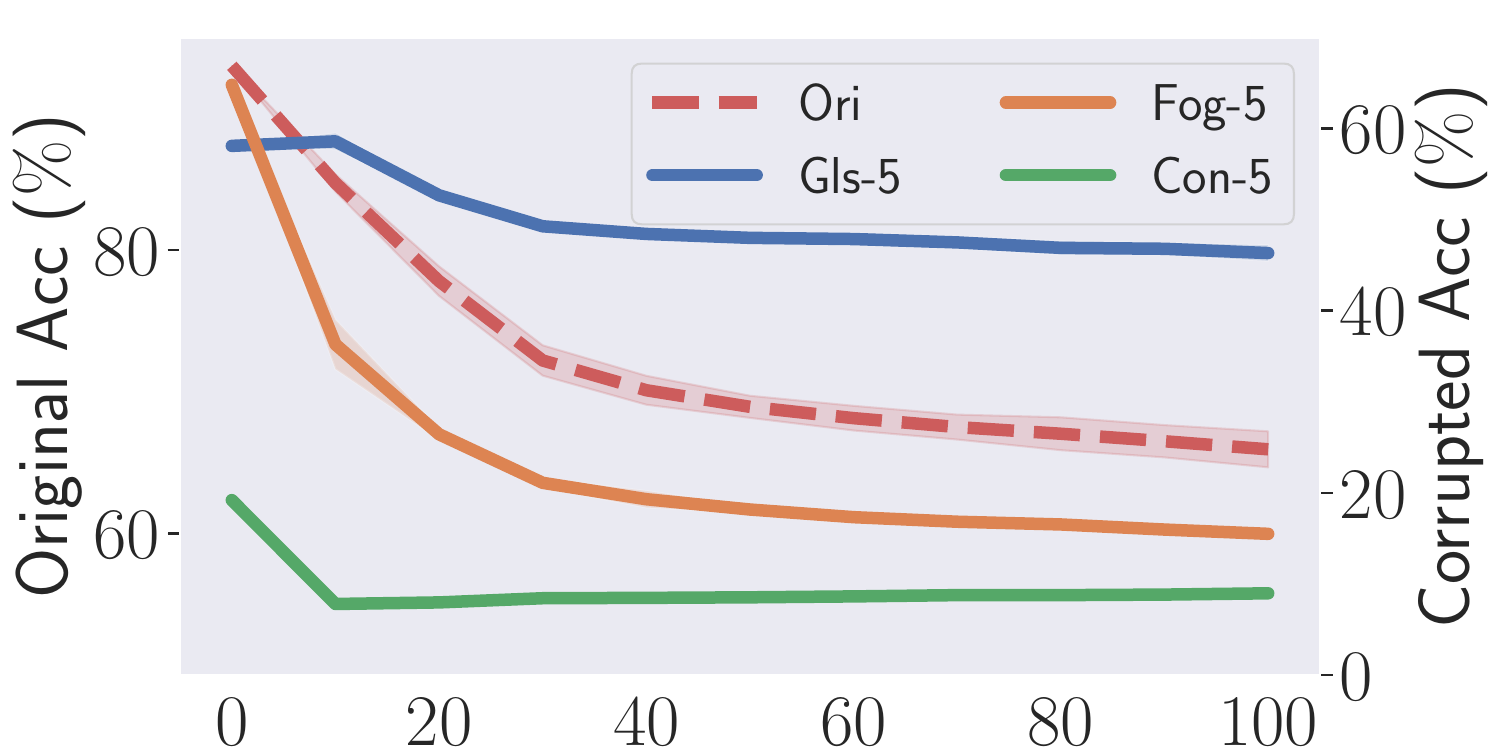}}
\hspace{3mm}
\subfloat[C10-Res50 \label{fig:attack2num_dua_b}]{%
\includegraphics[width=0.2\linewidth]{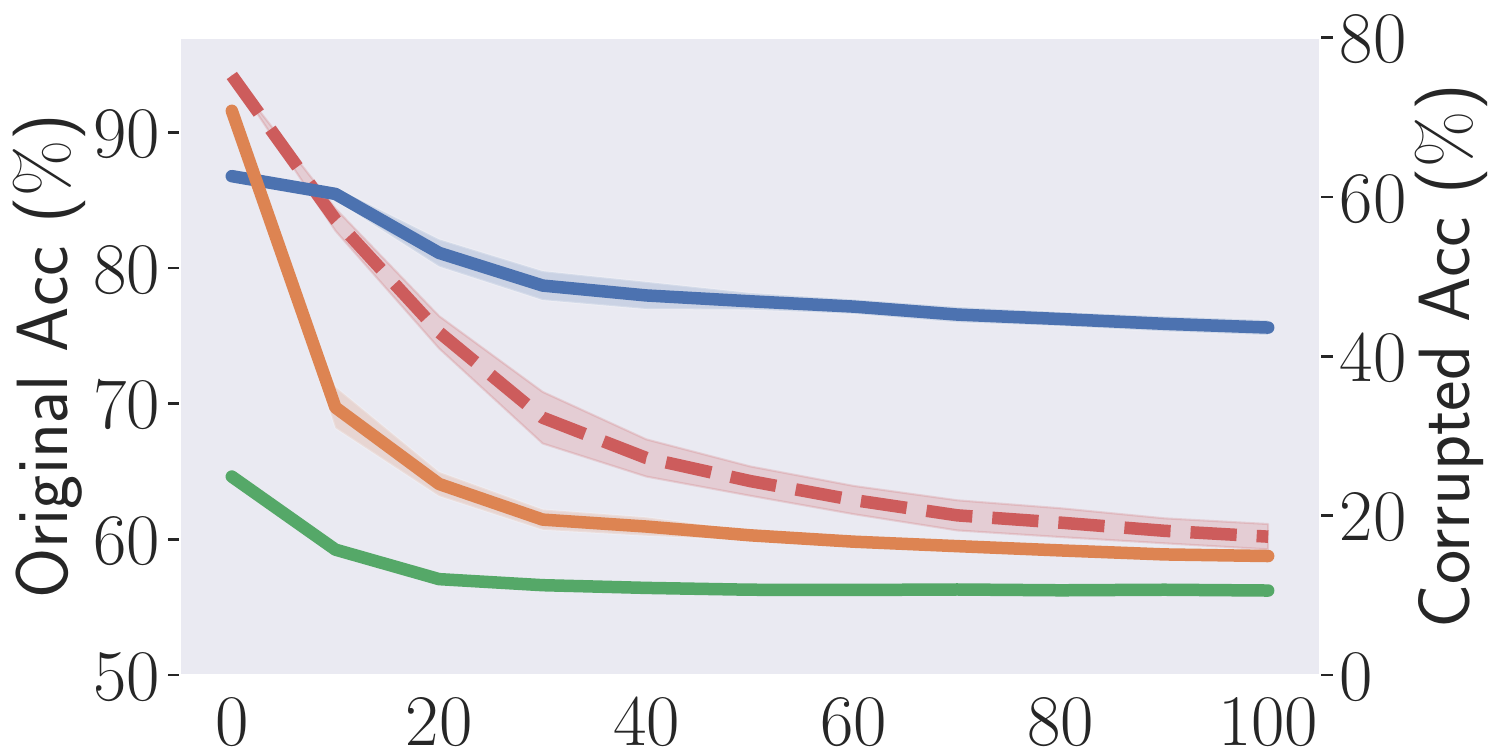}}
\hspace{3mm}
\subfloat[C100-Res18\label{fig:attack2num_dua_c}]{%
\includegraphics[width=0.2\linewidth]{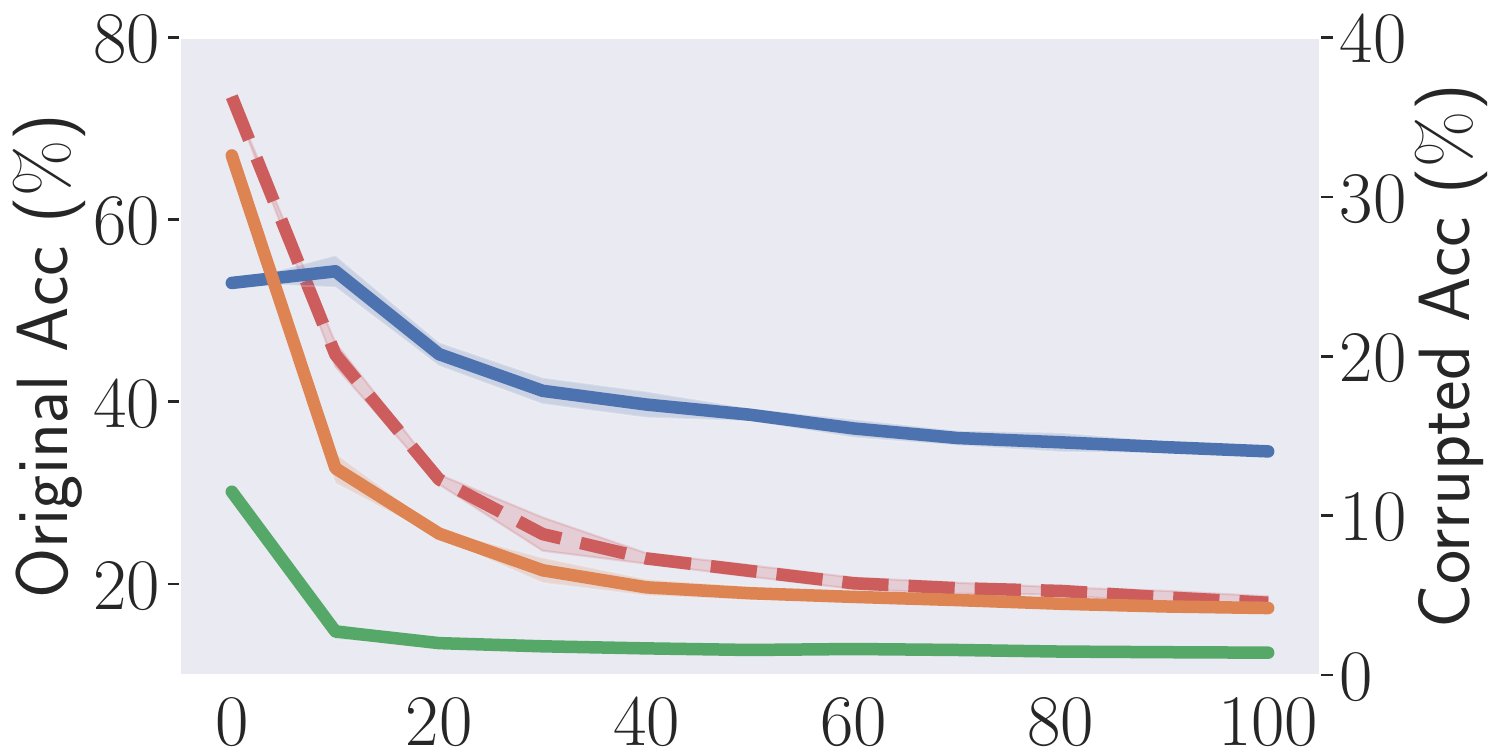}}
\hspace{3mm}
\subfloat[C100-Res50 \label{fig:attack2num_dua_d}]{%
\includegraphics[width=0.2\linewidth]{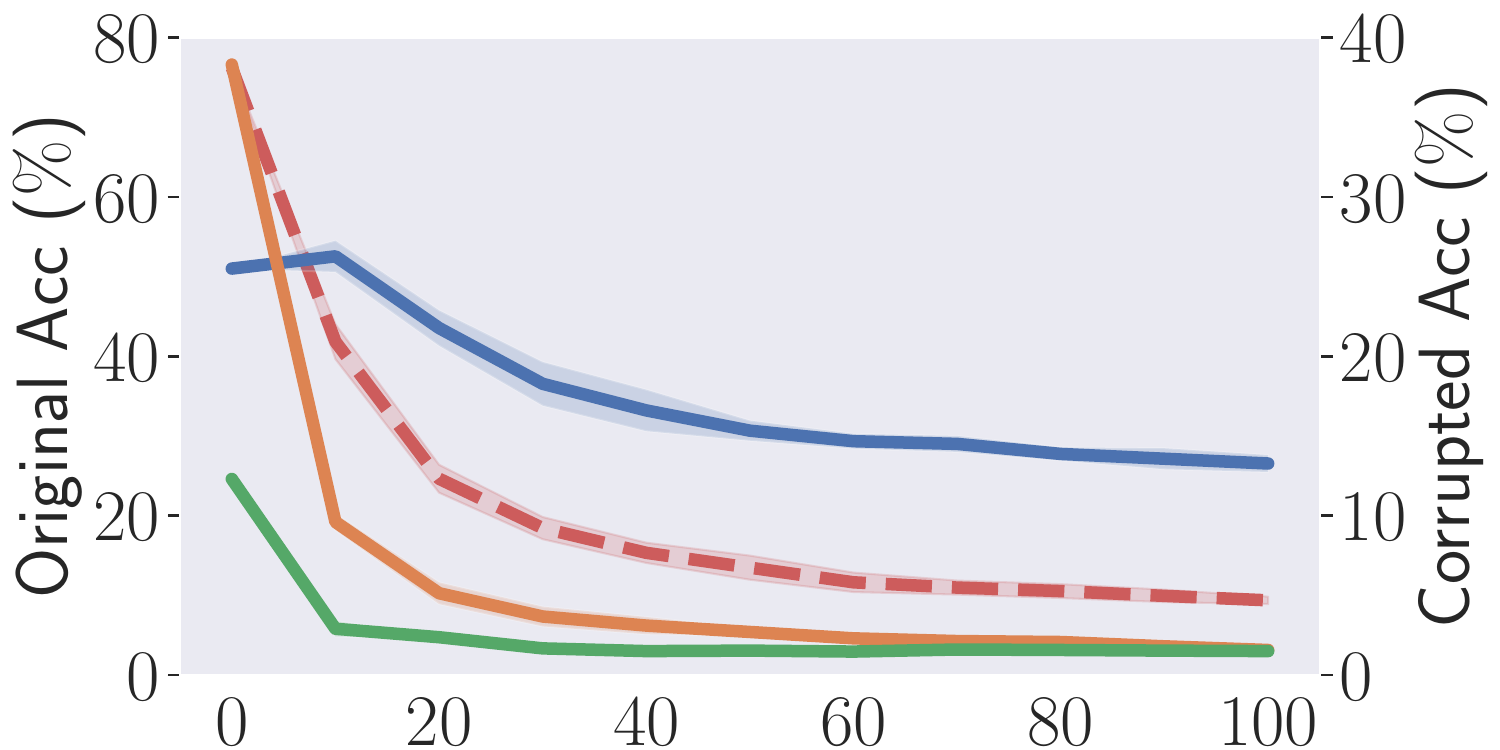}}
\caption{TePAs Against DUA-models. 
The left y-axis and the right y-axis represent the prediction accuracy on the original and corrupted evaluation datasets, respectively. 
The x-axis represents the number of poisoned samples.}
\label{fig:attack2num_dua} 
\end{figure*}

\begin{figure*}[!t] 
\centering
\subfloat[C10-Res18 \label{fig:attack2num_tent_a}]{%
\includegraphics[width=0.2\linewidth]{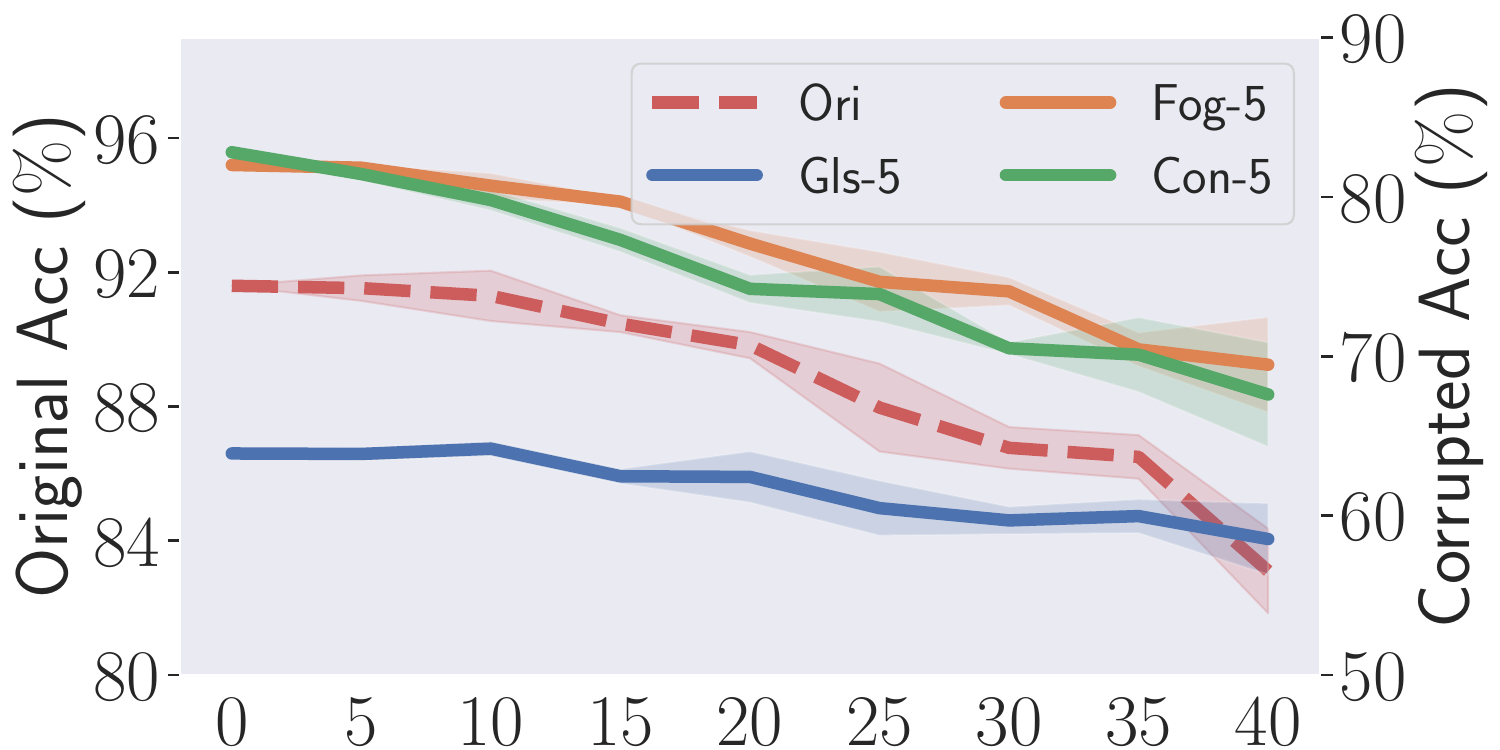}}
\hspace{3mm}
\subfloat[C10-Res50 \label{fig:attack2num_tent_b}]{%
\includegraphics[width=0.2\linewidth]{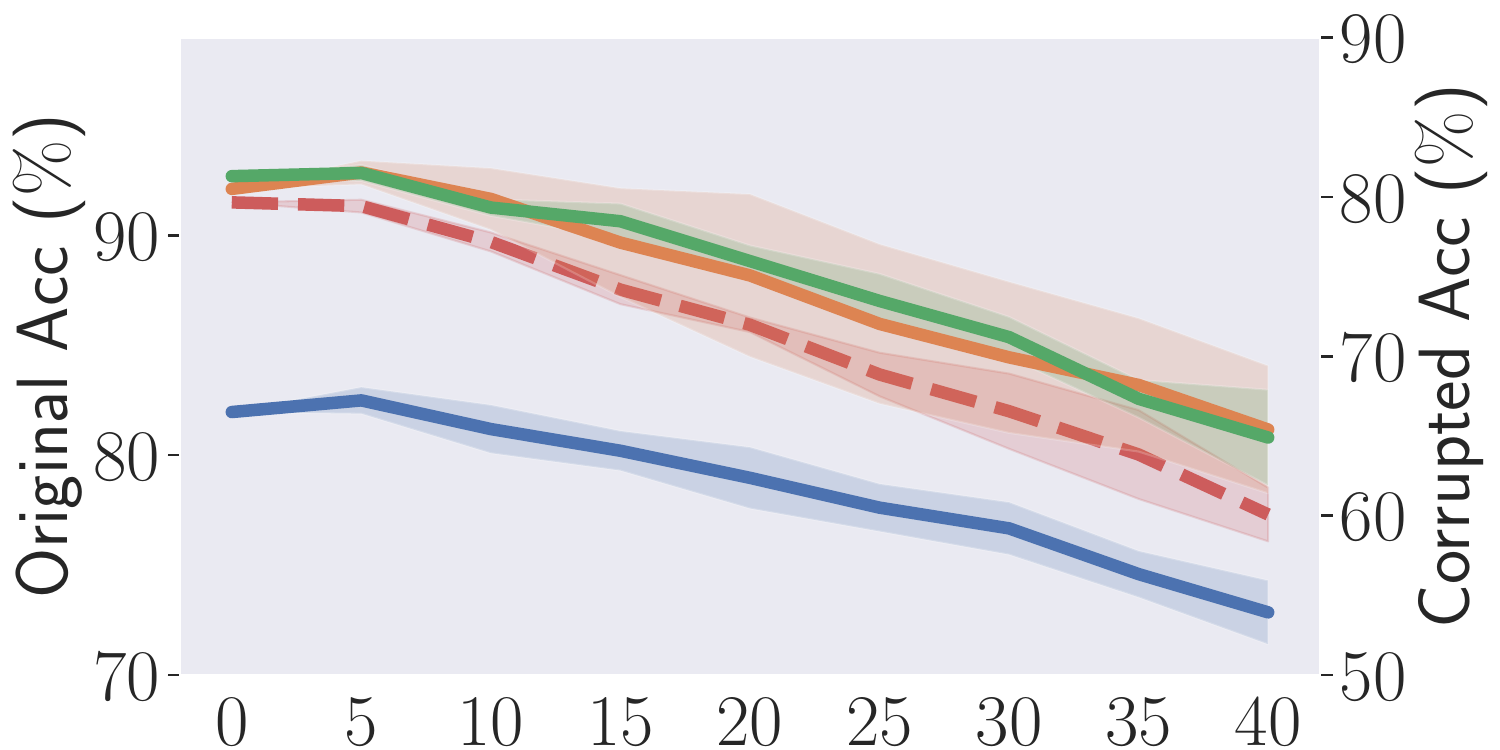}}
\hspace{3mm}
\subfloat[C100-Res18\label{fig:attack2num_tent_c}]{%
\includegraphics[width=0.2\linewidth]{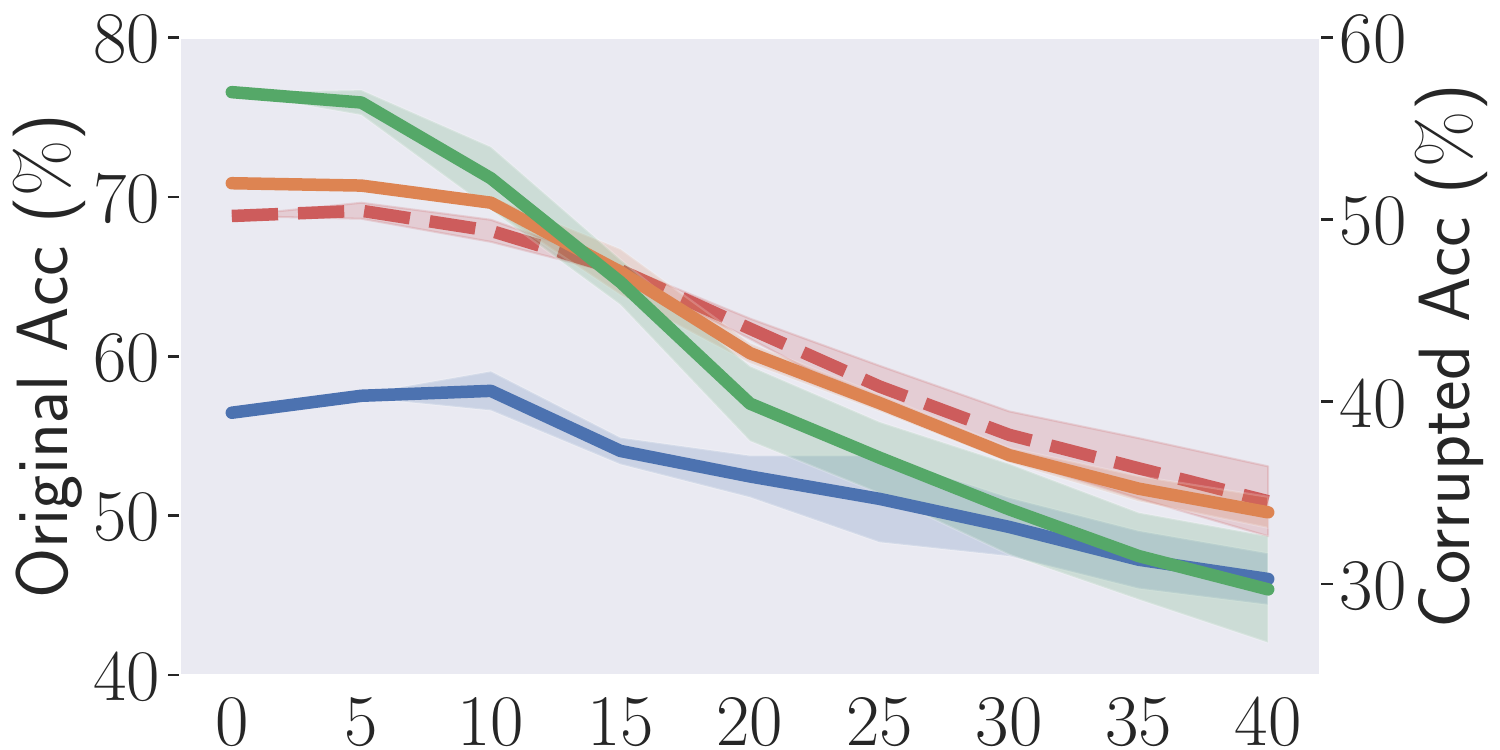}}
\hspace{3mm}
\subfloat[C100-Res50 \label{fig:attack2num_tent_d}]{%
\includegraphics[width=0.2\linewidth]{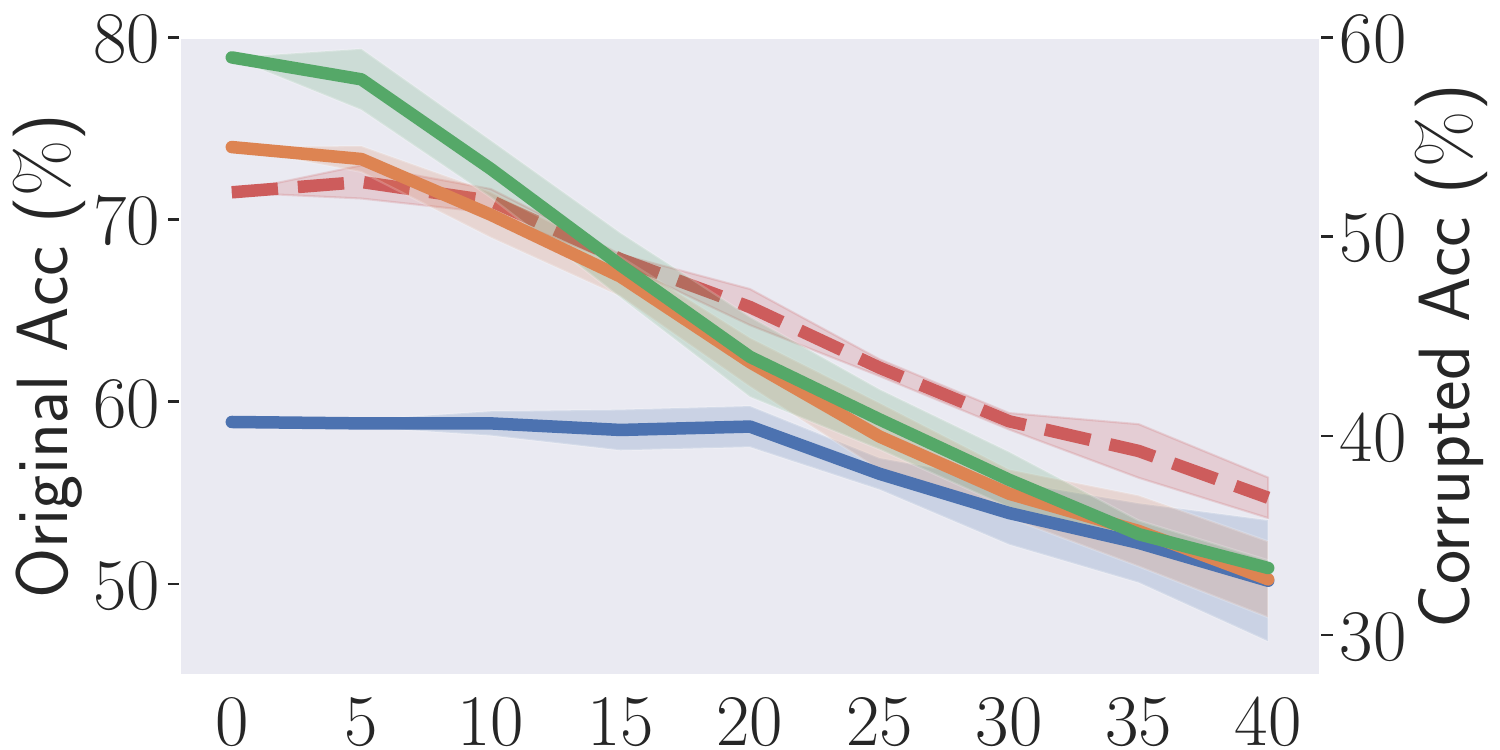}}
\caption{TePAs Against TENT-models. 
The left y-axis and the right y-axis represent the prediction accuracy on the original and corrupted evaluation datasets, respectively. 
The x-axis represents the number of poisoned samples.}
\label{fig:attack2num_tent} 
\end{figure*}

\begin{figure*}[!t] 
\centering
\subfloat[C10-Res18 \label{fig:attack2num_rpl_a}]{%
\includegraphics[width=0.2\linewidth]{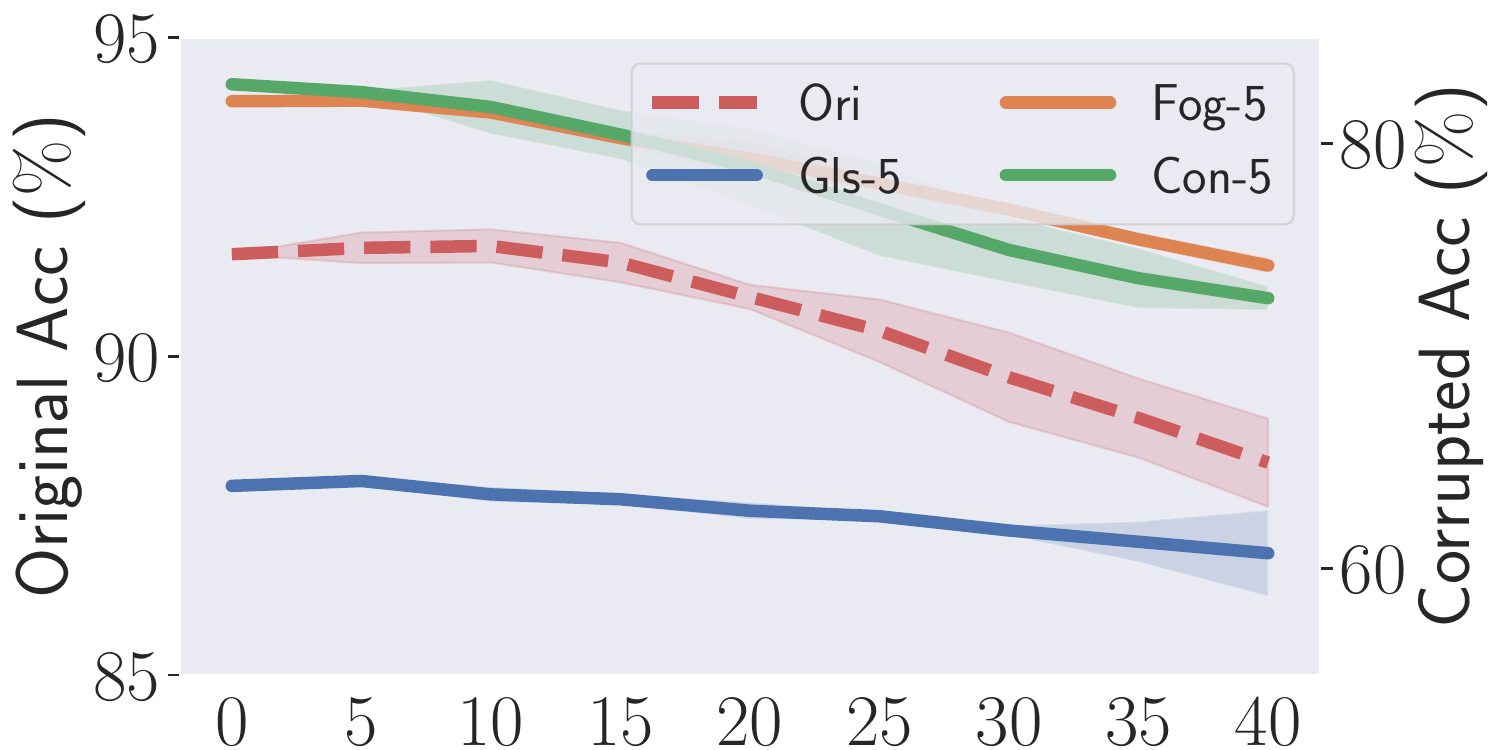}}
\hspace{3mm}
\subfloat[C10-Res50 \label{fig:attack2num_rpl_b}]{%
\includegraphics[width=0.2\linewidth]{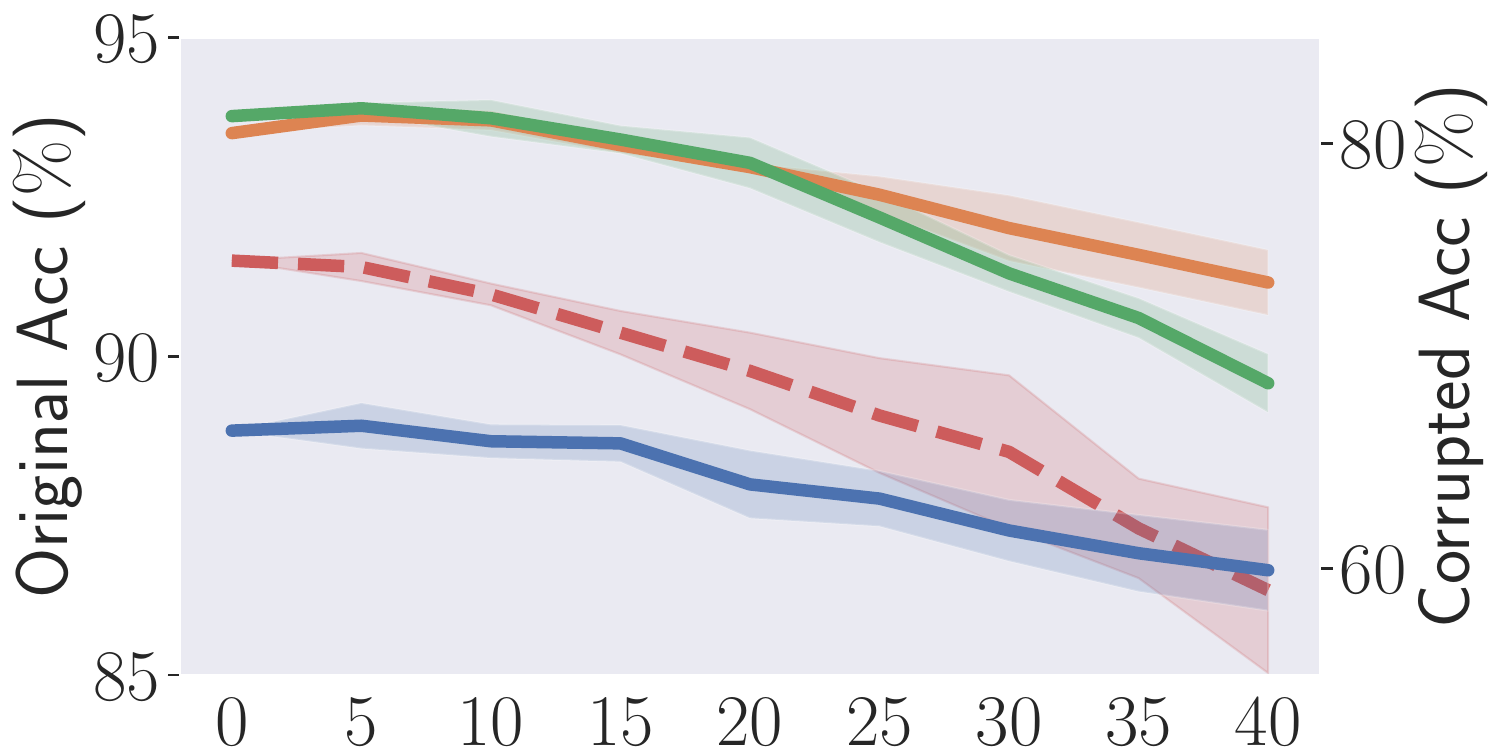}}
\hspace{3mm}
\subfloat[C100-Res18\label{fig:attack2num_rpl_c}]{%
\includegraphics[width=0.2\linewidth]{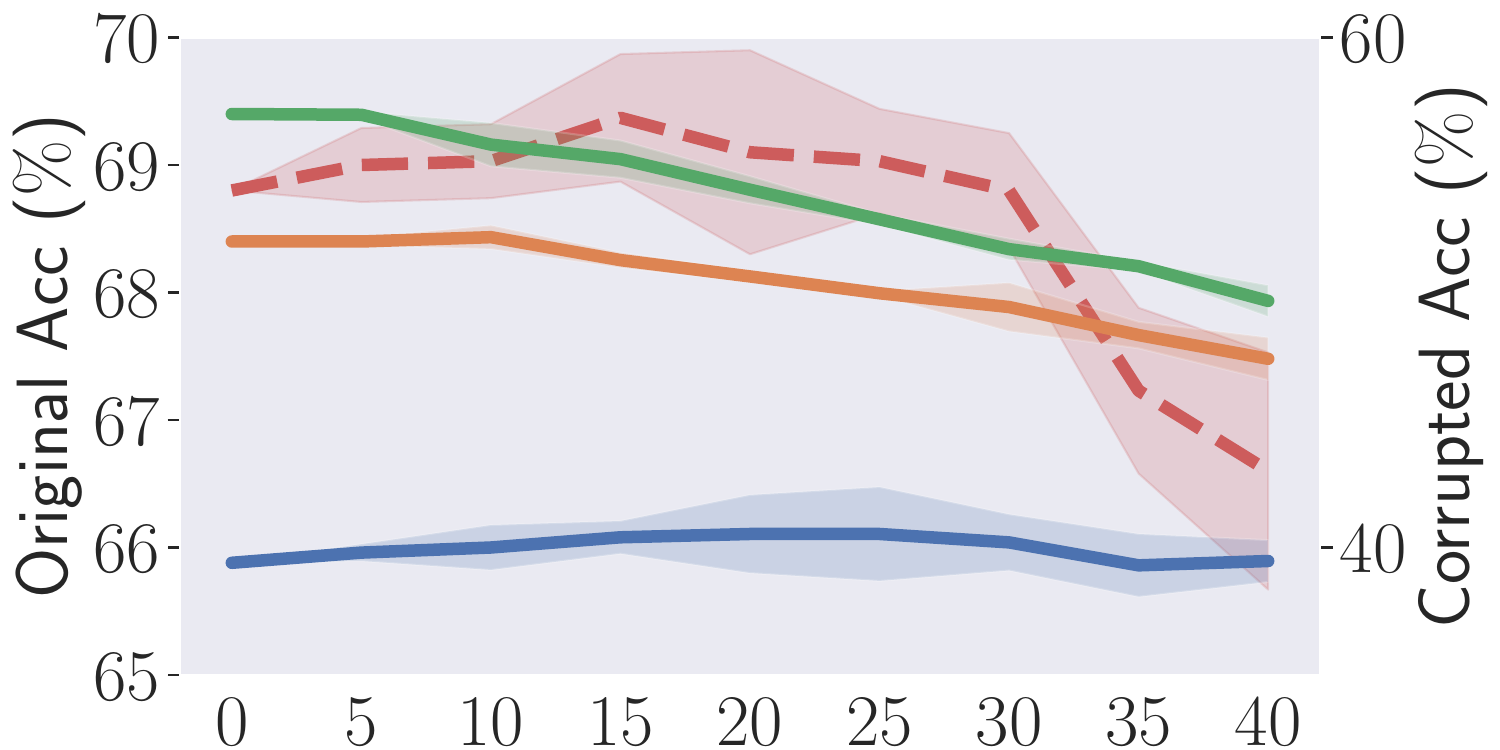}}
\hspace{3mm}
\subfloat[C100-Res50 \label{fig:attack2num_rpl_d}]{%
\includegraphics[width=0.2\linewidth]{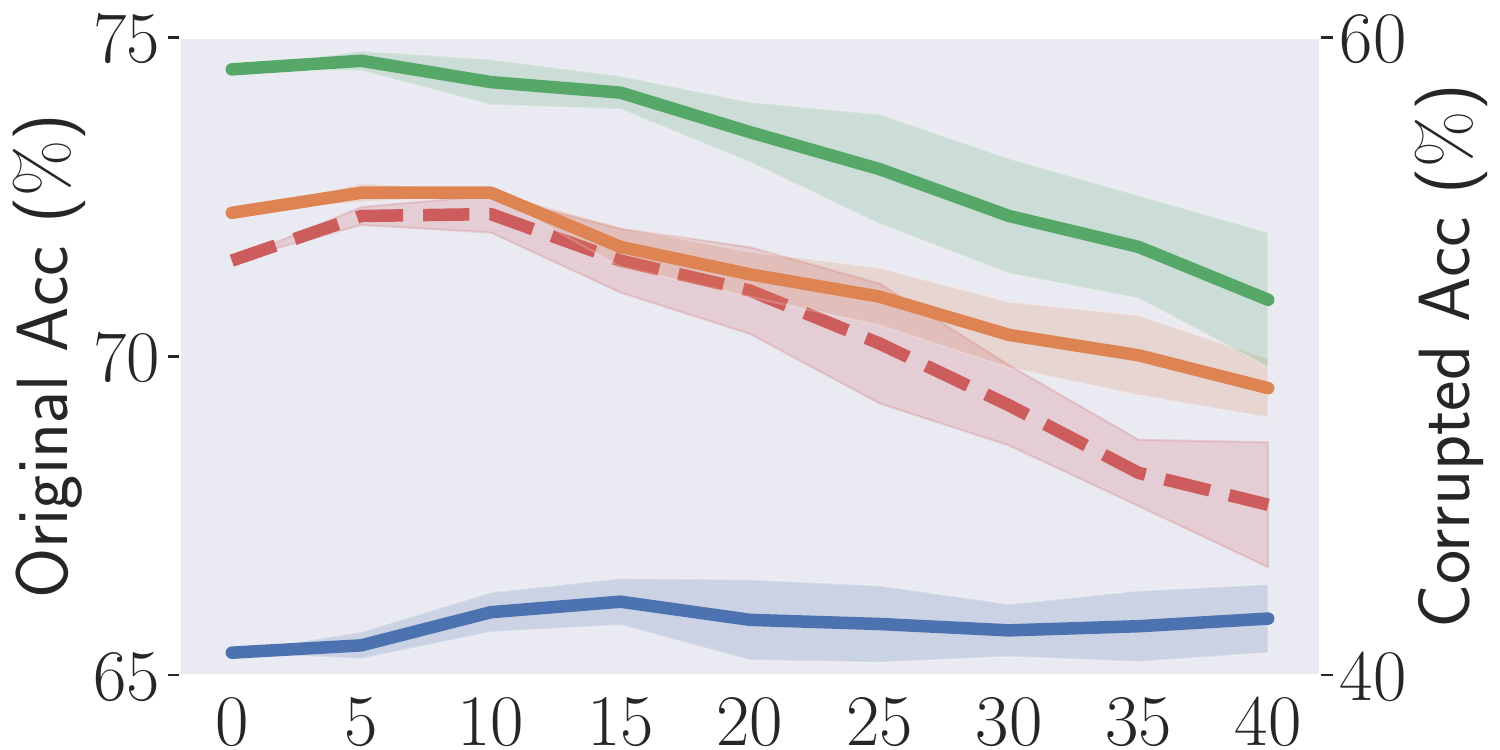}}
\caption{TePAs Against RPL-models. 
The left y-axis and the right y-axis represent the prediction accuracy on the original and corrupted evaluation datasets, respectively. 
The x-axis represents the number of poisoned samples.}
\label{fig:attack2num_rpl} 
\end{figure*}

\subsection{Utility of TTA Methods}
\label{sec::TTT_ut_tta}

We now show that TTA methods can improve the target models' performance on distribution shifts.
To better demonstrate the process of improving model performance with the TTA method, we divide the inference phase into two stages: the ``warming-up phase'' and the ``evaluation phase.''

In the warming-up phase, the model will be updated by the coming test samples from the warming-up dataset $\mathcal{D}_{w}$ through TTA methods.
Assuming the initial state of the target model is $f^0$, through being updated by $t$ (batches of) test samples, the model comes to $f^t$.
Then, if we would like to monitor the performance of $f^t$, we should input $f^t$ to the evaluation phase to calculate Acc.
Note that the incoming test samples are independent and identically distributed (i.i.d.) samples as the evaluation dataset $\mathcal{D}_{e}$.
For instance, if we use Gls-5 as the evaluation dataset, the warming-up samples should also come from Gls-5.
Also, we set the $\mathcal{D}_{w} \cap \mathcal{D}_{e} = \phi$.
Through our setting, we would like to evaluate how much the model will be boosted by learning distributional information from the i.i.d.\ samples.
To fully demonstrate the lifting power of the TTA methods, we adapt our target models by four TTA methods, including TTT, DUA, TENT, and RPL.
The evaluation results of these four TTA methods on ResNet-18 with CIFAR-10-C are shown in~\autoref{fig:utility_tta}.
The results for the ResNet-50 trained on CIFAR-100 are shown in~\autoref{fig:utility_tta_res50_CIFAR-100}.

Firstly, we can observe that the performance of the target models can be improved by the TTA methods.
Meanwhile, as the amount of i.i.d.\ samples increases, the model gains more performance improvement.
For instance, from the results shown in \autoref{fig:utility_tta_a} we can observe that the Acc of TTT-0 on Fog-5 and Con-5 are 73.93\% and 83.97\%, respectively.
However, after being updated by 50 i.i.d.\ samples (the model comes to TTT-50), the performances have been improved to 75.17\% and 84.43\%.
Meanwhile, the performance could be further improved to 81.9\% and 88.37\% when the model comes to TTT-1000.

Secondly, we compare DUA, TENT, and RPL together since they adapt the same target model.
Compared to DUA, TENT and RPL both have a greater ability to enhance the model.
For example, when the target model is C10-Res18 and $\mathcal{D}_e$ is Fog-5, the Acc of TENT-10 and RPL-10 are both higher than 80.00\%, but only 76.50\% for DUA-10.
This is because DUA processes one test sample at a time while TENT and RPL require the test samples to come in a ``batch-by-batch'' manner, which makes TENT and RPL learn normalization statistics information quickly.

Thirdly, we compare TENT and RPL since they both adapt the affine parameters in the BN layers but use different loss functions.
We observe that TENT can achieve better performance than RPL.
For instance, when the target model is C10-Res18, the Acc of RPL-40 on Ori, Gls-5, Fog-5 and Con-5 are 91.83\%, 64.87\%, 83.97\%, and 83.57\%, respectively, but the Acc of TENT-40 are 92.10\%, 68.27\%, 85.40\%, and 84.27\%, respectively.

\begin{figure*}[!t]
\centering
\includegraphics[width=12.5cm]{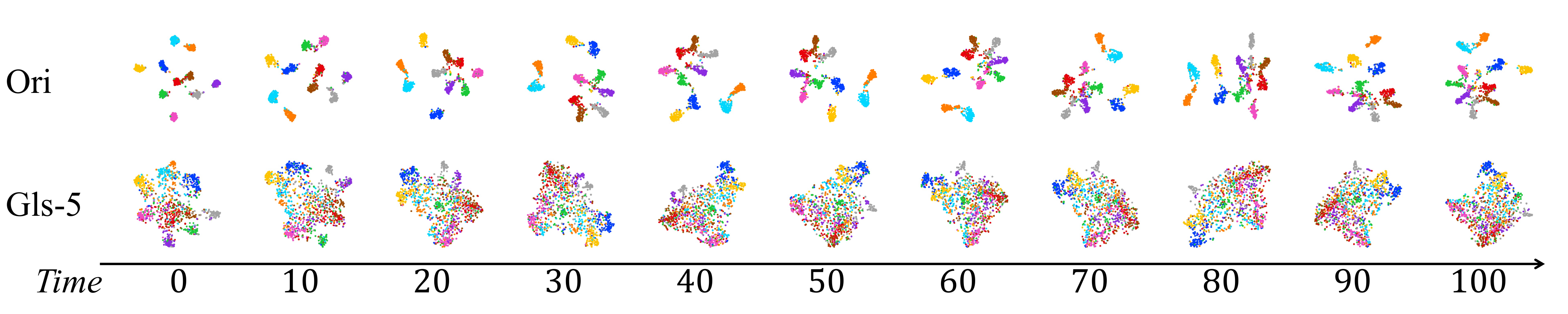}
\caption{The features are obtained from the evaluation dataset (1k evaluation samples) through the target TTT-model C10-Res18@Y4. 
We project them into a plane using t-SNE and arrange the t-SNE results in time order from left to right.}
\label{fig:tsne_ttt}
\end{figure*}

\subsection{TePAs Against TTA Models}

We here launch TePAs against TTA models.
To fully demonstrate the vulnerability of the TTA models against TePAs, we feed poisoned samples to adapt all eight target models and evaluate the impact on the prediction performance.
The results are shown in \autoref{fig:attack2num_ttt}, \autoref{fig:attack2num_dua}, \autoref{fig:attack2num_tent}, and \autoref{fig:attack2num_rpl}.

Firstly, we can observe that regardless of the network architecture or the dataset, our poisoned samples lead to a significant reduction in the prediction abilities of the target models.
The performance of the model gradually decreases as the number of poisoned samples increases.
For instance, as~\autoref{fig:attack2num_ttt_a} shows, when we feed 50 poisoned samples to C10-res18@Y4, the Acc on Gls-5 drops to 30.87\%, and the Acc further drops to 26.97\% with 100 poisoned samples.
Meanwhile, we can also observe that with TePAs, the model's performance decreases on both original and corrupted evaluation datasets.
For instance, from~\autoref{fig:attack2num_tent_c}, we can observe that when we feed 40 batches of poisoned samples, the Acc both drop about $20\%$ on Ori and Gls-5.
Note that we only need a few points to significantly reduce the target model's performance, e.g., when we feed just 10 poisoned samples, the Acc on Ori drops from 76.20\% to 41.83\% (\autoref{fig:attack2num_dua_d}).

Secondly, we can observe that even if the surrogate model has a different architecture and is trained on a different surrogate dataset, TePAs are still effective.
Recall that we use a Res18@Y3 as the surrogate model which is pre-trained on the ImageNet (from CINIC-10) to poison TTT-models, whose structure and training dataset are both different from the target C10-Res18@Y4 model.
Moreover, when poisoning C100-Res18@Y3, our surrogate dataset (CIFAR-10) only contains part of the distribution information compared to the training dataset of the target model (CIFAR-100).
When poisoning TENT- and RPL-models, we use a VGG-11 as the surrogate model.
In short, the adversary can always generate powerful poisoned samples based on the surrogate model even though they do not have adequate background knowledge about the target model.

Thirdly, through comparing~\autoref{fig:attack2num_tent} and \autoref{fig:attack2num_rpl}, we find that RPL is more robust against TePAs than TENT.
For instance, when the target model is C10-Res18, with 40 batches of poisoned samples, TENT's Acc drops 12.53\% and 15.20\% on Fog-5 and Con-5, respectively, and RPL's Acc drops 7.73\% and 10.07\%, respectively.
In conjunction with the discussion in \autoref{sec::TTT_ut_tta}, we can conclude that by minimizing entropy instead of using GCE loss, TENT can obtain a greater increase than RPL when the test samples are i.i.d.\ samples; however, its performance also decreases larger than RPL.
Nevertheless, RPL cannot resist TePAs perfectly.
For instance, when we feed 40 batches of poisoned samples (\autoref{fig:attack2num_rpl_b}), Acc drops from 81.30\% to 68.73\% on Con-5.

\mypara{t-SNE Visualization}
To better demonstrate the effect of TePAs, we feed the evaluation data (Ori and Gls-5) to the poisoned model and visualize the features with t-Distributed Neighbor Embedding (t-SNE)~\cite{MH08}.
The results are shown in~\autoref{fig:tsne_ttt}, in which different colors denote samples from different classes. 
We can observe that when $t=0$, the evaluation samples on Ori can be well distinguished by the model, and the clustering effect on Gls-5 is weak, i.e., some features are close to each other.
However, as poisoned samples are incrementally introduced to the model, the features become increasingly entangled.
For instance, at $t=100$, the model exhibits a significant reduction in distinguishability between different sample categories on Gls-5.

\begin{figure*}[!t]
\centering
\subfloat[TTT \label{fig:prob_ttt}]{%
\includegraphics[width=0.2\linewidth]{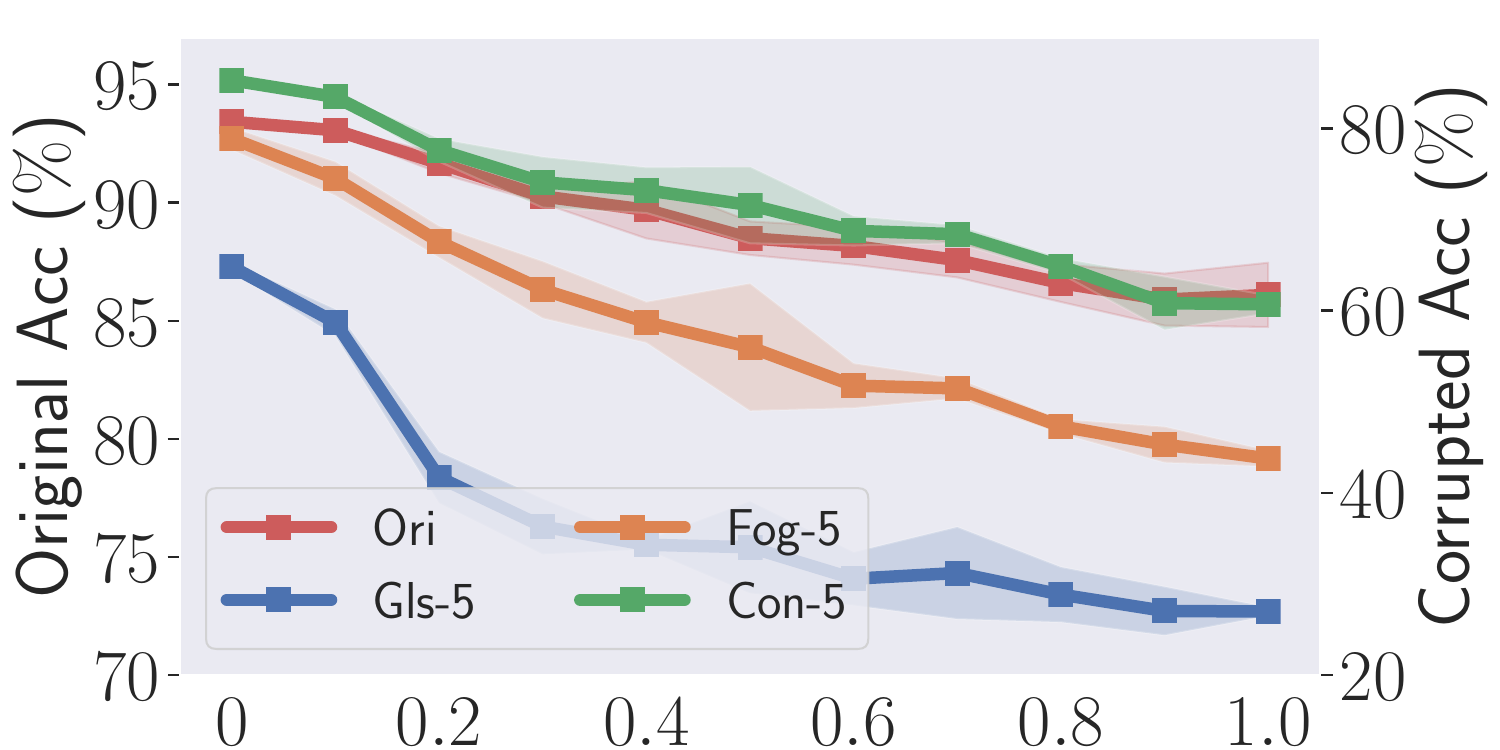}}
\hspace{3mm}
\subfloat[DUA \label{fig:prob_dua}]{%
\includegraphics[width=0.2\linewidth]{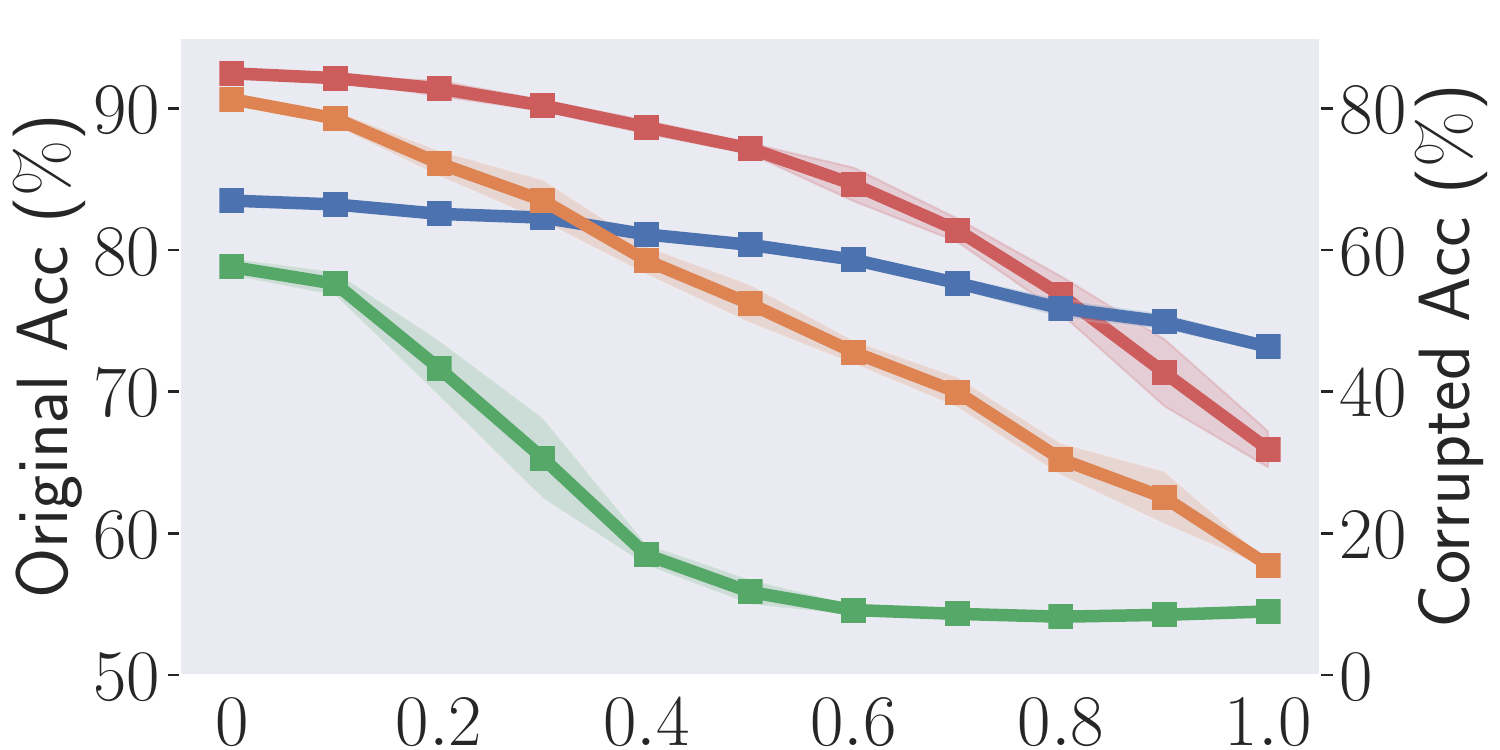}}
\hspace{3mm}
\subfloat[TENT\label{fig:prob_tent}]{%
\includegraphics[width=0.2\linewidth]{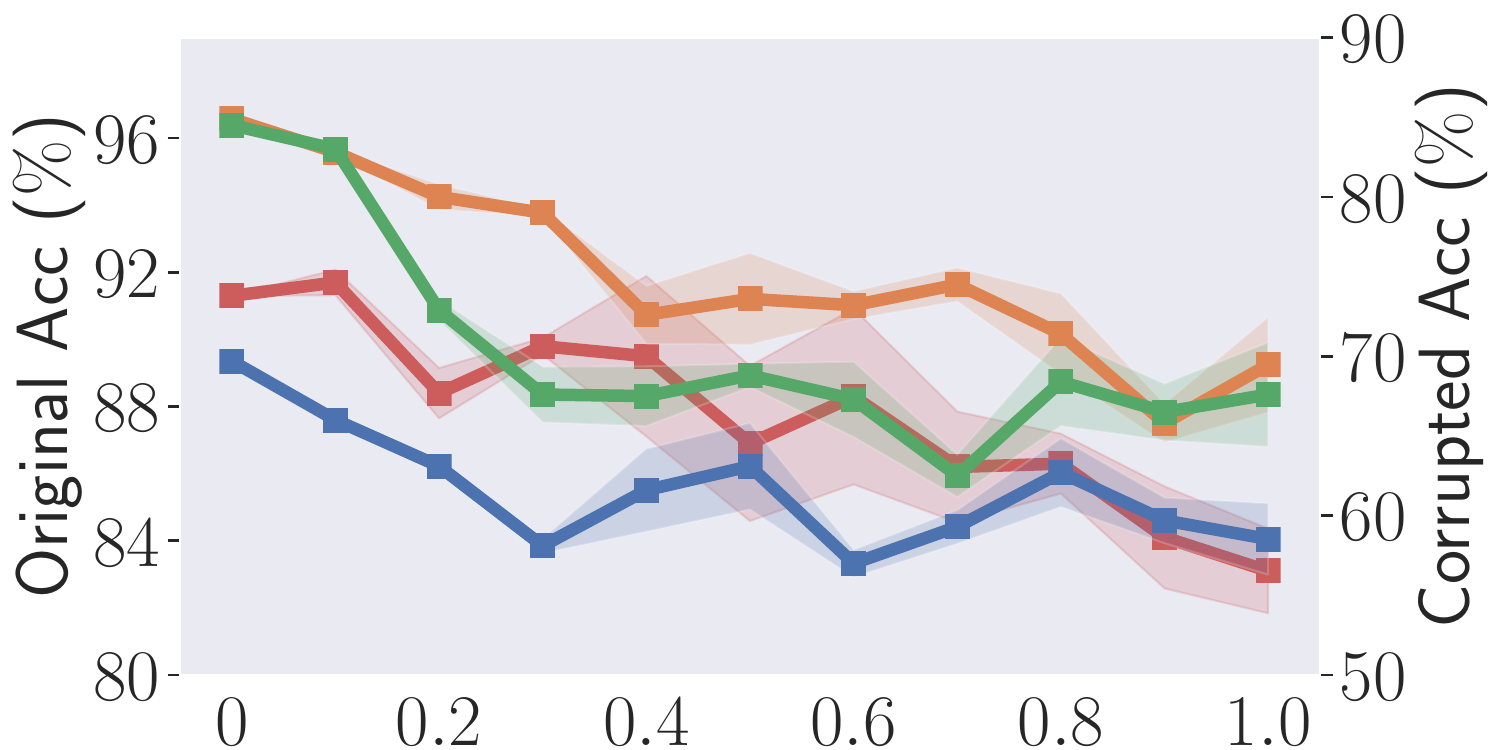}}
\hspace{3mm}
\subfloat[RPL \label{fig:prob_rpl}]{%
\includegraphics[width=0.2\linewidth]{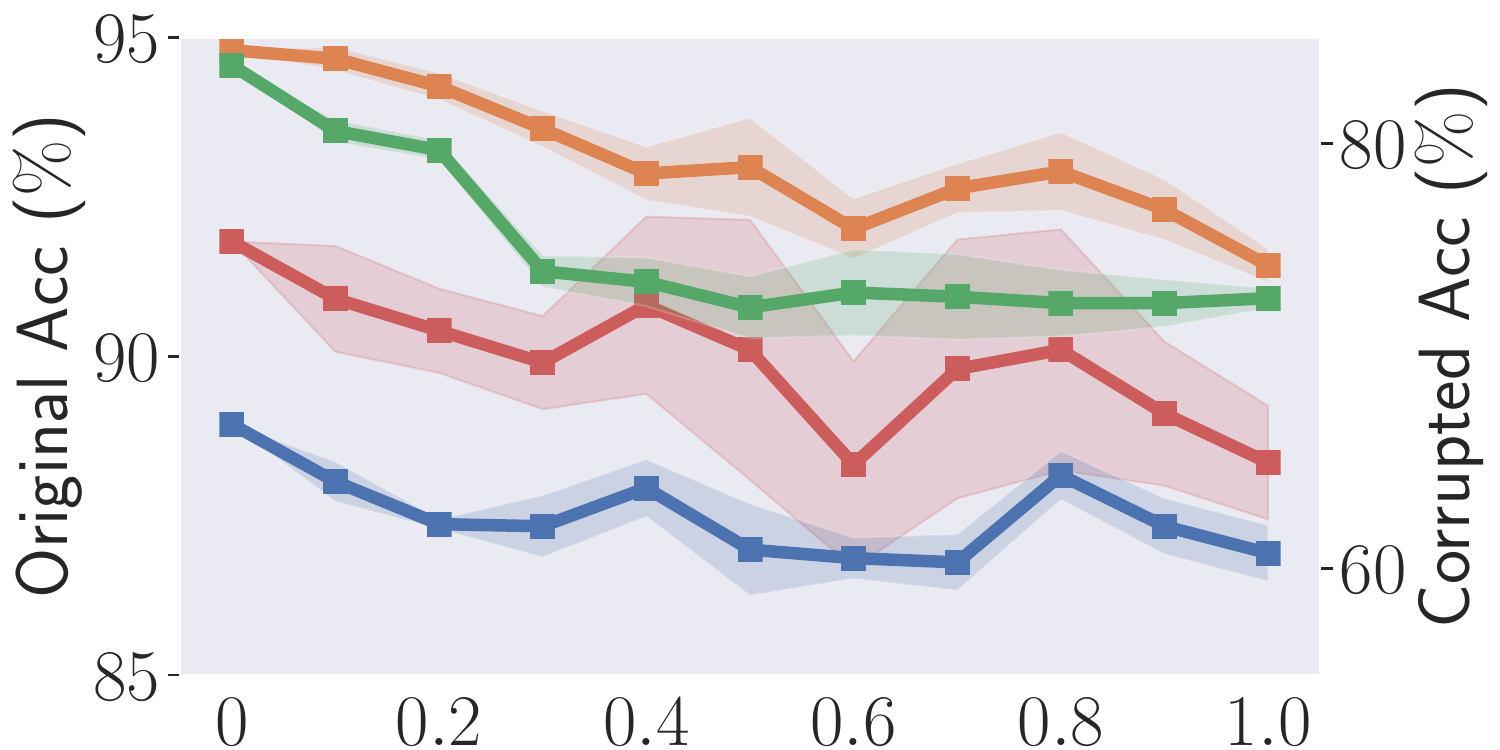}}
\caption{Uniformly Poisoning. 
The target model is ResNet-18 trained on CIFAR-10. 
The left y-axis and the right y-axis represent the prediction accuracy on the original and corrupted evaluation datasets, respectively. 
The x-axis represents the probability $P$ of being a poisoned sample for each test sample.}
\label{fig:attack_prob} 
\end{figure*}

\begin{figure*}[!t]
\centering
\subfloat[TTT \label{fig:clean_poison_gls5_a}]{%
\includegraphics[width=0.15\linewidth]{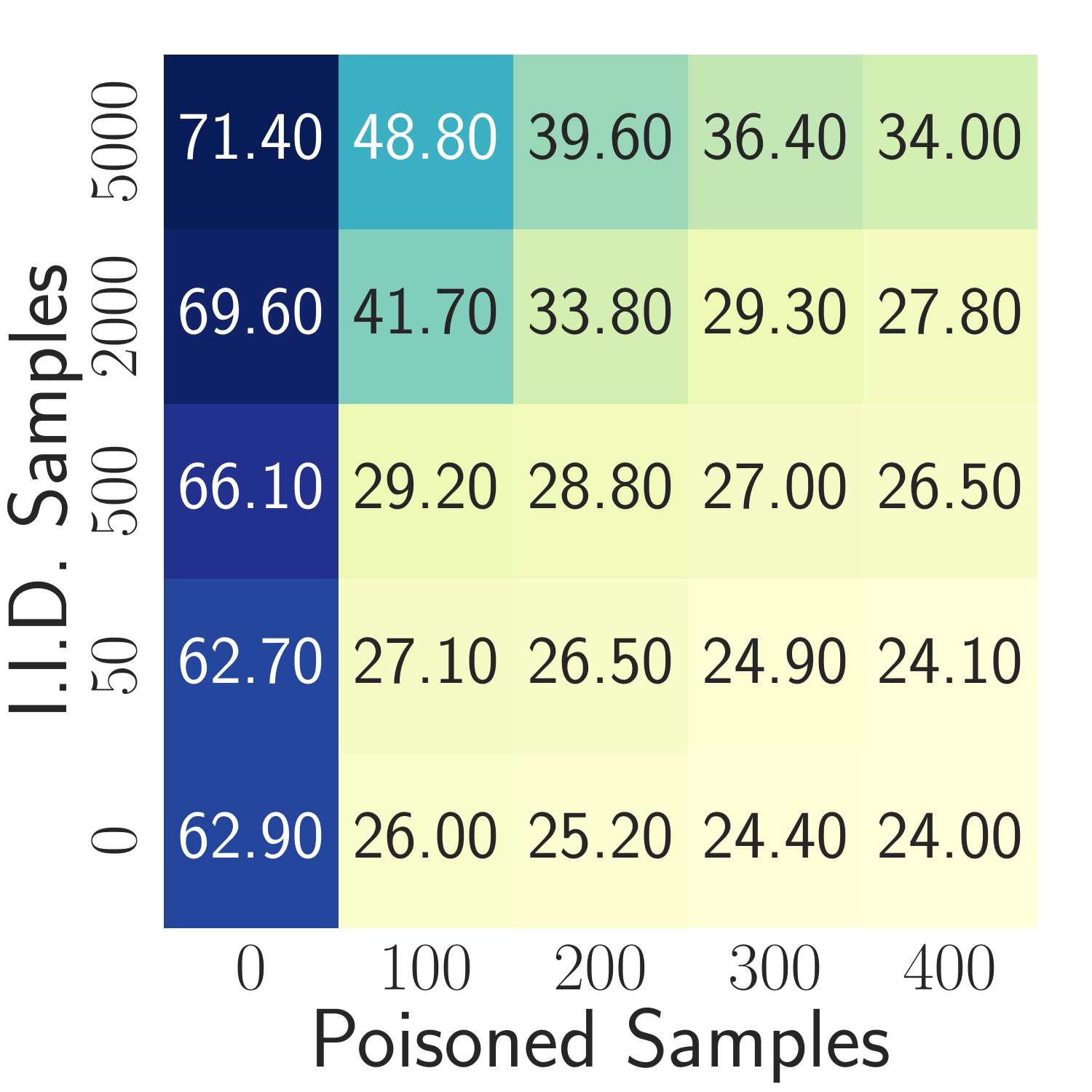 }}
\hspace{6mm}
\subfloat[DUA \label{fig:clean_poison_gls5_b}]{%
\includegraphics[width=0.15\linewidth]{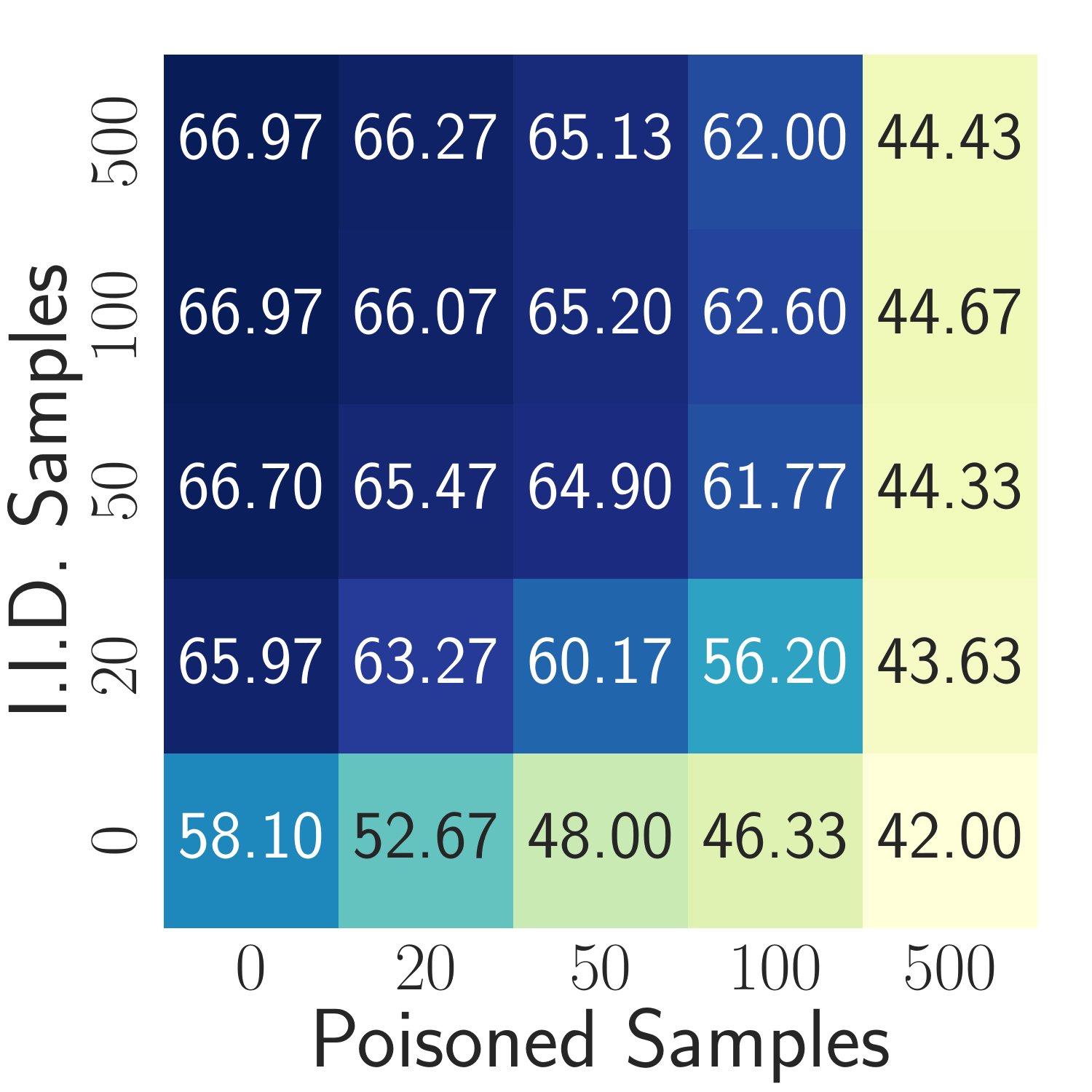}}
\hspace{6mm}
\subfloat[TENT\label{fig:clean_poison_gls5_c}]{%
\includegraphics[width=0.15\linewidth]{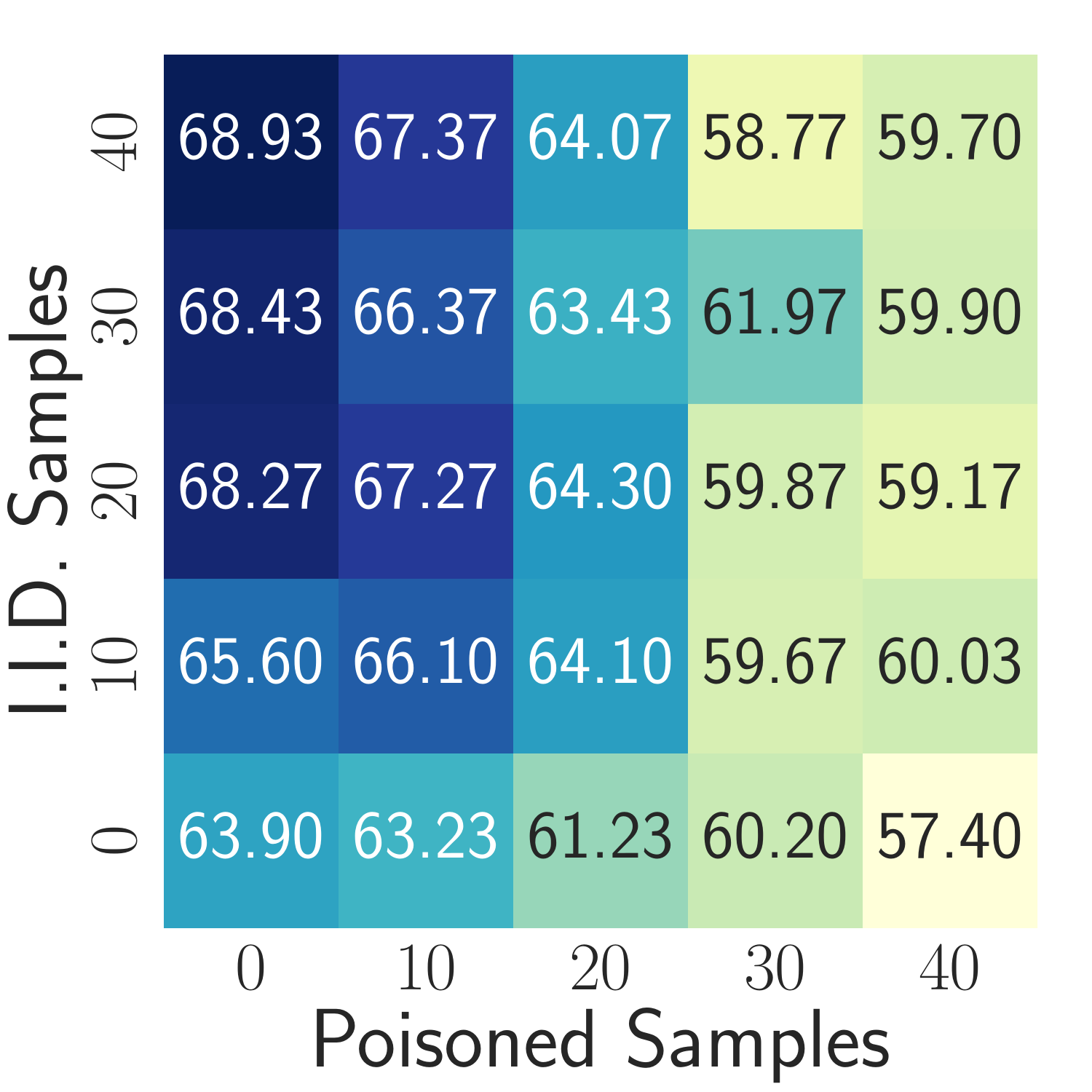}}
\hspace{6mm}
\subfloat[RPL \label{fig:clean_poison_gls5_d}]{%
\includegraphics[width=0.15\linewidth]{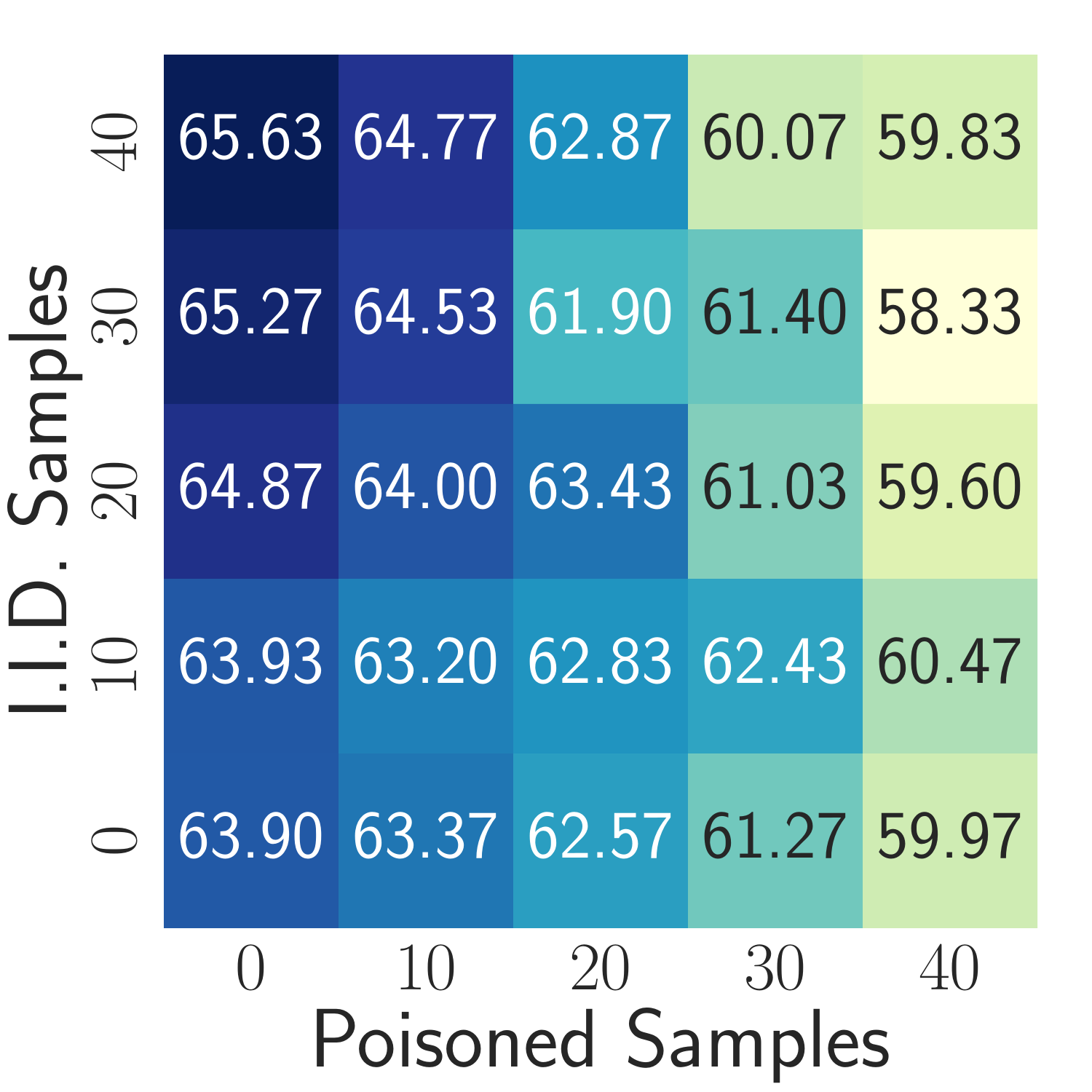}}
\caption{Warming-up before Poisoning. 
The target model is ResNet-18 trained on CIFAR-10. 
The y-axis and the x-axis represent the number of the i.i.d.\ samples and the poisoned samples, respectively. 
We fix the evaluation dataset to Gls-5.} 
\label{fig:clean_poison_gls5} 
\end{figure*}

\begin{figure*}[!t]
\centering
\subfloat[TTT \label{fig:poison_clean_gls5_a}]{%
\includegraphics[width=0.15\linewidth]{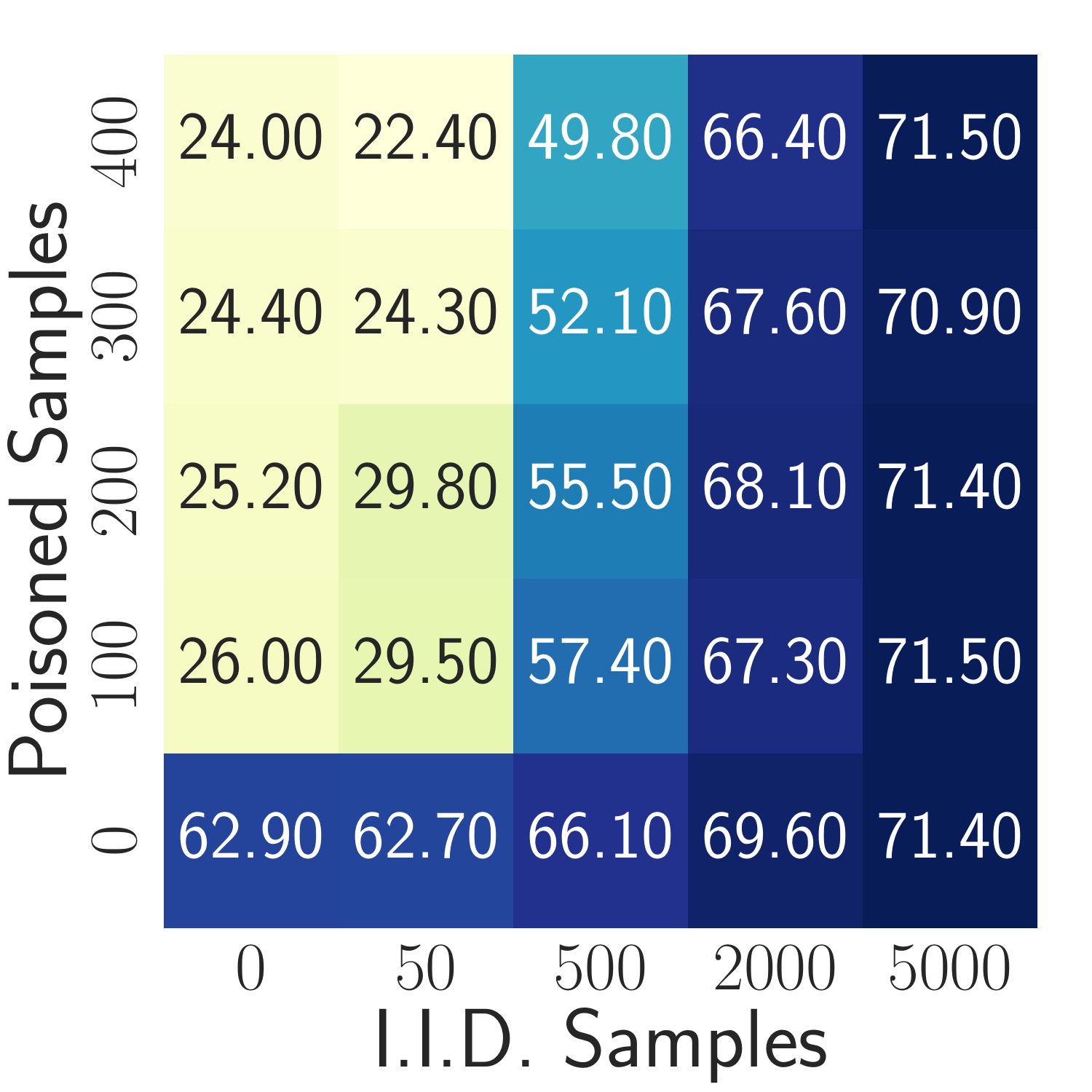}}
\hspace{6mm}
\subfloat[DUA \label{fig:poison_clean_gls5_b}]{%
\includegraphics[width=0.15\linewidth]{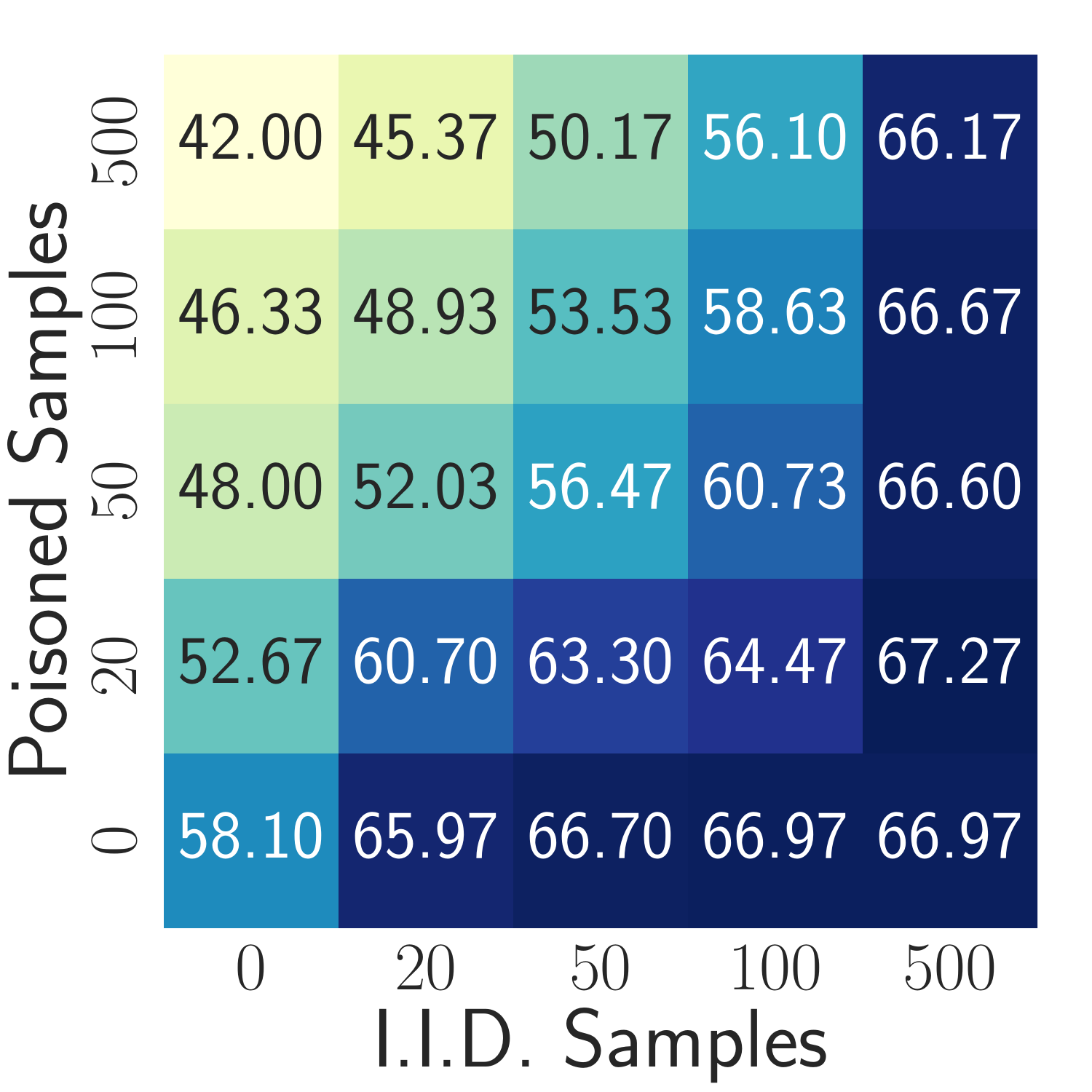}}
\hspace{6mm}
\subfloat[TENT\label{fig:poison_clean_gls5_c}]{%
\includegraphics[width=0.15\linewidth]{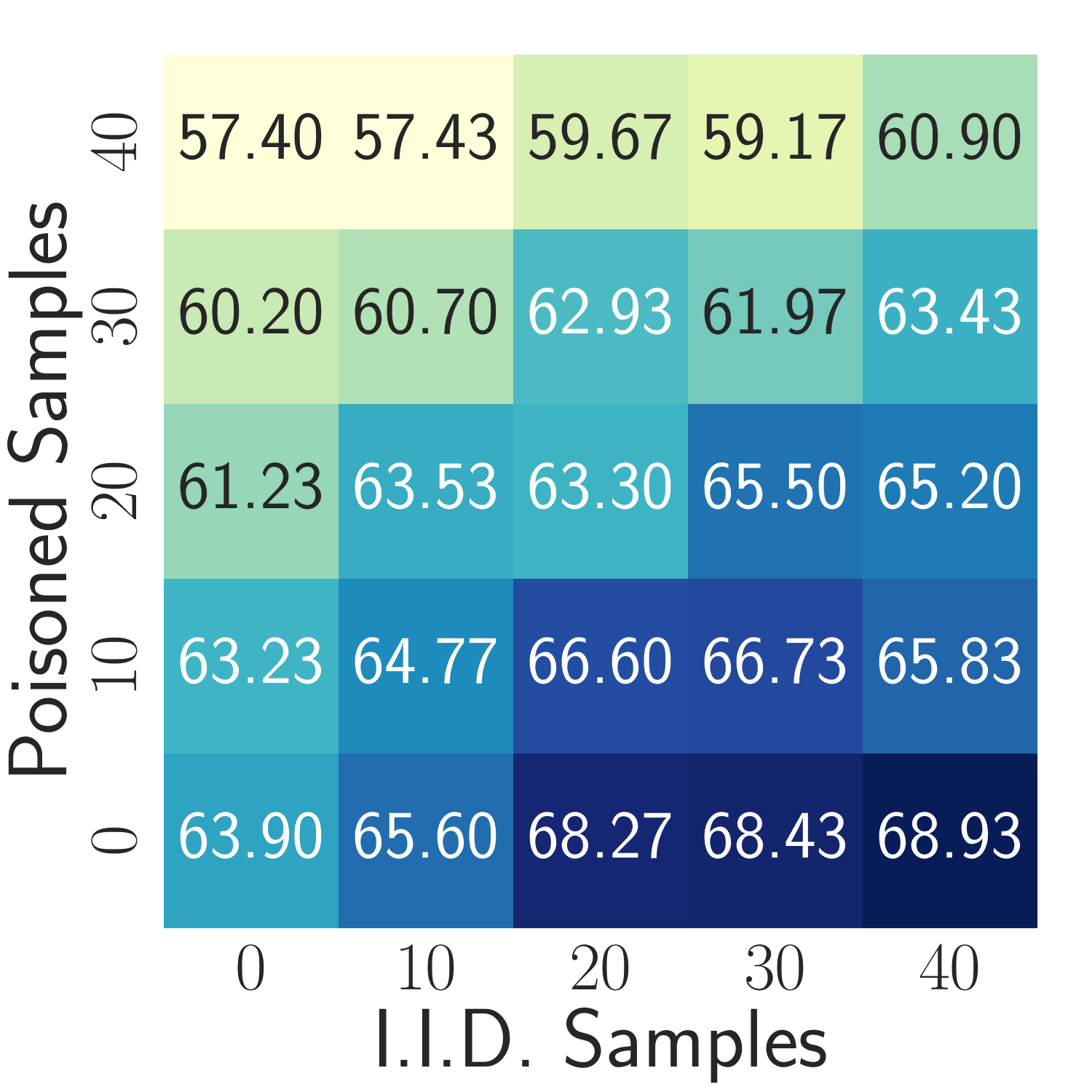}}
\hspace{6mm}
\subfloat[RPL \label{fig:poison_clean_gls5_d}]{%
\includegraphics[width=0.15\linewidth]{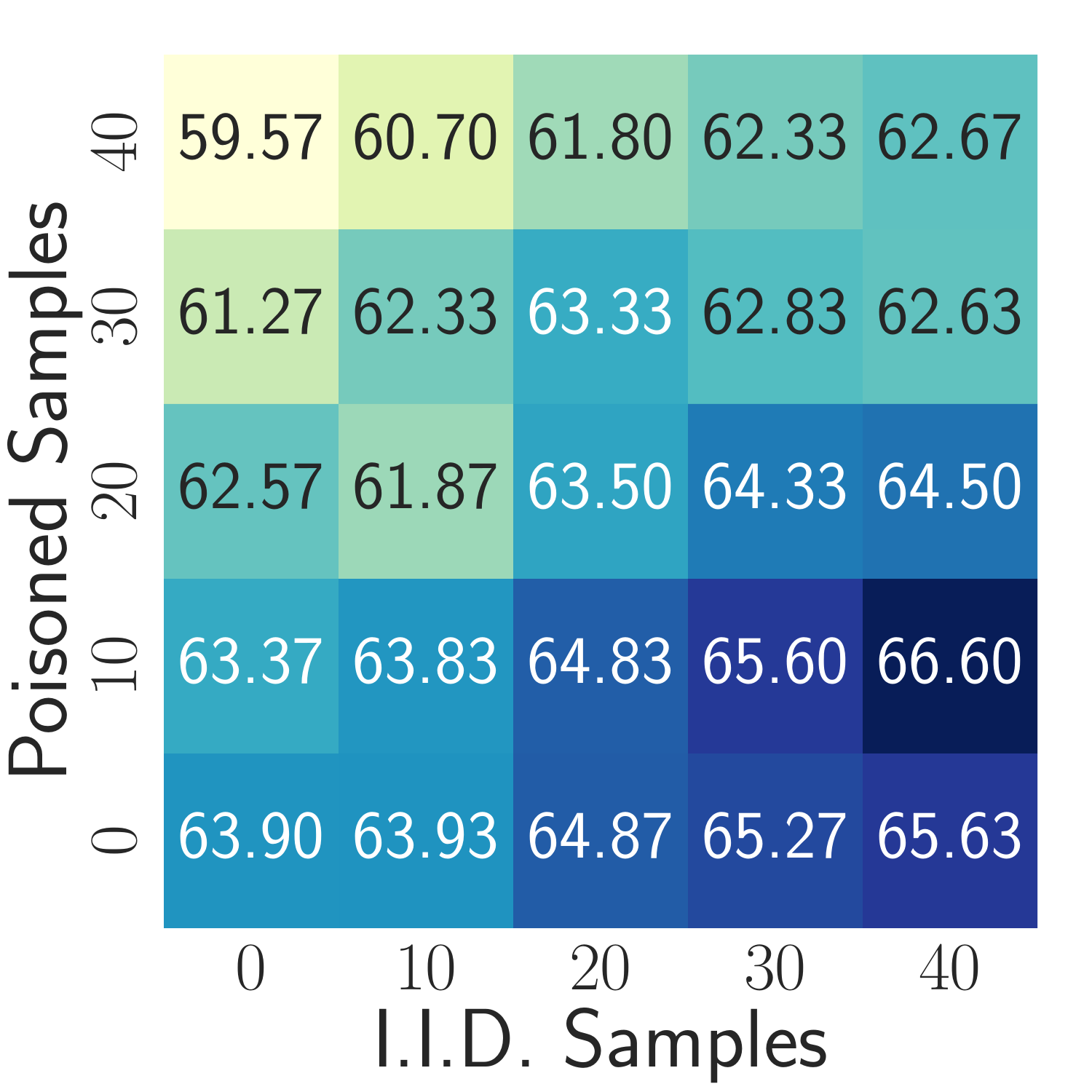}}
\caption{Warming-up after Poisoning. 
The target model is ResNet-18 trained on CIFAR-10. 
The y-axis and the x-axis represent the number of poisoned samples and the i.i.d.\ samples, respectively. 
We fix the evaluation dataset to Gls-5.}
\label{fig:poison_clean_gls5} 
\end{figure*}

\subsection{Impact of the Poisoning Strategies}
\label{section:exp_poisoning_strategy}

In the threat model, we consider the adversary cannot control the location of the poisoned samples appearing in the test data stream.
As such, we here discuss the effect of the poisoned samples' location on TePAs.
For instance, we focus on a relatively small attack window, in which the adversary can inject multiple poisoned samples. 
We consider three scenarios: 
(1) \emph{Uniformly poisoning}: poisoned samples appear in the test data stream ``uniformly.''
(2) \emph{Warming-up before poisoning}: The target model has been fine-tuned by several i.i.d.\ samples before the arrival of the poisoned samples.
(3) \emph{Warming-up after poisoning}: After the poisoning process, the target model will be further fine-tuned by several i.i.d.\ samples.
In this part, we use the ResNet-18 trained on CIFAR-10 as the target model.

\mypara{Uniformly Poisoning}
Given a test data stream, we first consider that each test sample has a probability $P$ to be a poisoned sample, and a probability $1-P$ to be an i.i.d.\ sample.
For TTT- and DUA-models, we feed 100 test samples in total.
For TENT- and RPL-models, we feed 40 batches of test samples.
We traverse $P$ from $0.0$ to $1.0$.
The results of the target model's performance are shown in~\autoref{fig:attack_prob}.
First, we can observe a general phenomenon: Acc drops as the probability of the poisoned samples increases.
Moreover, this trend is reflected in all TTA methods and all evaluation datasets (Original and other corrupted datasets).
Concretely, when we poison TTT-models and the evaluation dataset is Ori, Acc is 93.43\%, 88.50\%, and 86.60\% if $P$ is $0.0$, $0.5$, and $0.8$, respectively (shown in \autoref{fig:prob_ttt}).
Meanwhile, we notice that our poisoning attacks can degrade the target model's performance significantly with a low poisoning ratio.
For example, when the target model is a TENT-model and $P=0.2$ (see \autoref{fig:prob_tent}), we can degrade the performance of the target model by 2.90\%, 6.60\%, 4.90\% and 11.60\% on Ori, Gls-5, Fog-5, and Con-5, respectively, which further demonstrates the efficacy of our attacks.

\mypara{Warming-up before Poisoning}
In this part, we aim to evaluate TePAs in the following scenario: the target model has learned distributional information (about the evaluation samples) before the poisoning process.
In other words, the target models have received several i.i.d.\ samples in advance.
We feed i.i.d.\ samples first and then feed poisoned samples to TTT-, DUA-, TENT-, and RPL-models in order, respectively.
Then, we evaluate the target model's performance on Gls-5 (\autoref{fig:clean_poison_gls5}) and Ori (\autoref{fig:clean_poison_ori}).
Take the TTT-model's performance on Gls-5 as an example (\autoref{fig:clean_poison_gls5_a}).
Firstly, we can observe that given a certain number of i.i.d.\ samples ($\texttt{\#} x$), as the number of poisoned samples ($\texttt{\#}x'$) increases, Acc is gradually decreasing.
For instance, when $\texttt{\#} x=0$, Acc drops by 36.90\% (38.90\%) if $\texttt{\#} x'$ is 100 (400).
Another observation is that increasing $\texttt{\#} x$ helps to mitigate the performance decrease.
For instance, when $\texttt{\#} x'$=200, Acc drops by 37.70\% if $\texttt{\#} x$ is 0, and Acc drops only 32.80\% if $\texttt{\#} x$ is 5,000.
Secondly, although i.i.d.\ samples lead to a more robust target model against TePAs, the improved Acc through large amounts of i.i.d.\ samples can be quickly degraded by a few poisoned samples.
For instance, TTT increases 8.50\% Acc by warming up with $5,000$ i.i.d.\ samples but drops by 22.60\% when we feed only 100 poisoned samples.
In general, the TTA models are still vulnerable to TePAs even if being adjusted with i.i.d.\ samples beforehand.

\mypara{Warming-up after Poisoning}
Besides feeding the i.i.d.\ samples in advance, the target model can also continue to receive i.i.d.\ samples after the poisoning process.
In this part, we feed poisoned samples first and then feed i.i.d.\ samples.
We evaluate the model's performance on Gls-5 (\autoref{fig:poison_clean_gls5}) and Ori (\autoref{fig:poison_clean_ori}).
The trends in experimental results for the four TTA algorithms are generally consistent.
Here we take the DUA-model's performance on Gls-5 as an example.
First, we observe that, regardless of the number of poisoned samples, the model's utility will recover to the normal level with i.i.d.\ samples.
For instance, when $\texttt{\#}x$ is 500, Acc are 66.60\% and 66.17\% when $\texttt{\#}x'$ is 50 and 500.
Therefore, the performance drop caused by the poisoned samples can be largely eliminated.
However, we find that the cost required for this resilience is relatively expensive, i.e., the recovery is less sufficient with fewer i.i.d.\ samples.
For instance, if we leverage 20 poisoned samples to launch attacks first, and 100 i.i.d.\ samples can only recover the model to 64.47\%.

\mypara{Note}
Combining the results in \autoref{fig:clean_poison_gls5} and \autoref{fig:poison_clean_gls5}, we notice that TTA methods have ``instant response'' to the relative location between poisoned samples and i.i.d.\ samples.
Take the DUA-model as an example.
Concretely, when $\texttt{\#} x'$ and $\texttt{\#} x$ are both $500$.
If we feed poisoned samples earlier than i.i.d.\ samples, the Acc will be 66.17\%.
However, it drops to $44.43\%$ if we feed the i.i.d.\ samples beforehand.
Meanwhile, we notice that attacking TTT is easier than the other three TTA methods by comparing the poisoning percentage and the degree of performance degradation, this is because TTT updates the whole parameters of the feature extractor, but other TTA methods only update the parameters in the BN layers.

\begin{figure*}[!t]
\centering
\subfloat[TTT \label{fig:loss_comp_a}]{%
\includegraphics[width=0.22\linewidth]{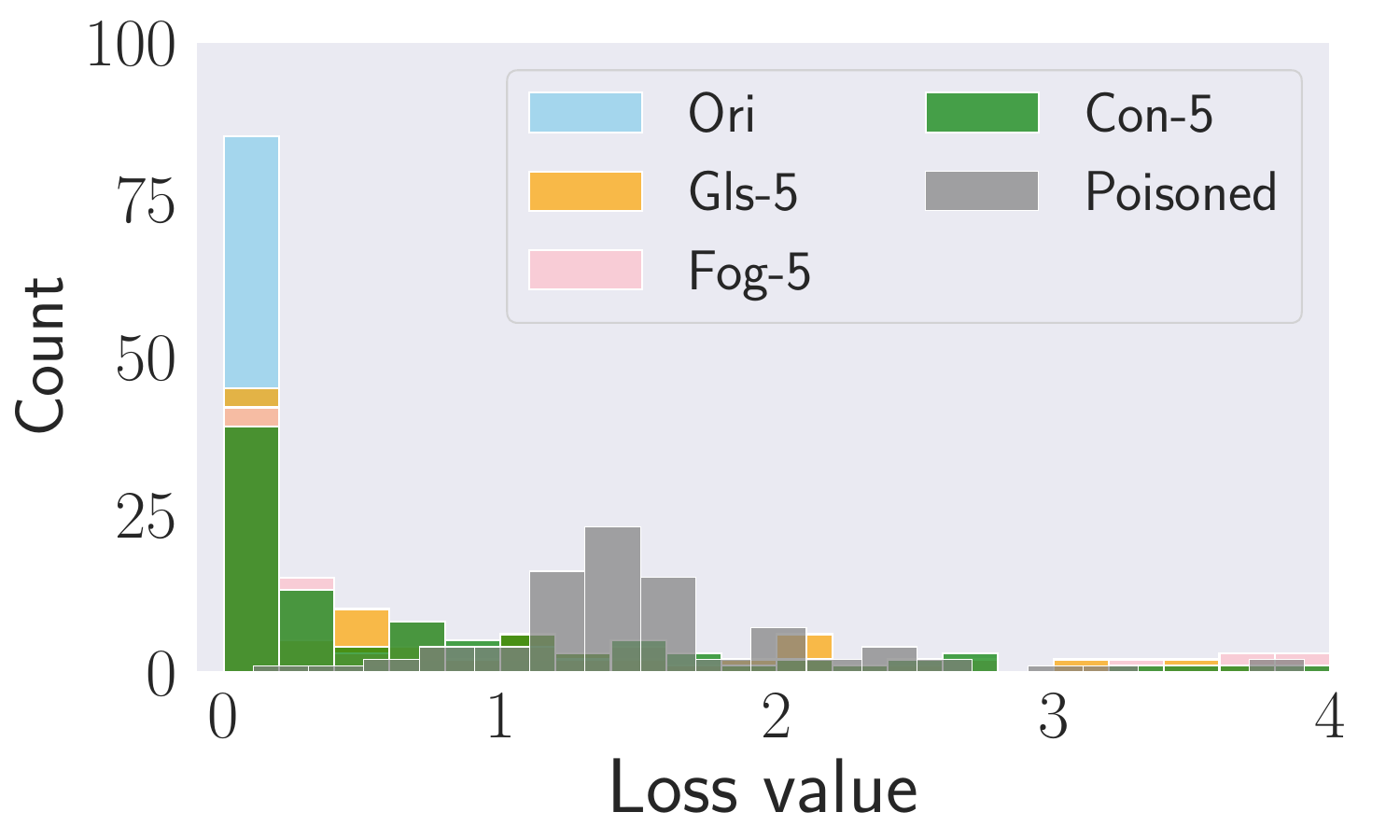}}
\hspace{5mm}
\subfloat[TENT \label{fig:loss_comp_b}]{%
\includegraphics[width=0.22\linewidth]{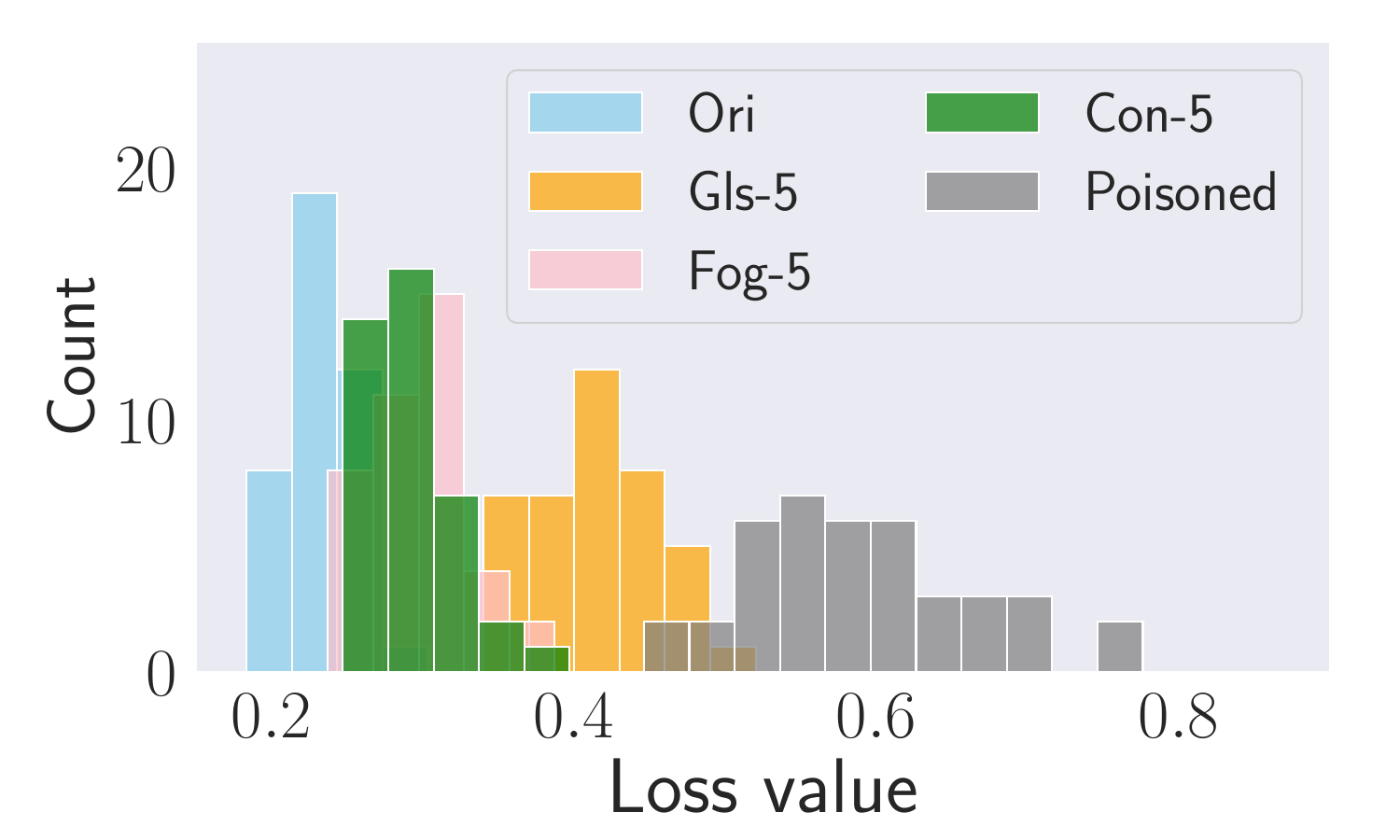}}
\hspace{5mm}
\subfloat[RPL \label{fig:loss_comp_c}]{%
\includegraphics[width=0.22\linewidth]{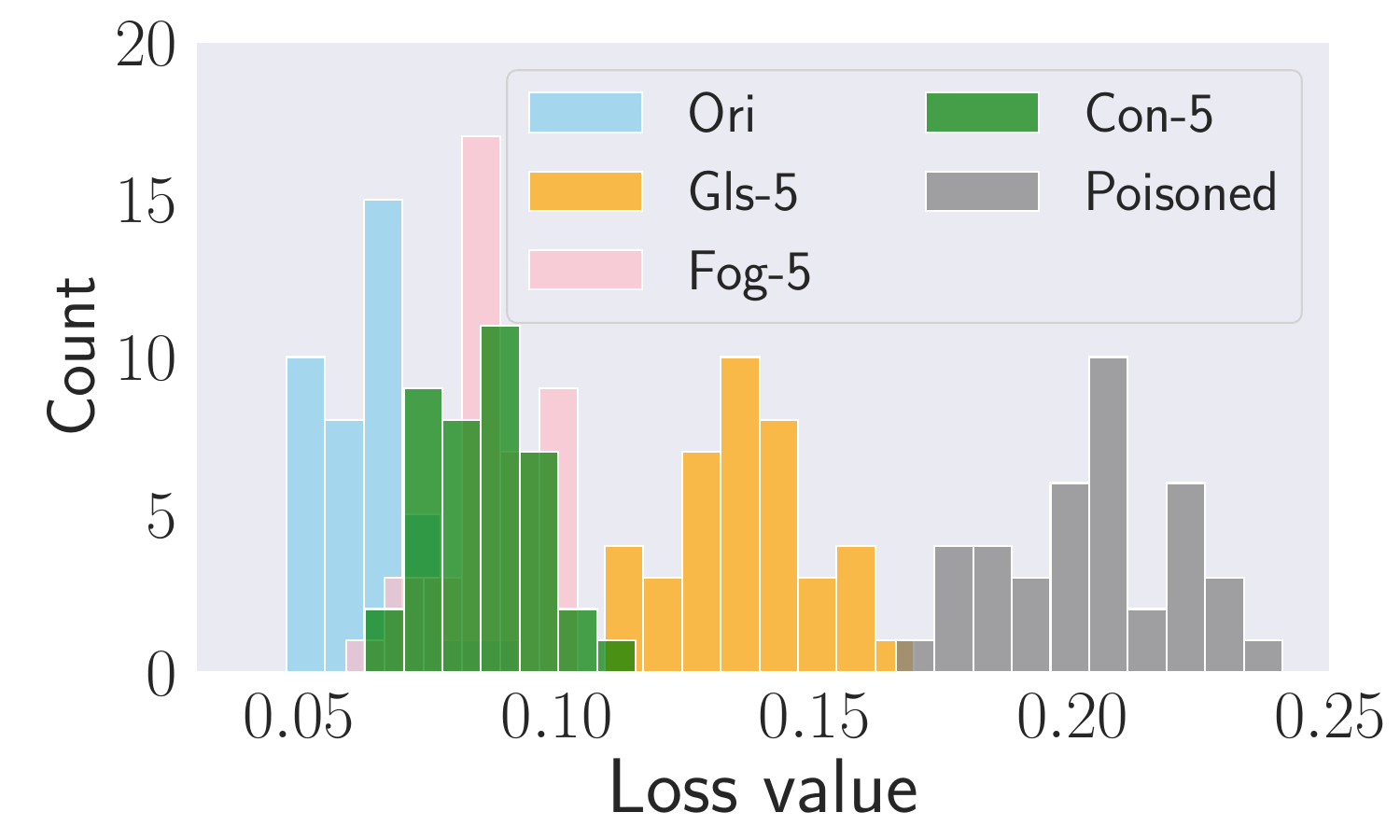}}
\caption{The statistics results of the loss values. 
The target model is ResNet-18 trained on CIFAR-10. 
Different colored bars indicate different types of arrived test samples.}
\label{fig:loss_comp} 
\end{figure*}

\subsection{Discussion}

\mypara{Loss Value}
To better explain why TePAs are successful, we visualize the statistics of the loss values, which are shown in~\autoref{fig:loss_comp}.
For instance, we feed 100 test samples (poisoned samples or benign samples) to the TTT-model.
For TENT- and RPL-models, we feed 40 batches of test samples.
First, we can observe that the poisoned samples indeed have greater loss values, which means the poisoned samples that have larger losses on the surrogate model can be transferred to the target model as well.
For instance, as shown in~\autoref{fig:loss_comp_a}, the losses of the benign samples are concentrated around 0.0.
However, the losses caused by poisoned samples are around 1.5.
Second, by comparing the results of TENT and RPL (\autoref{fig:loss_comp_b} and \autoref{fig:loss_comp_c}), we observe that RPL takes a smaller range of losses than TENT, which is the reason why RPL has less fluctuation of the performance caused by benign samples or poisoned samples.

\mypara{Non-i.i.d.\ Samples}
From~\autoref{fig:loss_comp} we also notice that the corrupted benign samples (non-adversarial) also have larger loss values than the original samples.
Therefore, we are curious that what is the impact of using non-i.i.d.\ samples to warm up the model.
In this part, we use one kind of corrupted data (e.g., Gls-5) to update the model first, then we evaluate the model on another kind of corrupted data (e.g., Fog-5).
The results are shown in \autoref{fig:non_iid}.
First, we can observe that i.i.d.\ samples are the most effective samples in improving performance.
Second, there is uncertainty about the impact of non-i.i.d.\ samples (increasing or decreasing), but poisoned samples can consistently reduce the performance to the largest extent.

\begin{figure*}[!t]
\centering
\subfloat[AT \label{fig:defense_poison_a}]{%
\includegraphics[width=0.2\linewidth]{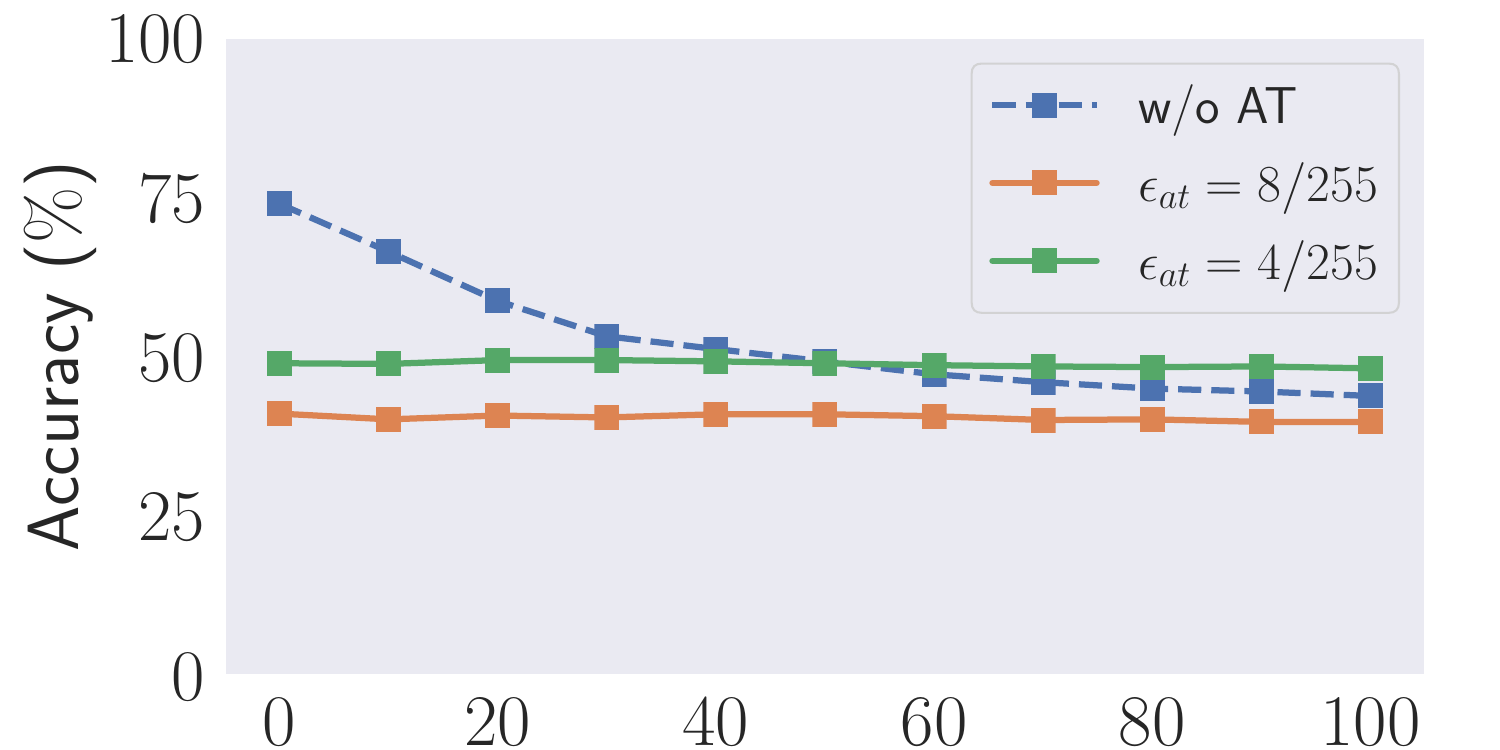}}
\hspace{2mm}
\subfloat[BDR \label{fig:defense_poison_b}]{%
\includegraphics[width=0.2\linewidth]{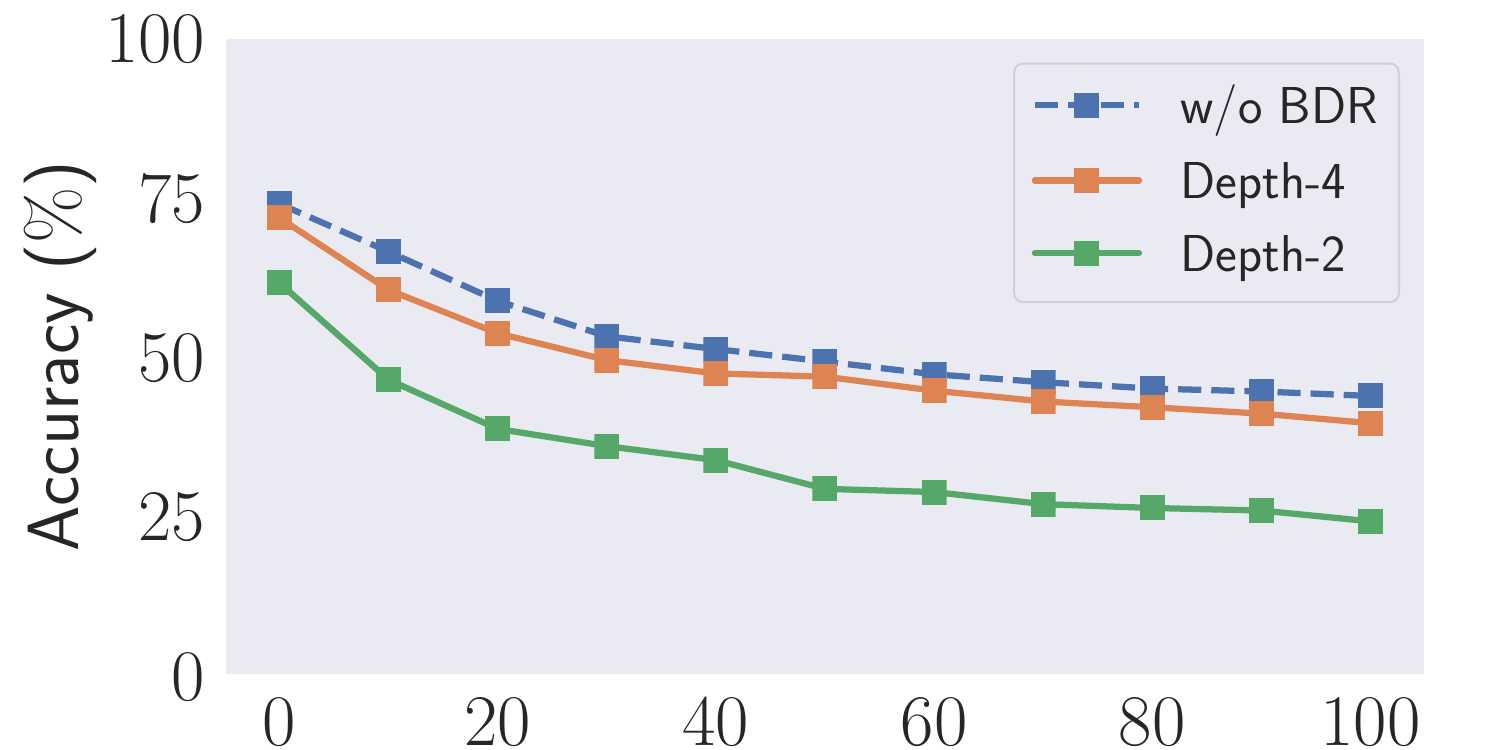}}
\hspace{2mm}
\subfloat[RRP\label{fig:defense_poison_c}]{%
\includegraphics[width=0.2\linewidth]{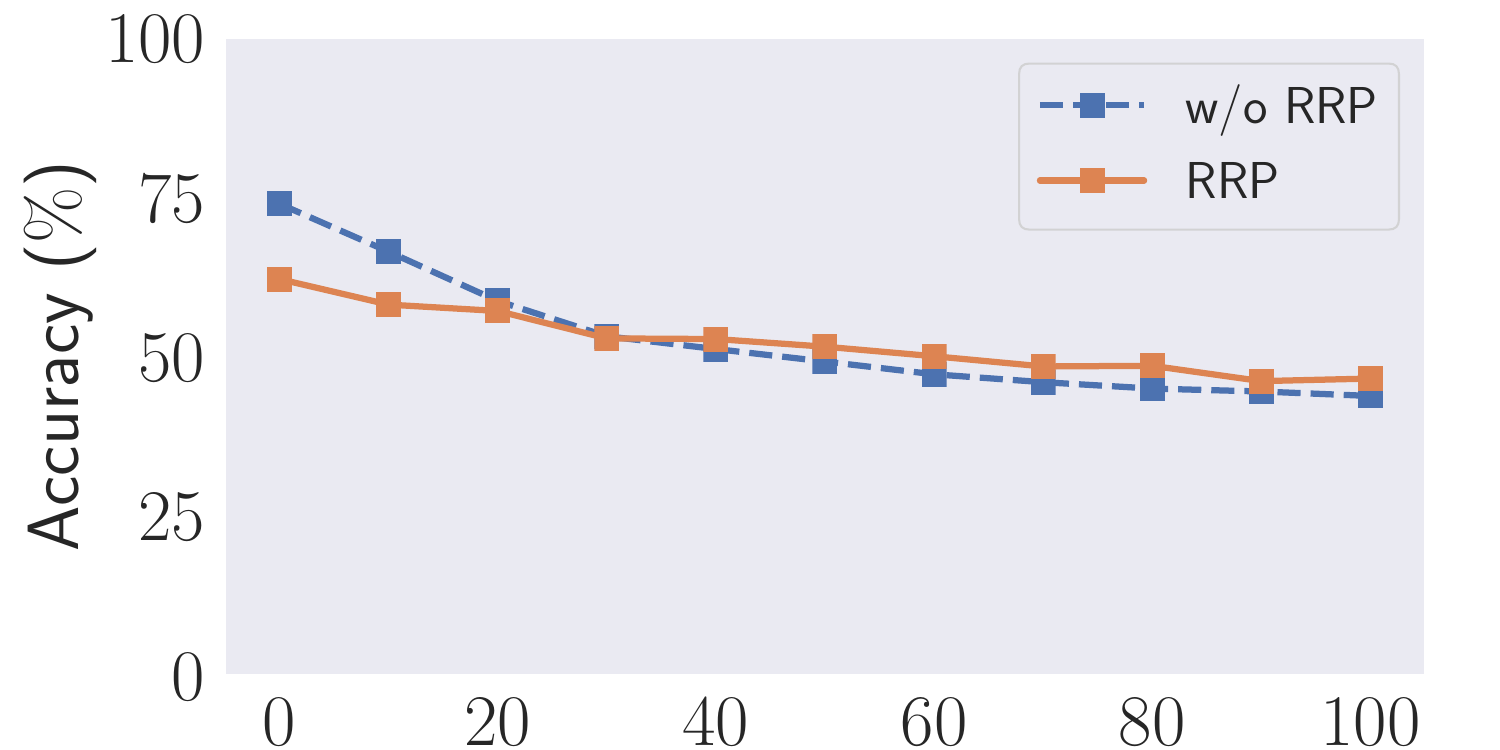}}
\hspace{2mm}
\subfloat[JC \label{fig:defense_poison_d}]{%
\includegraphics[width=0.2\linewidth]{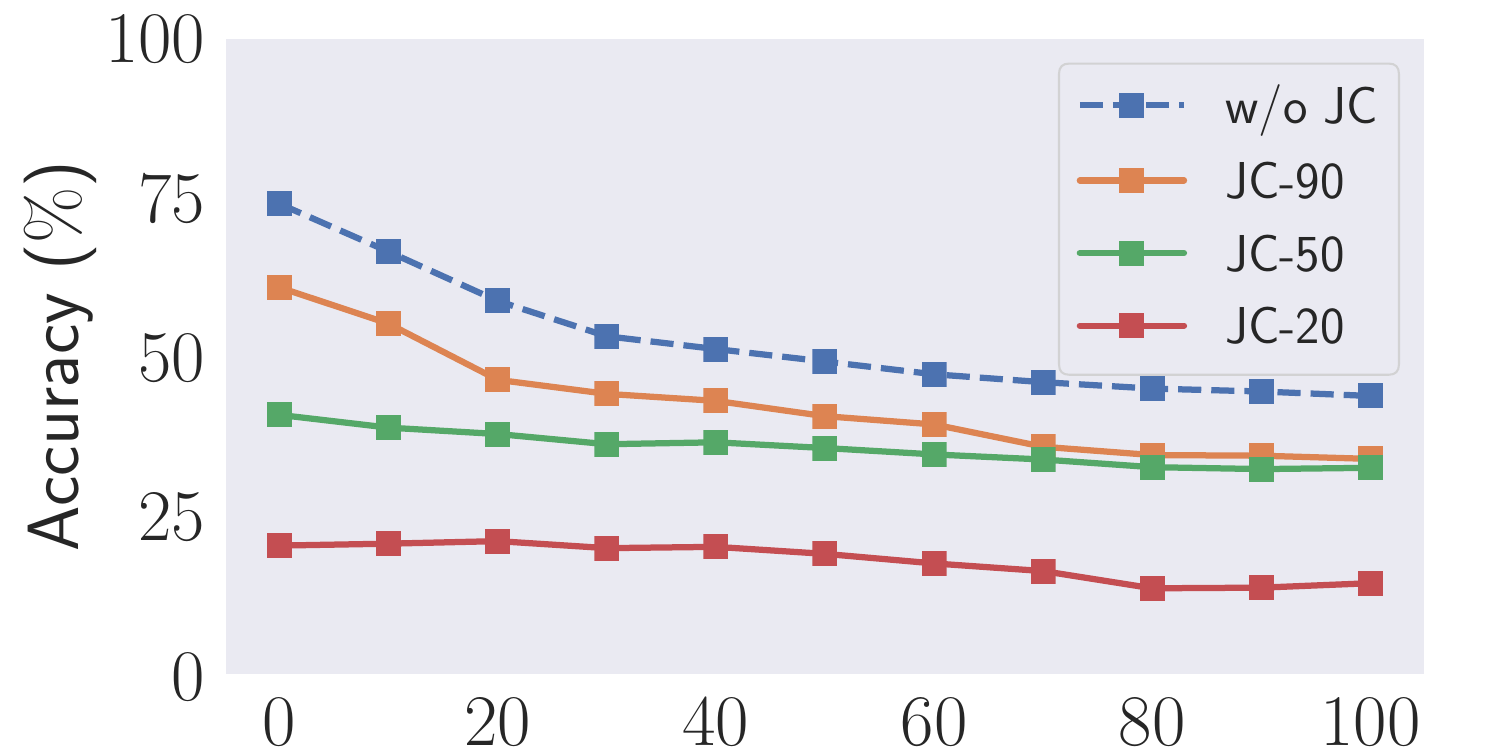}}
\caption{The impact of the four defense mechanisms on the poisoned samples. 
The x-axis represents the number of poisoned samples. 
We fix the target model to C10-Res18@Y4 and the evaluation dataset to Fog-5 of CIFAR-10-C.}
\label{fig:defense_poison} 
\end{figure*}

\section{Defense}

As we have shown before, TTA methods are vulnerable to TePAs. 
To mitigate the attacks, we discuss possible defenses against TePAs.
In this part, we take C10-Res18@Y4 adapted with TTT as the target model and take Fog-5 as the evaluation dataset.
We feed 100 poisoned samples to the target model, the experimental results are shown in \autoref{fig:defense_poison}.
Meanwhile, we discuss these defenses' impact on the benign samples in \autoref{fig:defense_iid}.
Other TTA methods show similar trends.

\mypara{Adversarial Training (AT)}
Since TePAs generate poisoned samples with adversarial attacks, one possible defense method is using adversarial training (AT)\cite{MMSTV18} to improve the target model's robustness on poisoned samples.
The detailed AT process for TTT-model is shown in Algorithm~\ref{alg:at_ttt}.
For instance, we train the target robust TTT-model $f_{at}$ using both the original training samples and adversarial examples generated by \texttt{PoiGen}.
Meanwhile, we use different perturbation budgets $\epsilon_{at}$ to generate adversarial examples.
From \autoref{fig:defense_poison_a}, we can observe that poisoned samples cannot reduce the target TTT-model's performance after AT.
However, AT can cause a reduction in the model's performance, which may be because the adversarial examples introduce a negative impact on the feature extractor of the model.
Meanwhile, another disadvantage of AT is its high computational cost.
For instance, the training time of 1 epoch for AT is $\sim 13$ times more than that of training the model without AT (We set $I_{iter}=2$ and $I_{adv}=5$ in AT), which means it is unrealistic to apply AT to larger-scale models.

\mypara{Bit-depth Reduction (BDR)}
Color bit-depths are used to present image pixels.
For instance, an 8-bit value can represent a pixel value in [0, 255].
Xu~\etal~\cite{XEQ18} found that reducing the bit depth is effective against adversarial attacks.
To verify if bit-depth reduction (BDR) can defend against TePAs, we reduce the bit-depths of the input poisoned samples to $4$ and $2$.
The results of the target model's performance are shown in \autoref{fig:defense_poison_b}.
We note that the evaluation data in calculating Acc are also crafted by BDR as the target model cannot distinguish the poisoned samples and benign samples (The same applies below defense methods).
We can observe that as the number of poisoned samples increases, the performance of the target model gradually decreases, which means BDR cannot defend against TePAs effectively.

\mypara{Random Resizing \& Padding (RRP)}
Adding a random resizing (RR) layer and a random padding (RP) layer is an effective way to build DNNs that are robust to the poisoned samples~\cite{XWZRY18}.
The first RR layer resizes the original input $x$ with size $W \times W \times C$ to a newly resized $x'$ with random size $W' \times W' \times C$, where $|W'-W|$ should be in a small range.
After that, the second RP layer outputs a new padding image $x''$ with padding zero pixels around the resized images.
For instance, it pads random $w$ zeros on the left and $h$ zeros on the top of $x'$, respectively.
We set $W'$ as the random numbers in $[32,40]$, and $w,h$ are the random numbers in $[0,W'-W]$.
We then leverage RRP to the poisoned samples.
From \autoref{fig:defense_poison_c}, we can observe that RRP can also not attenuate the impact of TePAs on model performance degradation.

\mypara{JPEG Compression (JC)}
JPEG Compression (JC) is another typical defense method to mitigate poisoning attacks~\cite{DGR16}.
In this part, we discuss the effect of JPEG quality on TePAs.
We choose $90$, $50$, and $10$ (out of $100$) as the JPEG quality values.
We use JC on the poisoned samples and show the results of the target model's performance in \autoref{fig:defense_poison_d}.
We can observe that poisoned samples after JC can still degrade the target model's performance.
Meanwhile, as the quality of the compressed image decreases, the performance of the target model decreases.

\section{Related Work}

\mypara{Domain Adaptation}
Improving the robustness of ML models under shifted distribution data is a longstanding problem~\cite{KSMXZBHYPGLDSGEHBLKPLFL21}.
Besides TTA, there are other methods to improve the ML model's robustness.
Domain-invariant methods~\cite{ZLLJ19,XLYL19} aim to learn embeddings that are invariant across different domains.
Transfer learning~\cite{ZYMK16,YJBK17,WYVZZ18} leverages embeddings from a pre-trained (teacher) model to train a new (student) model, which can work well on the new distribution data.
Semi-supervised learning~\cite{SBCZZRCKL20,BCCKSZR20,HLGZ22} trains the model on a mixed dataset with labeled and unlabeled data, and the use of the unlabeled data can improve the model's performance on shifted distribution data.
Self-supervised learning~\cite{CKNH20, HCXLDG22,CHZ22,SHYBZ23} trains models by large-scale unlabeled datasets, and the learned embeddings can be applied to the downstream tasks in different domains.

\mypara{Poisoning Attacks}
Poisoning attacks are one of the most exploited threats to modern data-driven ML models~\cite{Q22,WMWHQR22,LWHSZBCFZ22}. 
The attackers inject a small number of poisoned samples during the training process to sabotage the prediction performance of the target model at test time.
Poisoning attacks have been successfully applied to many ML settings, such as supervised learning~\cite{YWLC17,SHNSSDG18}, self-supervised learning~\cite{CT22}, federated machine learning~\cite{TTGL20,PYDSZ212}, etc.
The closest work to our attack is poisoning attacks against online learning (where data becomes available in sequential order and is used to update the best predictor for future data at each step)~\cite{S12,ZRWRCHZ21}.
Though the target model sequentially updates its parameters in both online settings and TTA settings, poisoning attacks against online learning still assume that the adversaries have partial knowledge of testing data and are white-box adversaries.
Our poisoning attack assumes neither.
We only assume the attackers have query access to the target model.

\section{Conclusion}

In this paper, we perform the first untargeted test-time poisoning attacks (TePAs) against four prominent TTA methods - TTT, DUA, TENT, and RPL.
Concretely, we propose a poisoned samples generation framework, \texttt{PoiGen}, which relies on surrogate models and transfer-based adversarial attacks to build adversarial examples and transfer such samples as poisoned samples to degrade the performance of the target TTA models.
Empirical evaluations show that TePAs can successfully break the target TTA models by degrading their performance to a large extent.
To mitigate the attacks, we investigate four defense mechanisms, i.e., adversarial training, bit-depth reduction, JPEG compression, and random resizing \& padding.
We observe that adversarial training (AT) can avoid the effect of poison attacks on target model degradation.
However, AT affects the performance of the target model.
Also, it enlarges the training effort, making it less possible to train the target models on large-scale datasets.
Moreover, we notice that the recovery of the target model’s performance is inevitable for our attacks. Even though the performance degradation is sufficient to cause a safety incident. We leave how to irreversibly degrade the target model’s performance as an interesting future work.
In summary, we demonstrate that the current TTA methods are prone to test-time poisoning attacks,
and we advocate for the integration of defenses against test-time poisoning attacks into the design of future TTA methods.

\section*{Acknowledgments}

We thank all anonymous reviewers and our shepherd for their constructive comments and valuable feedback.
This work is partially funded by the National Key R\&D Program of China (2018YFA0704701, 2020YFA0309705), Shandong Key Research and Development Program (2020ZLYS09), the Major Scientific and Technological Innovation Project of Shandong, China (2019JZZY010133), the Major Program of Guangdong Basic and Applied Research (2019B030302008), and the European Health and Digital Executive Agency (HADEA) within the project ``Understanding the individual host response against Hepatitis D Virus to develop a personalized approach for the management of hepatitis D'' (D-Solve) (grant agreement number 101057917).
Tianshuo Cong is also supported by Shuimu Tsinghua Scholar Program. 

\begin{small}
\bibliographystyle{plain}
\bibliography{normal_generated_py3}    
\end{small}

\appendix
\label{sec:appendix}

\section{Dataset}
\label{sec:appendix_dataset}

We leverage the following five datasets in our experiments:
\begin{itemize}
\item {\bf CIFAR-10~\cite{CIFAR} \& CIFAR-100~\cite{CIFAR}} both have 50k training images and 10k test images.
However, the images in CIFAR-10 are in 10 classes, and CIFAR-100 contains 100 types of images.
The size of each image is $32 \times 32 \times 3$.
We use the training datasets of CIFAR-10 and CIFAR-100 to train the target models.
Meanwhile, we use their test datasets to evaluate the model's performance on the original domain, which is denoted as ``Original (Ori).'' 
\item {\bf CIFAR-10-C~\cite{CIFAR10-C} \& CIFAR-100-C~\cite{CIFAR100-C}} are the corrupted version of CIFAR-10 and CIFAR-100, respectively.
For instance, they contain $15$ types of corruption in different categories, e.g., blur, weather, digital, etc.
Each corruption has $5$ different levels of severity, in which level $5$ is the severest level.
We choose three different types of corruptions, including Glass blur (Gls), Fog, and Contrast (Con) to evaluate the model's robustness on distribution shifts.
Note that Fog-5 stands for fog corruption with level 5. 
An illustration of CIFAR-10-C is shown in \autoref{fig::CIFAR-10_c}.
\item {\bf CINIC-10~\cite{DCAS18}} collects images from CIFAR-10 and ImageNet~\cite{DDSLLF09}.
For instance, the training dataset of CINIC-10 contains 20k images from CIFAR-10 and 70k images from ImageNet.
We use the 70k images from ImageNet as our surrogate dataset to train the surrogate model.
The size of each image in CINIC-10 is $32 \times 32 \times 3$.
\end{itemize}

\section{Notations}

\begin{table}[ht]
\centering
\customTableFont
\caption{Summary of the notations used in this paper.}
\begin{tabular}{l p{5cm}}
\toprule
{\bf Notation}   & {\bf Description} \\ \midrule
TePAs & Test-time Poisoning Attacks \\ \midrule
TrPAs & Training-time Poisoning Attacks \\ \midrule
C10-Res18  &  A ResNet-18 trained on CIFAR-10 dataset \\ \midrule
C10-Res18@Y4  &  A Y-structured ResNet-18 whose splitting point is the end of the 4th ResBlock and the training dataset is CIFAR-10 \\ \midrule
$\mathcal{D}_t$ & The training dataset of the target model \\ \midrule
$\mathcal{D}_s$ & The surrogate dataset \\ \midrule
$\mathcal{D}_e$ & The evaluation dataset \\ \midrule
$\mathcal{D}_w$ & The warming-up dataset  \\
\bottomrule
\end{tabular}
\label{tab:notation}
\end{table}

\section{Illustration of TTA Methods}
\label{sec:appendix_tta_method}

\begin{figure}[!t]
\centering
\includegraphics[width=5cm]{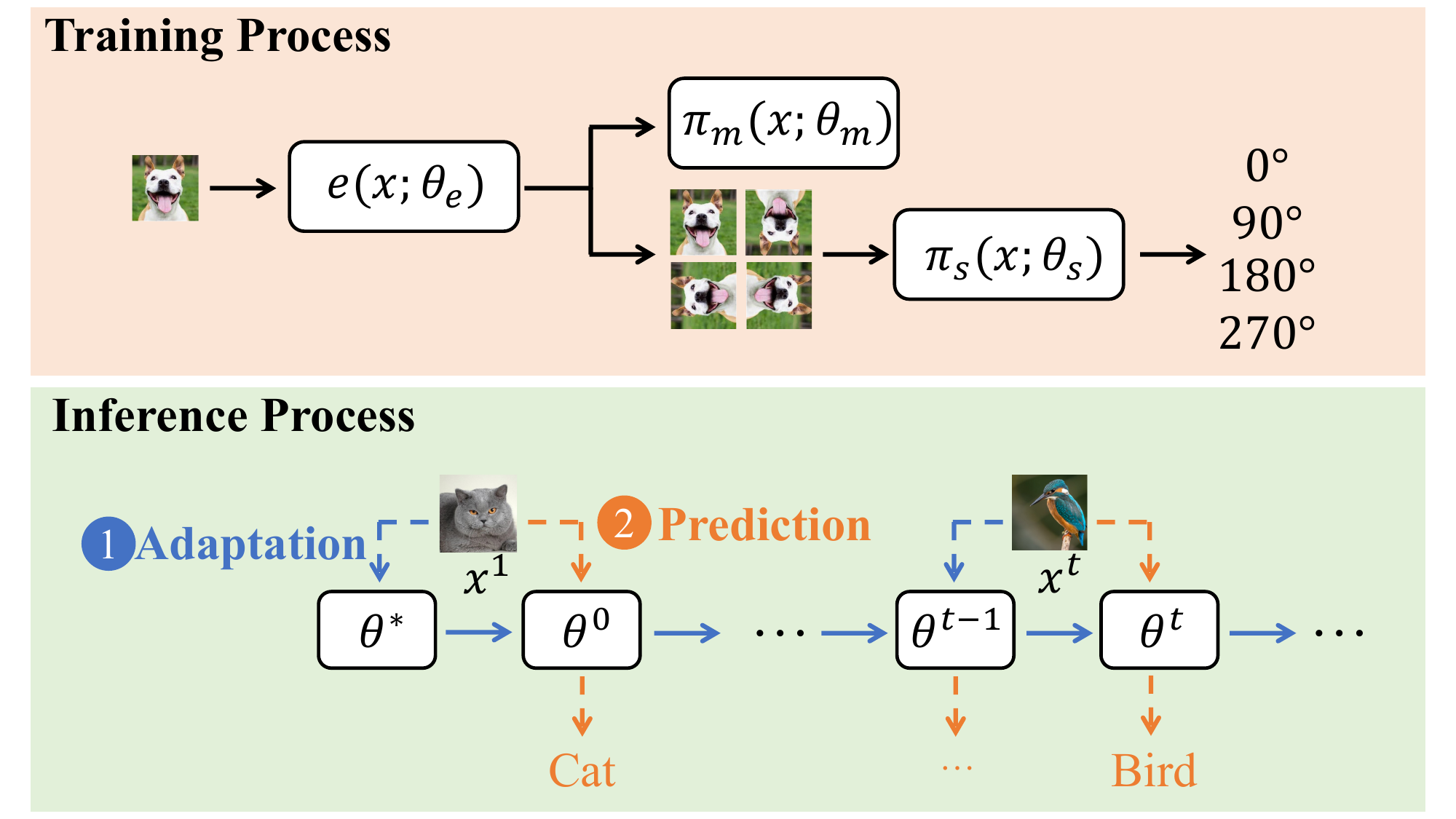}
\caption{Overview of TTT. 
First, TTT trains a Y-structured model with a rotation prediction task and a classification task.
During the inference process, TTT uses the rotation prediction loss to update the model first (colored in blue) and then uses the updated parameters to output the result (colored in orange).}
\label{fig:workflow_TTT}
\end{figure}

\begin{figure}[!t]
\centering
\includegraphics[width=5cm]{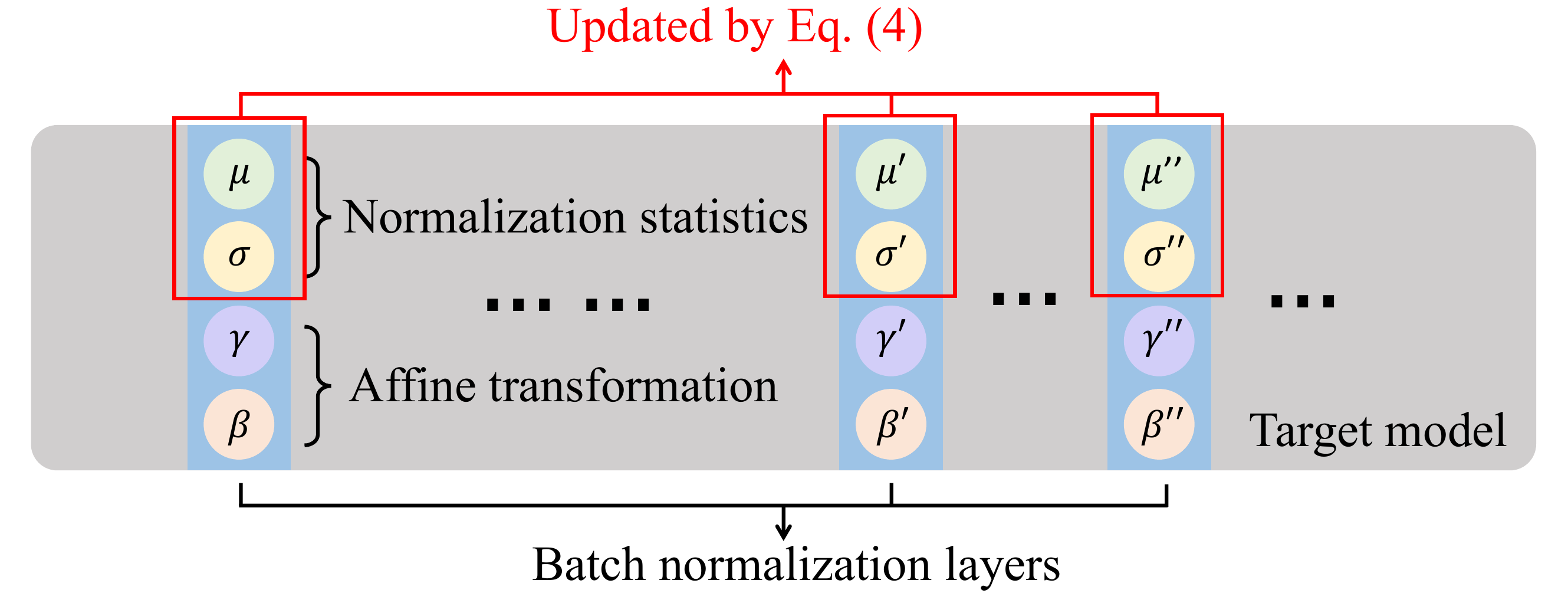}
\caption{Overview of DUA. 
DUA only adapts the normalization statistics in the BN layers of the target model, and all other parameters are frozen.}
\label{fig:workflow_DUA}
\end{figure}

\begin{figure}[!t]
\centering
\includegraphics[width=5cm]{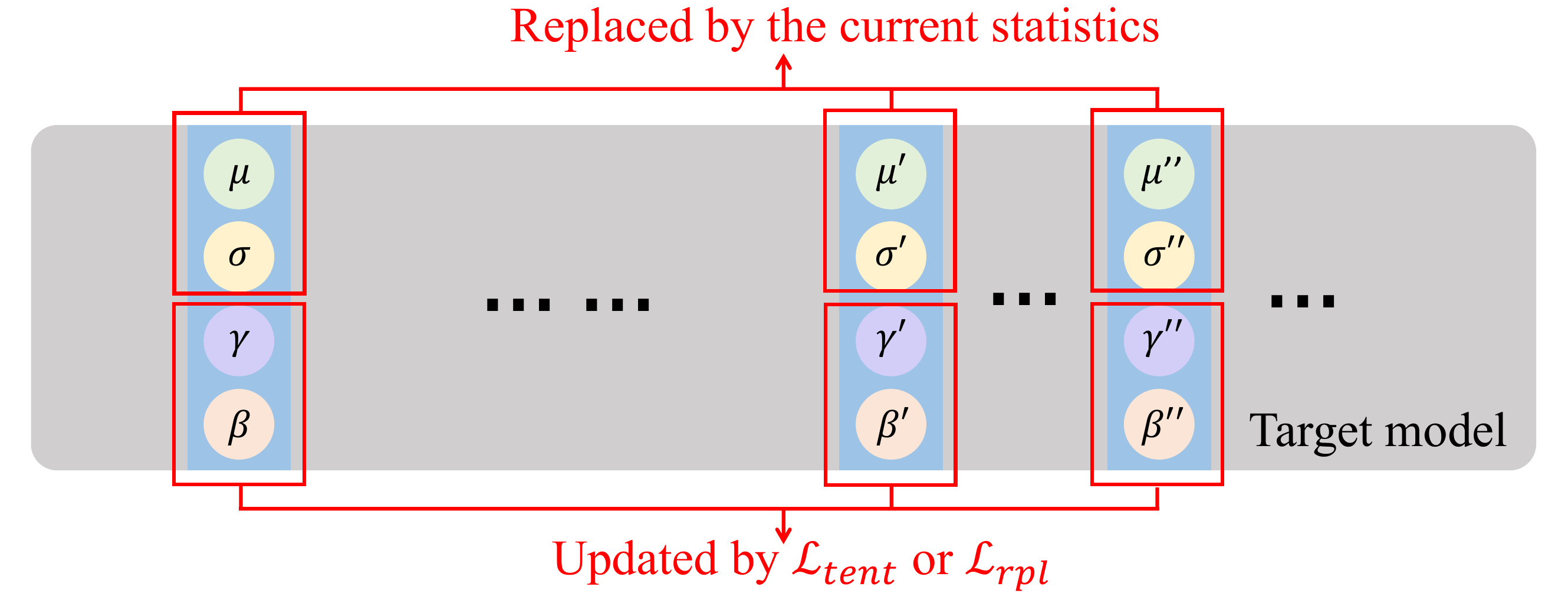}
\caption{Overview of TENT and RPL. 
TENT (RPL) adapts the affine transformation parameters of the BN layers by $\mathcal{L}_{tent}$ ($\mathcal{L}_{rpl}$), and they both replace the normalization statistics obtained from the original domain with the current transfer domain statistics.}
\label{fig:workflow_TENT}
\end{figure}

The illustrations of TTT, DUA, TENT, and RPL are shown in \autoref{fig:workflow_TTT}, \autoref{fig:workflow_DUA}, and \autoref{fig:workflow_TENT}.

\section{Ablation Study}

In this part, we discuss how the hyperparameters affect the performance of our TePAs.
We consider the perturbation budget $\epsilon$ and the severity level of corruption.
Here we use the ResNet18 trained on CIFAR-10 as the target model and use Fog-5 as the evaluation dataset for default.

\begin{figure}[!t] 
\centering
\subfloat[Perturbation Budget $\epsilon$ \label{fig:ablation_eps}]{%
\includegraphics[width=0.43\linewidth]{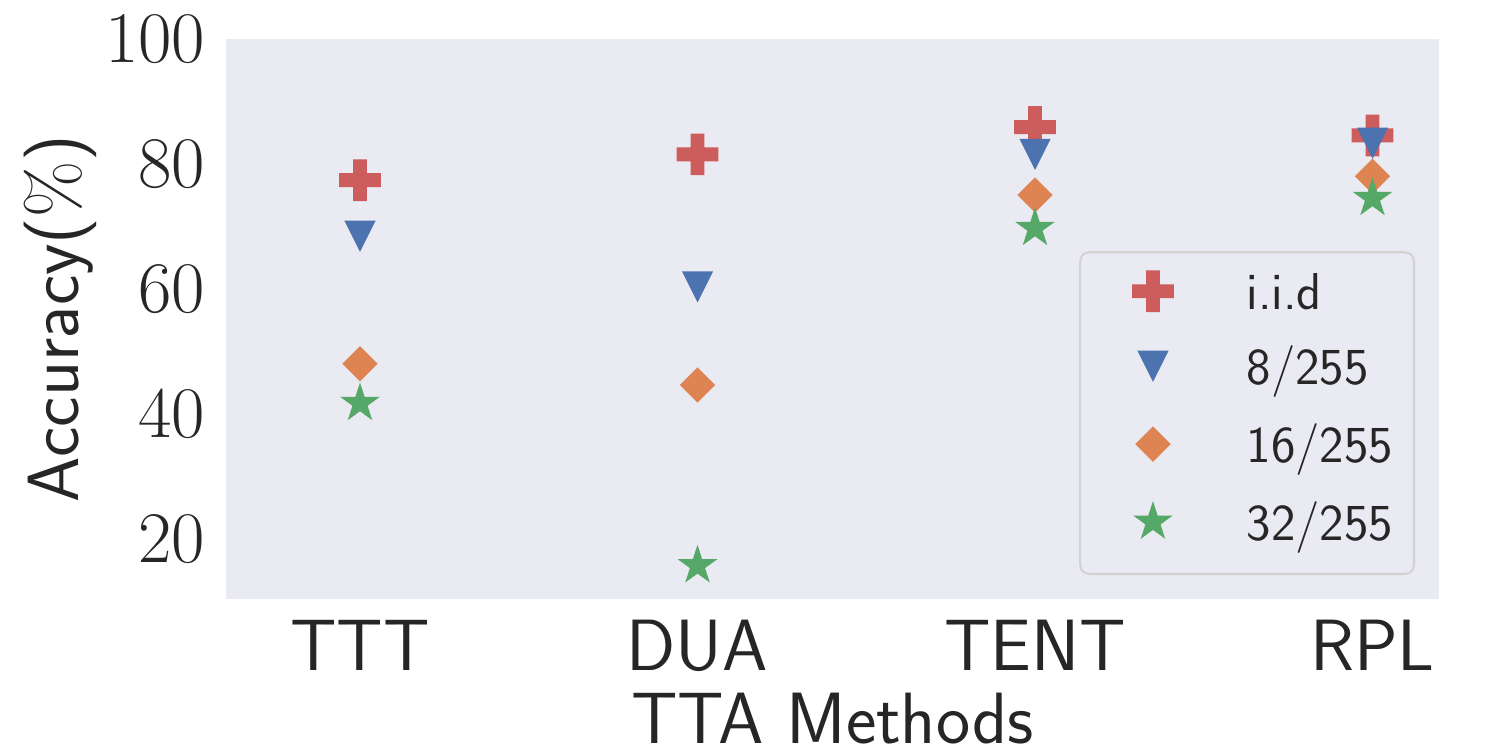}}
\hspace{2mm}
\subfloat[Corruption Level\label{fig:ablation_level}]{%
\includegraphics[width=0.43\linewidth]{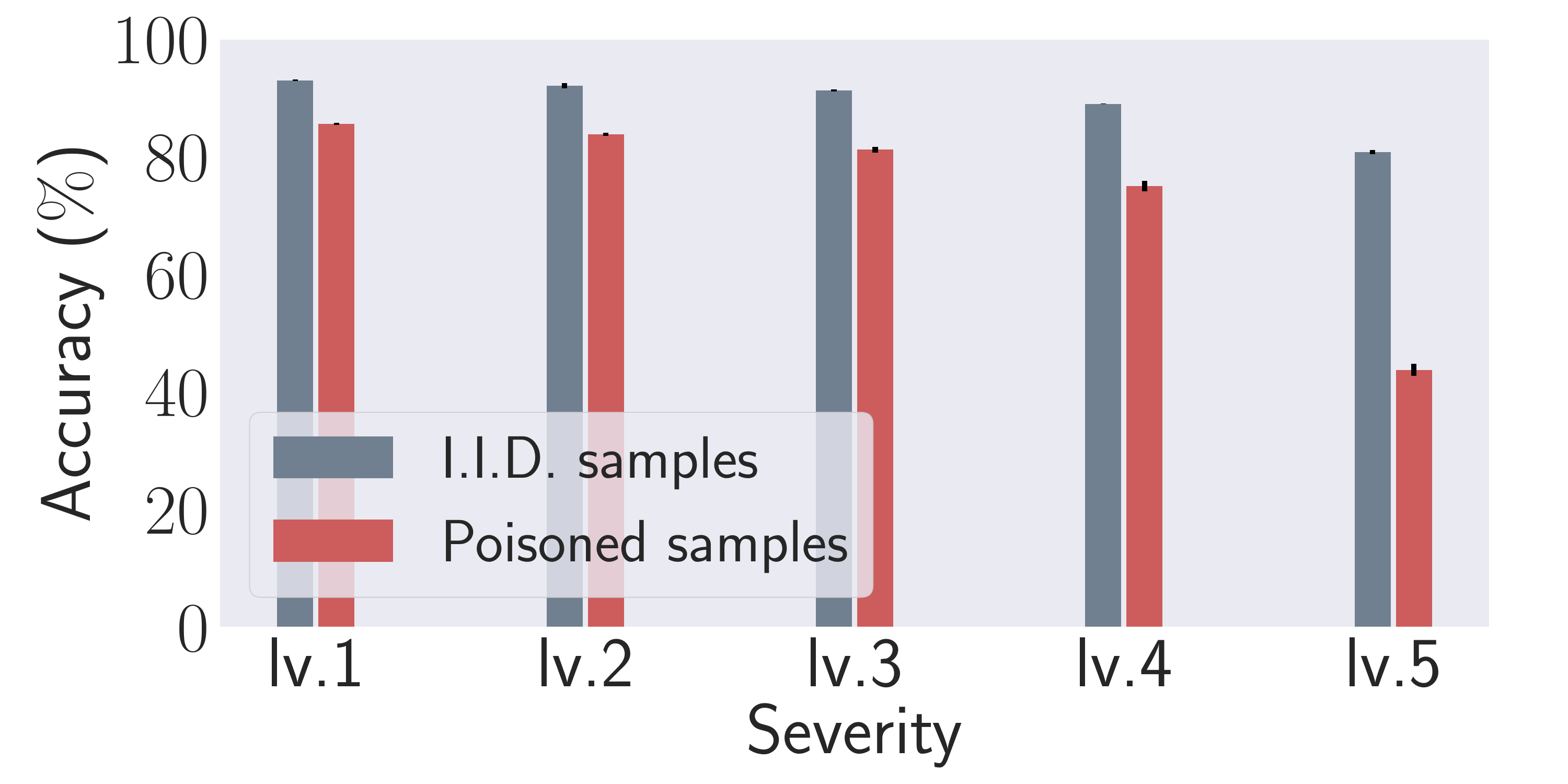}}
\caption{Ablation Study. 
The target model is ResNet-18 trained on CIFAR-10. We fix $\mathcal{D}_e$ to Fog-5 of CIFAR-10-C.}
\label{fig:ablation_study} 
\end{figure}

\mypara{The Impact of $\epsilon$}
We first show how $\epsilon$ affects the strength of TePAs.
We use different perturbation budgets, i.e., $\sfrac{8}{255}$, $\sfrac{16}{255}$ and $\sfrac{32}{255}$ to generate poisoned samples.
For TTT and DUA, we feed 100 test samples (i.i.d.\ samples or poisoned samples).
For TENT and RPL, we feed 40 batches of test samples.
The results are shown in \autoref{fig:ablation_eps}.
We can observe that for the four TTA methods, the greater the intensity of the perturbation contained in the poisoned samples, the more significant the degradation in model performance.

\mypara{The Impact of Corruption Level}
In the main body of our paper, we use the highest severity of corruption (Level 5) to evaluate the model's performance.
We here discuss the performance of TePAs on varying degrees of corruption.
We feed 100 i.i.d.\ or poisoned samples to the target TTT-model.
The results are shown in~\autoref{fig:ablation_level}.
First, we can observe that the performance of the model decreases as the level increases.
Second, TePAs could reduce the ability of the model on all levels of corruption.

\section{Additional Experimental Results}

\begin{figure}[!t]
\centering
\includegraphics[width=5cm]{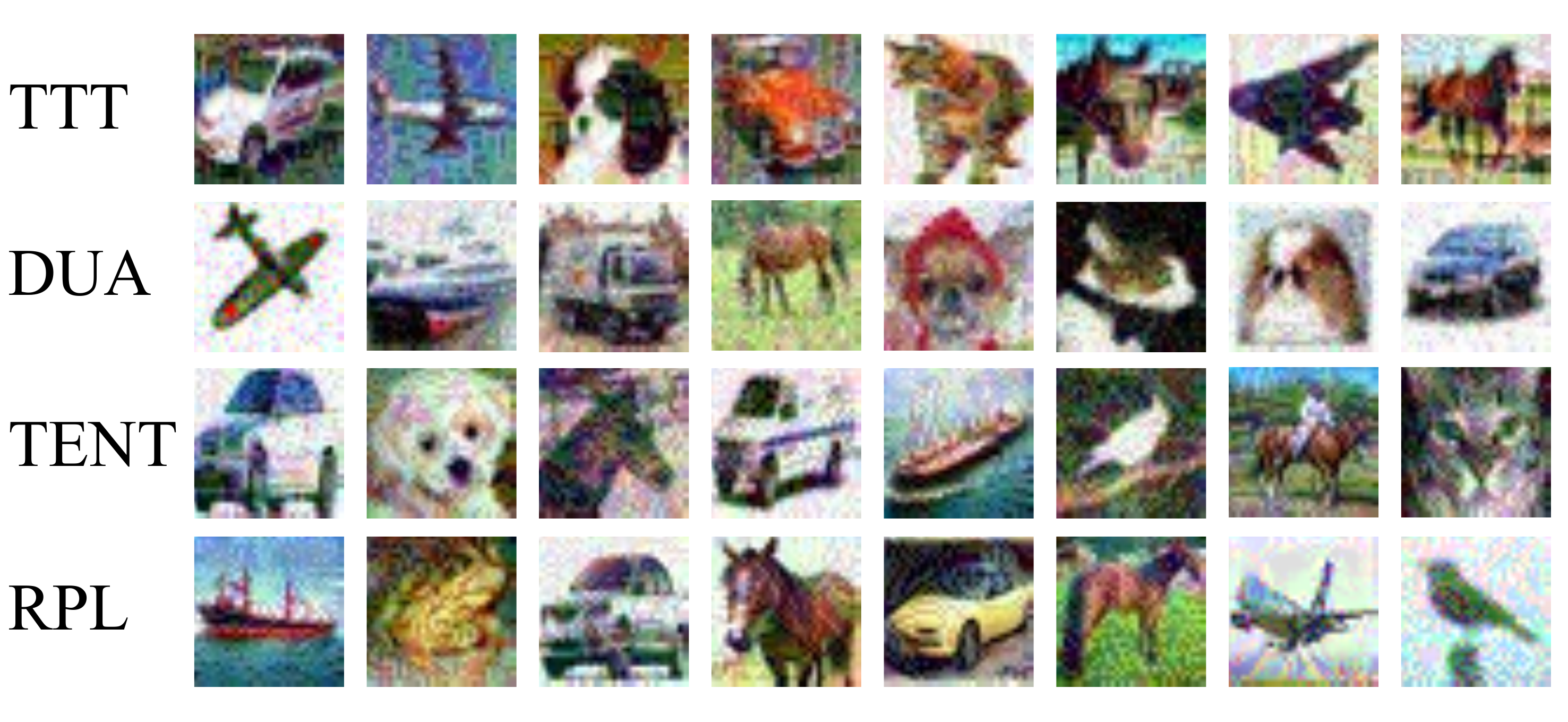}
\caption{Visualization results of the poisoned samples.}
\label{fig::poison_eps32}
\end{figure}

\begin{figure}[!t]
\centering
\includegraphics[width=6cm]{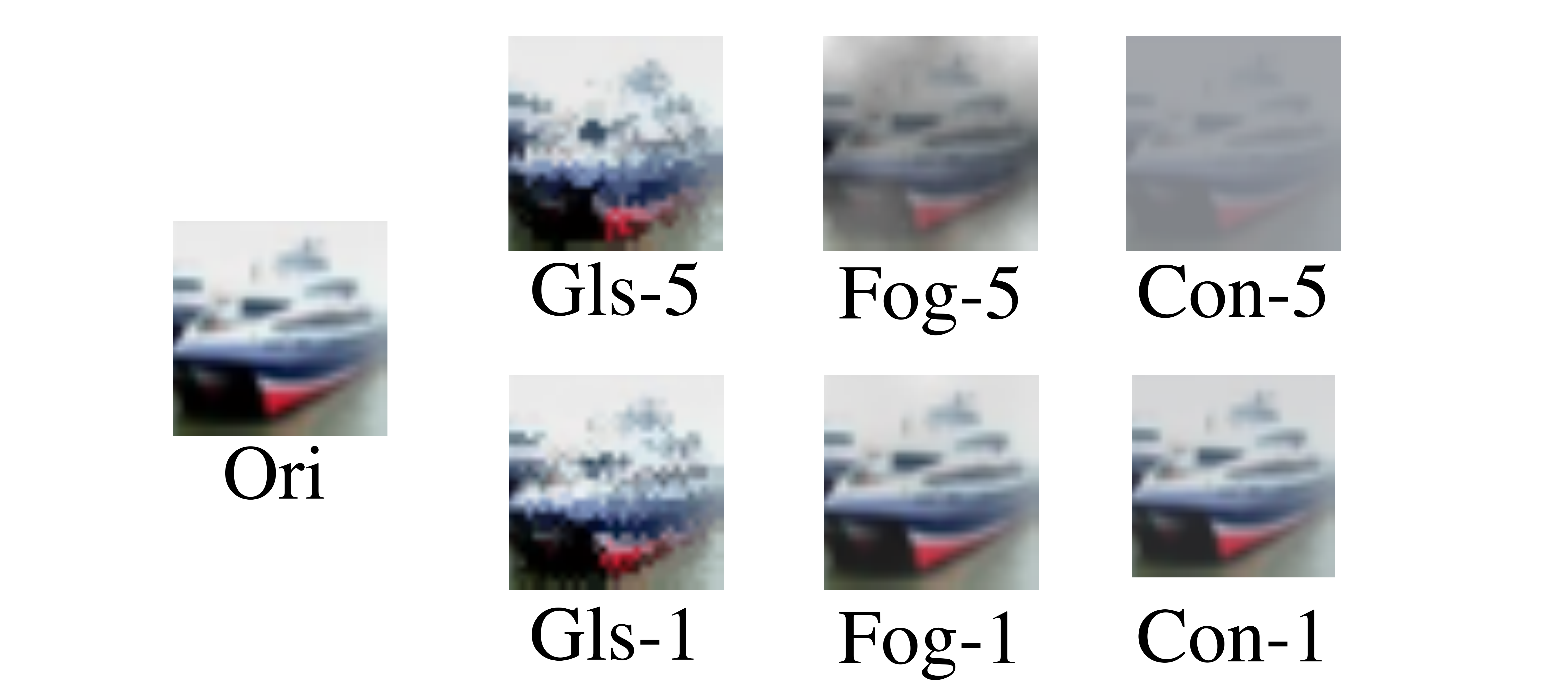}
\caption{Visualization results of the benign images from CIFAR-10-C.}
\label{fig::CIFAR-10_c}
\end{figure}

\begin{figure}[!t]
\centering
\subfloat[TTT \label{fig:non_iid_a}]{%
\includegraphics[width=0.3\linewidth]{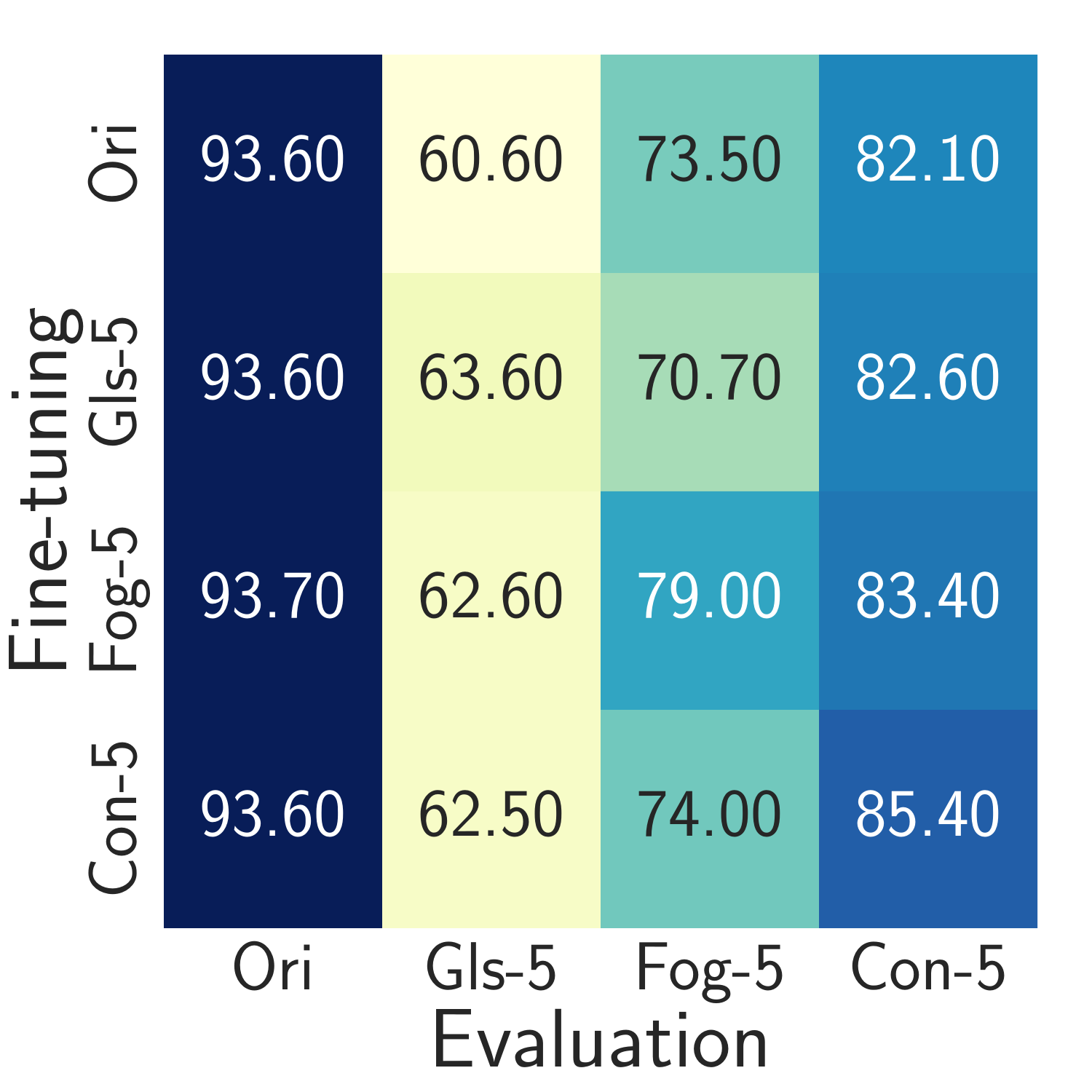 }}
\hspace{3mm}
\subfloat[DUA \label{fig:non_iid_b}]{%
\includegraphics[width=0.3\linewidth]{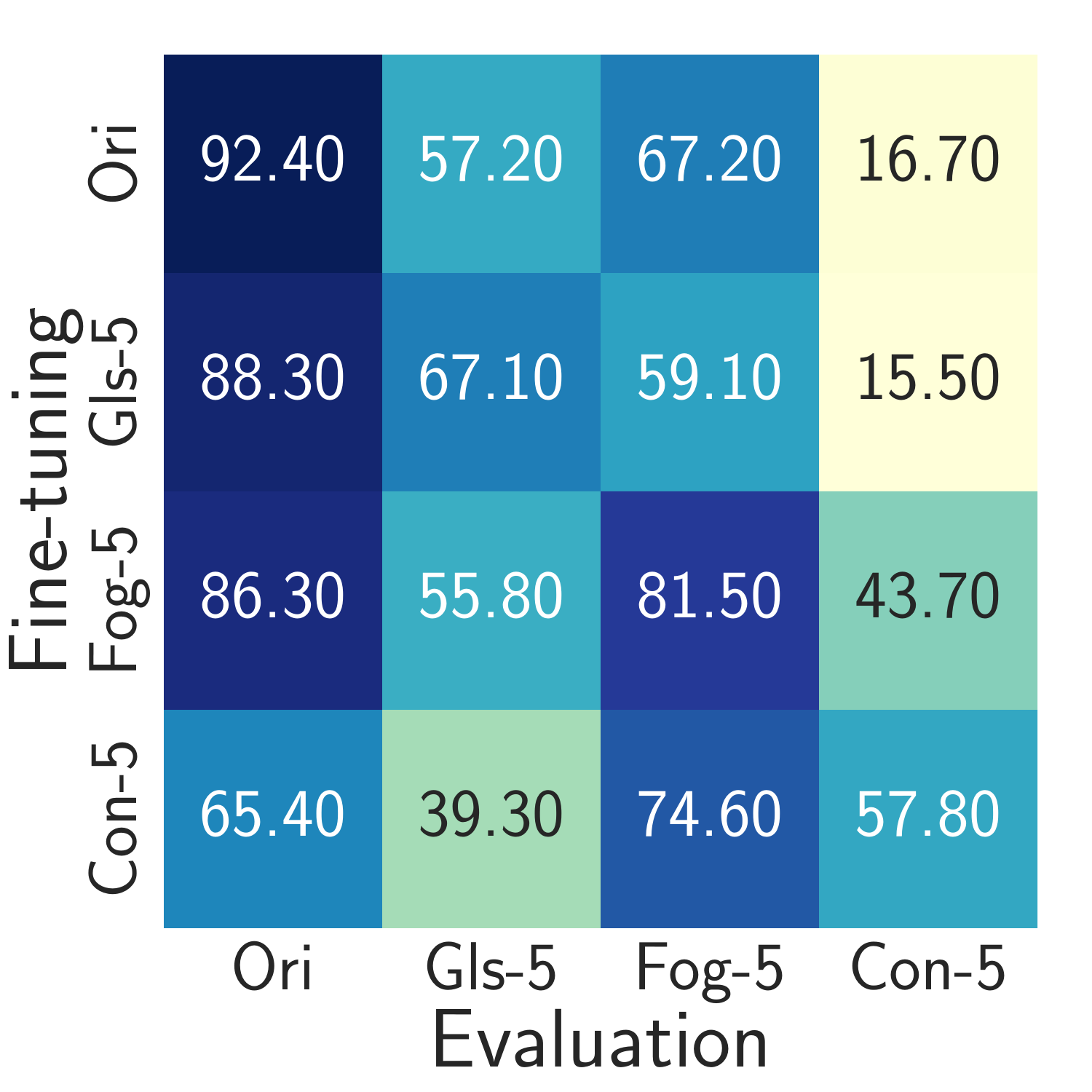 }}
\\
\subfloat[TENT\label{fig:non_iid_c}]{%
\includegraphics[width=0.3\linewidth]{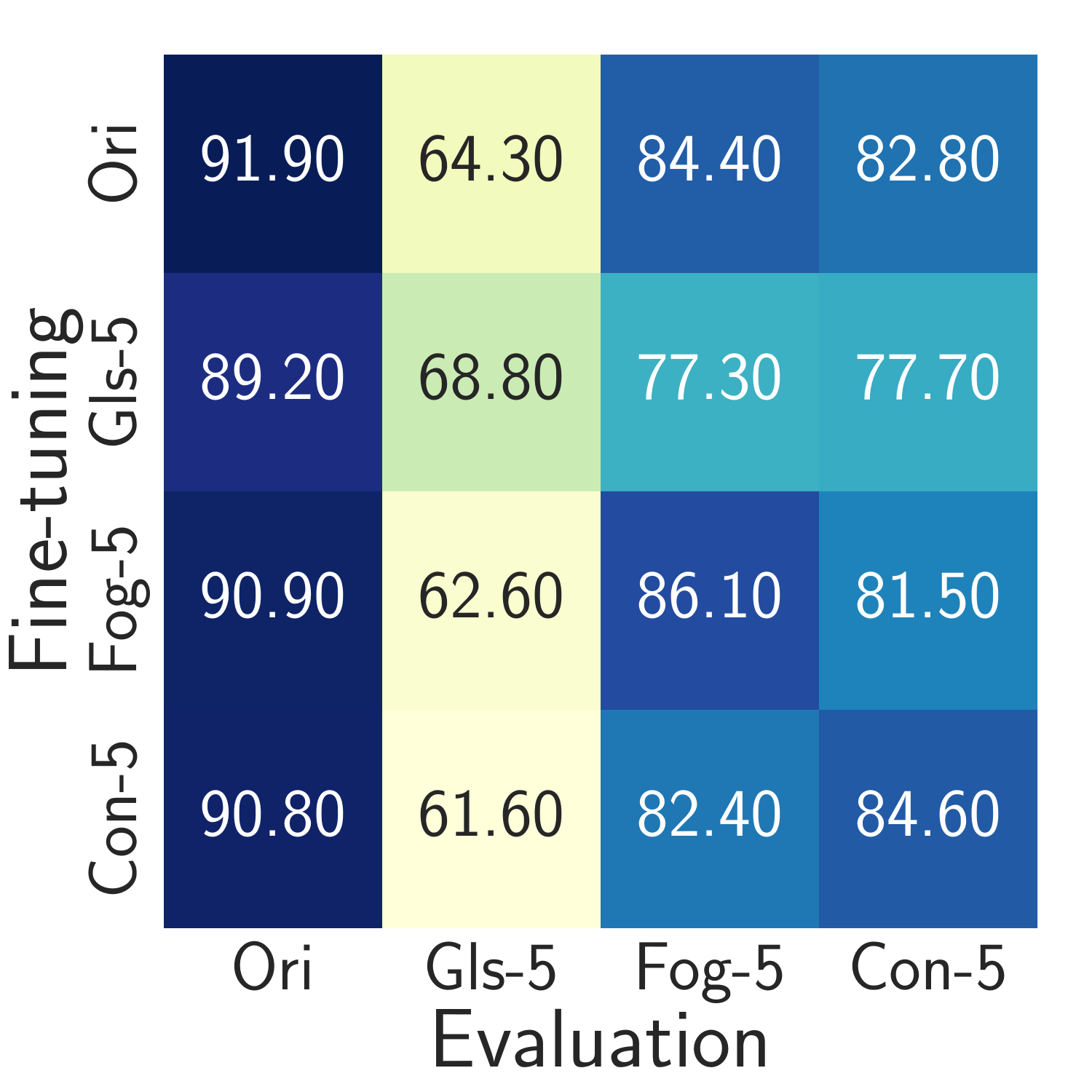 }}
\hspace{3mm}
\subfloat[RPL \label{fig:non_iid_d}]{%
\includegraphics[width=0.3\linewidth]{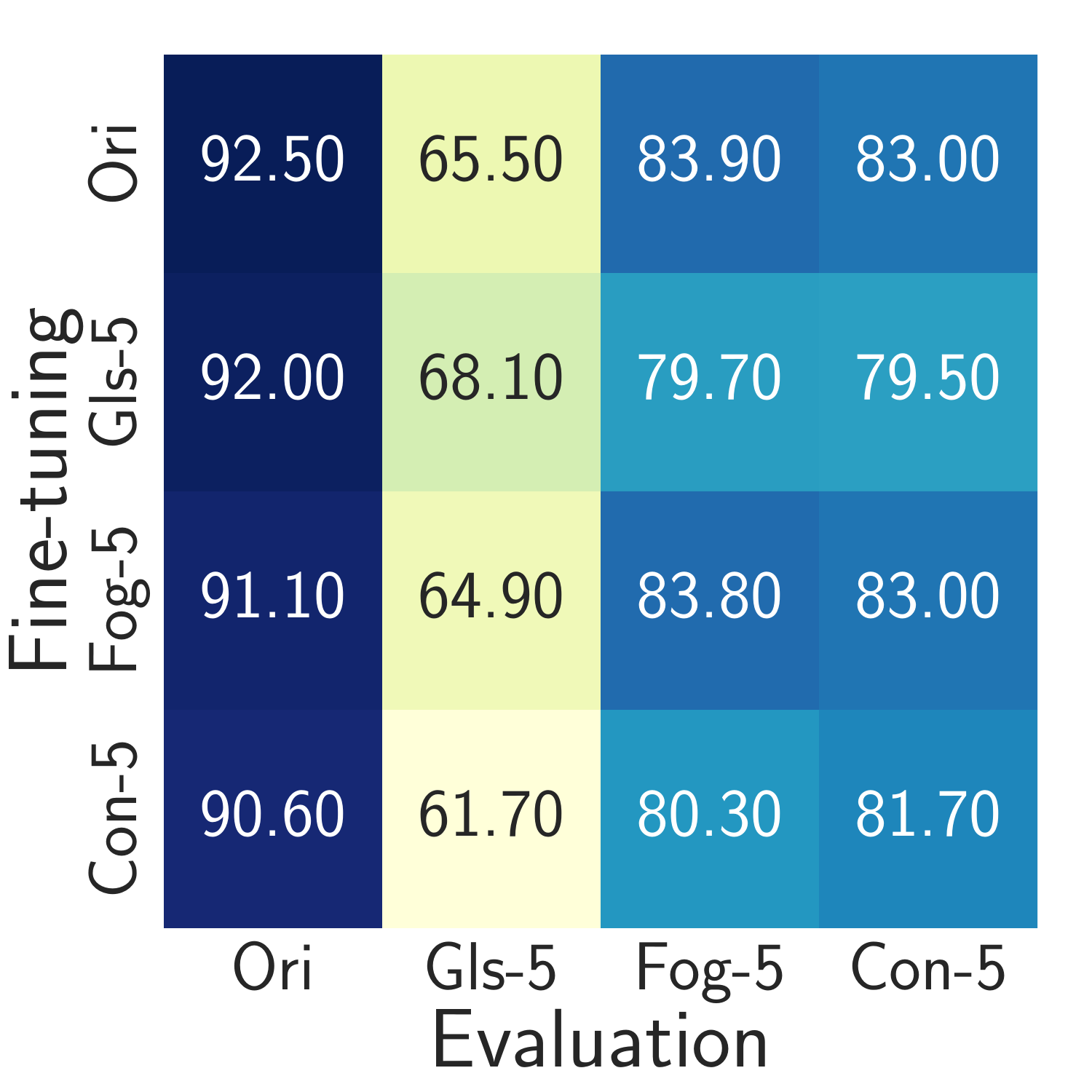 }}
\caption{The test samples used to fine-tune the target model and the evaluation samples are non-i.i.d..
The y-axis represents the corruption types of the fine-tuning samples. 
The x-axis stands for the evaluation dataset.}
\label{fig:non_iid} 
\end{figure}

\begin{figure*}[!t]
\centering
\subfloat[TTT \label{fig:utility_tta_res50_CIFAR-100_a}]{%
\includegraphics[width=0.2\linewidth]{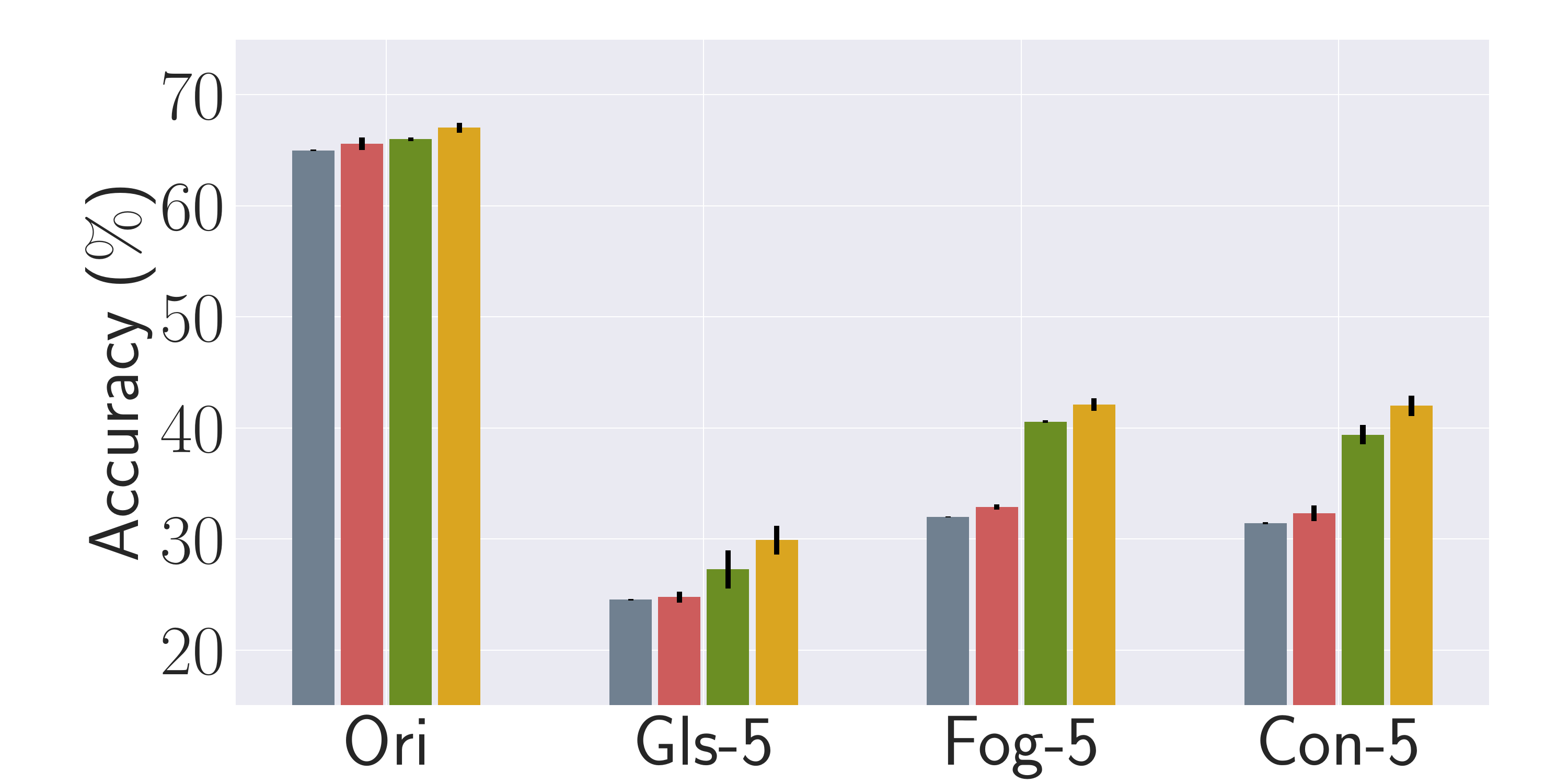}}
\hspace{3mm}
\subfloat[DUA \label{fig:utility_tta_res50_CIFAR-100_b}]{%
\includegraphics[width=0.2\linewidth]{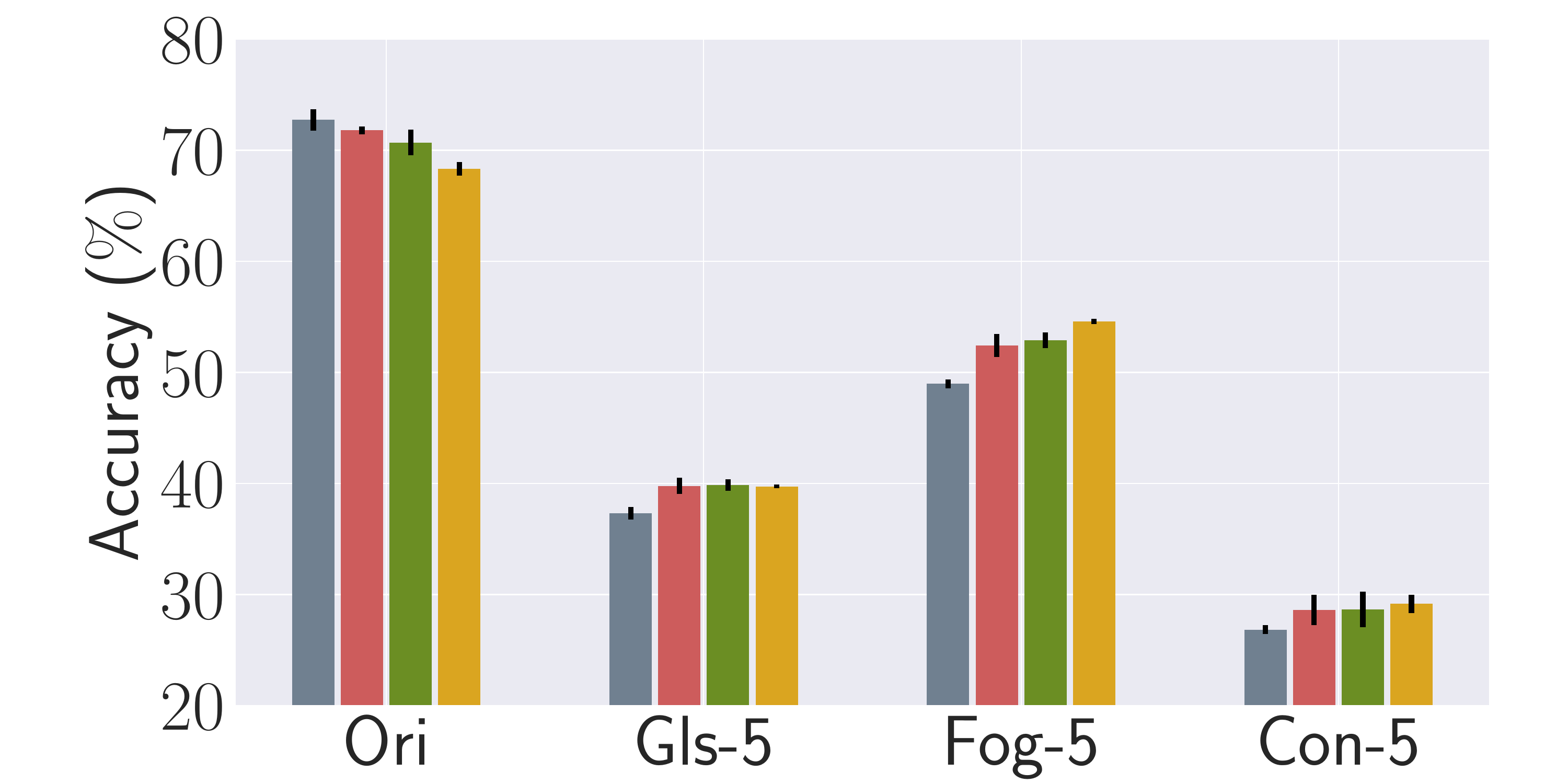}}
\hspace{3mm}
\subfloat[TENT\label{fig:utility_tta_res50_CIFAR-100_c}]{%
\includegraphics[width=0.2\linewidth]{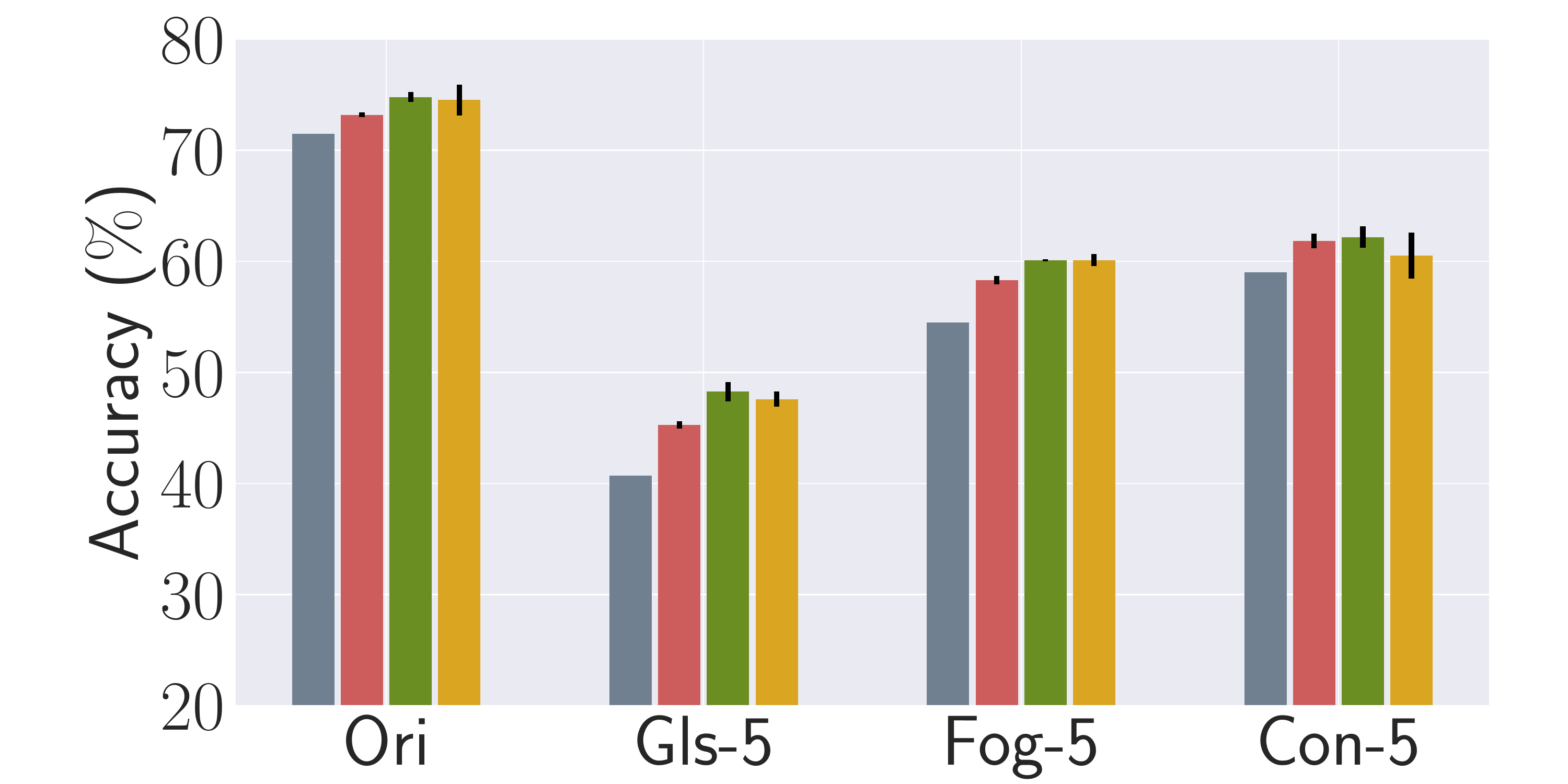}}
\hspace{3mm}
\subfloat[RPL \label{fig:utility_tta_res50_CIFAR-100_d}]{%
\includegraphics[width=0.2\linewidth]{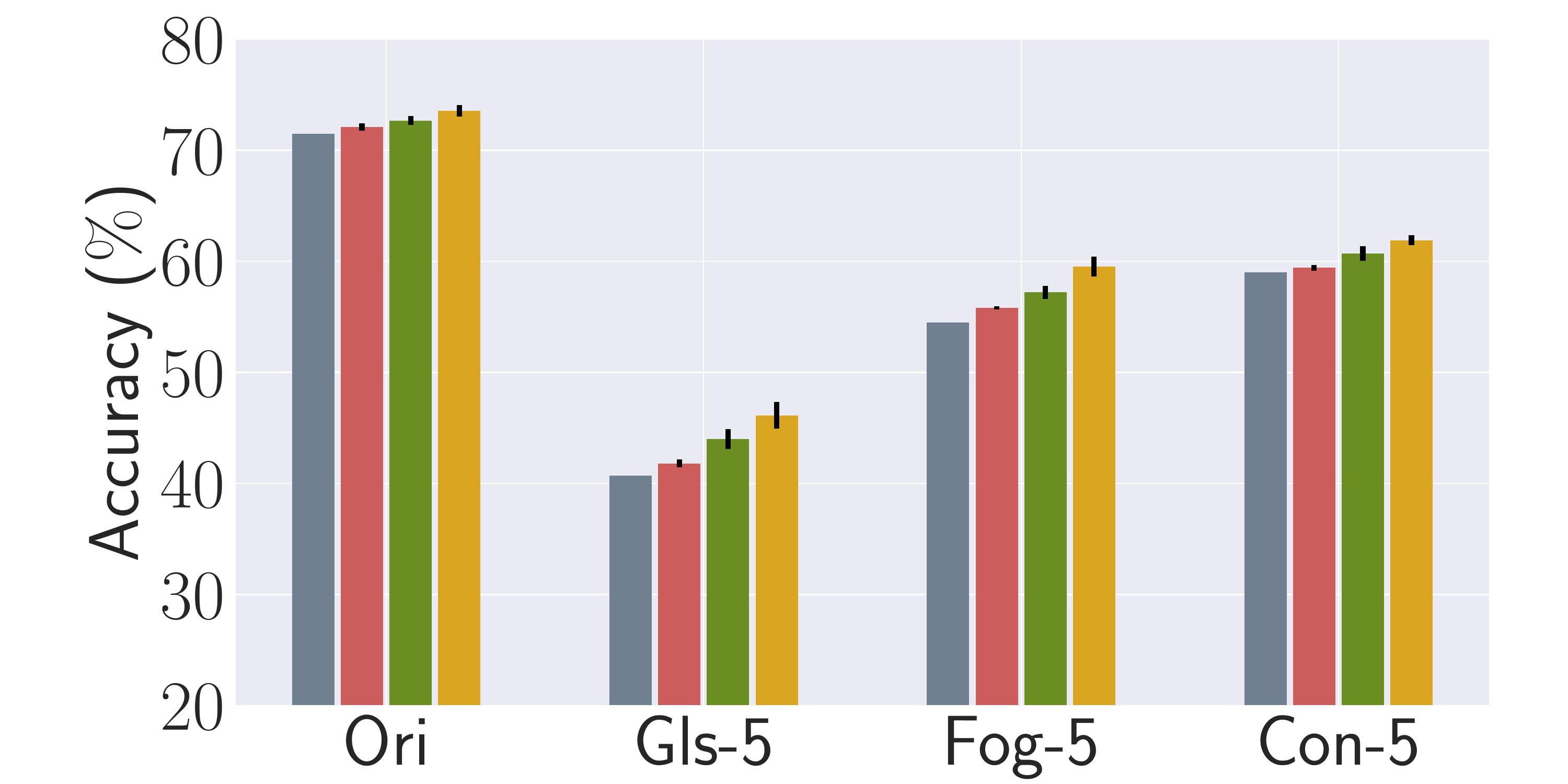}}
\caption{Utility of TTA methods. 
The target model is ResNet-50 trained on CIFAR-100. 
The x-axis represents different evaluation datasets. 
The y-axis represents the prediction accuracy.}
\label{fig:utility_tta_res50_CIFAR-100} 
\end{figure*}

\begin{figure*}[!t]
\centering
\subfloat[TTT \label{fig:clean_poison_ori_a}]{%
\includegraphics[width=0.15\linewidth]{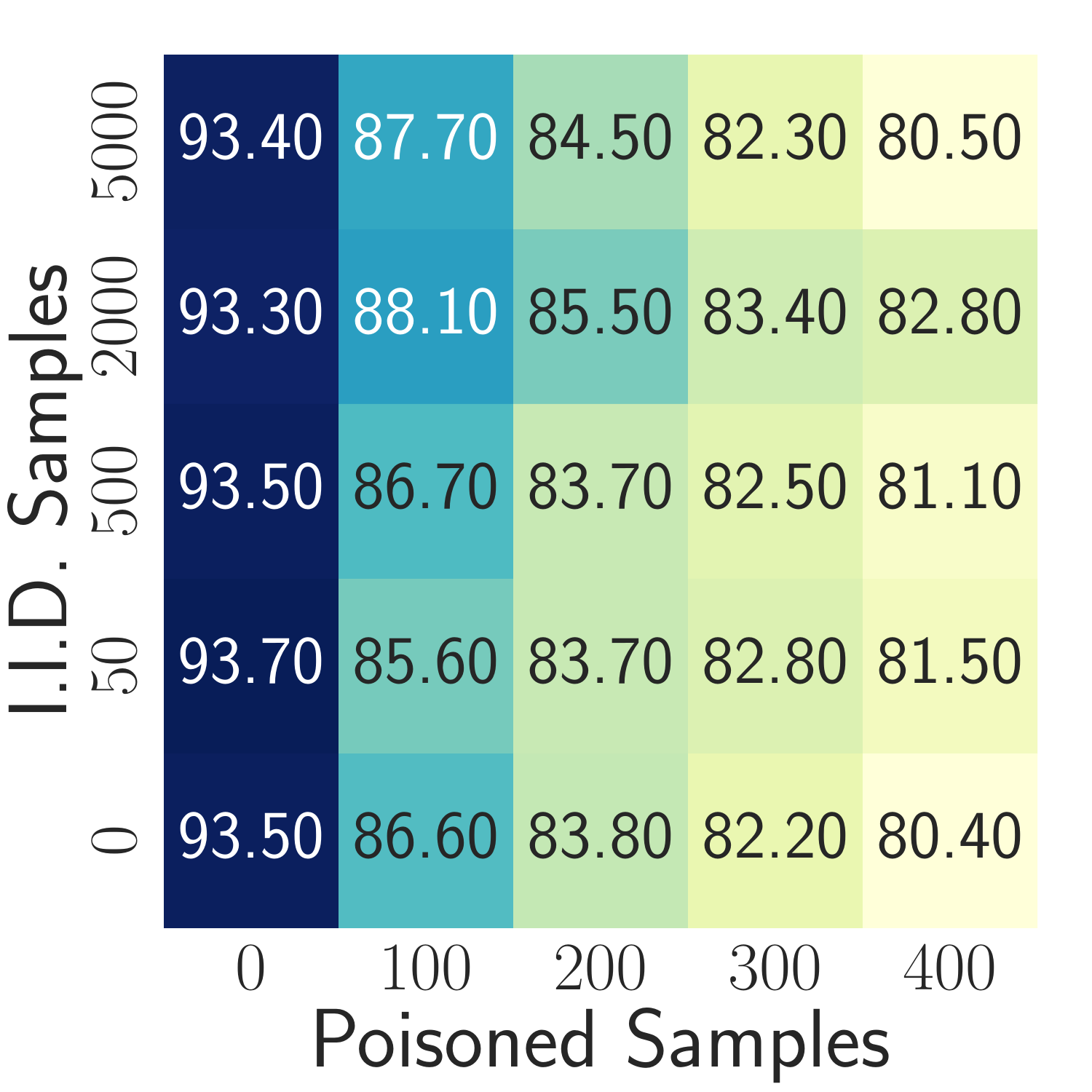}}
\hspace{6mm}
\subfloat[DUA \label{fig:clean_poison_ori_b}]{%
\includegraphics[width=0.15\linewidth]{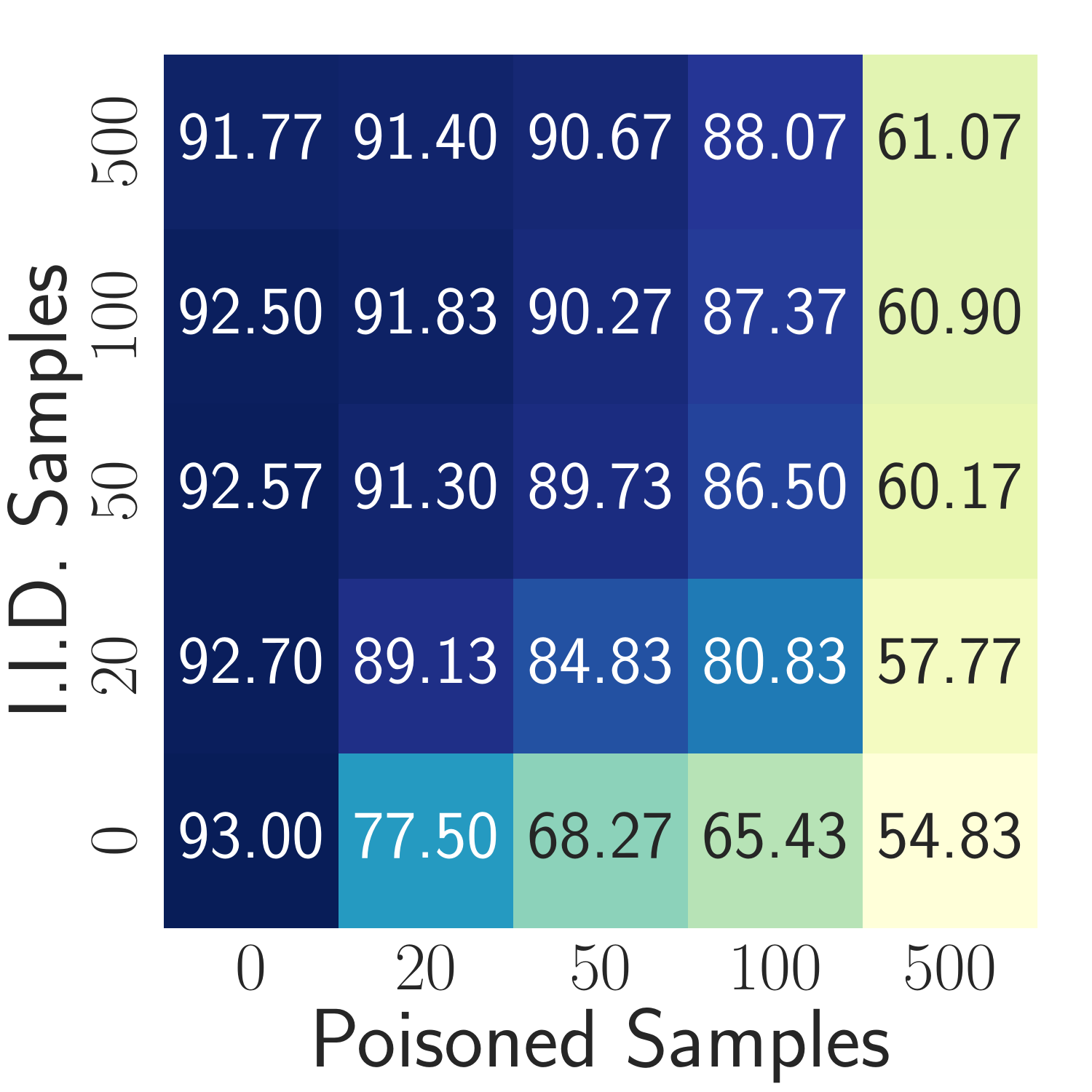}}
\hspace{6mm}
\subfloat[TENT\label{fig:clean_poison_ori_c}]{%
\includegraphics[width=0.15\linewidth]{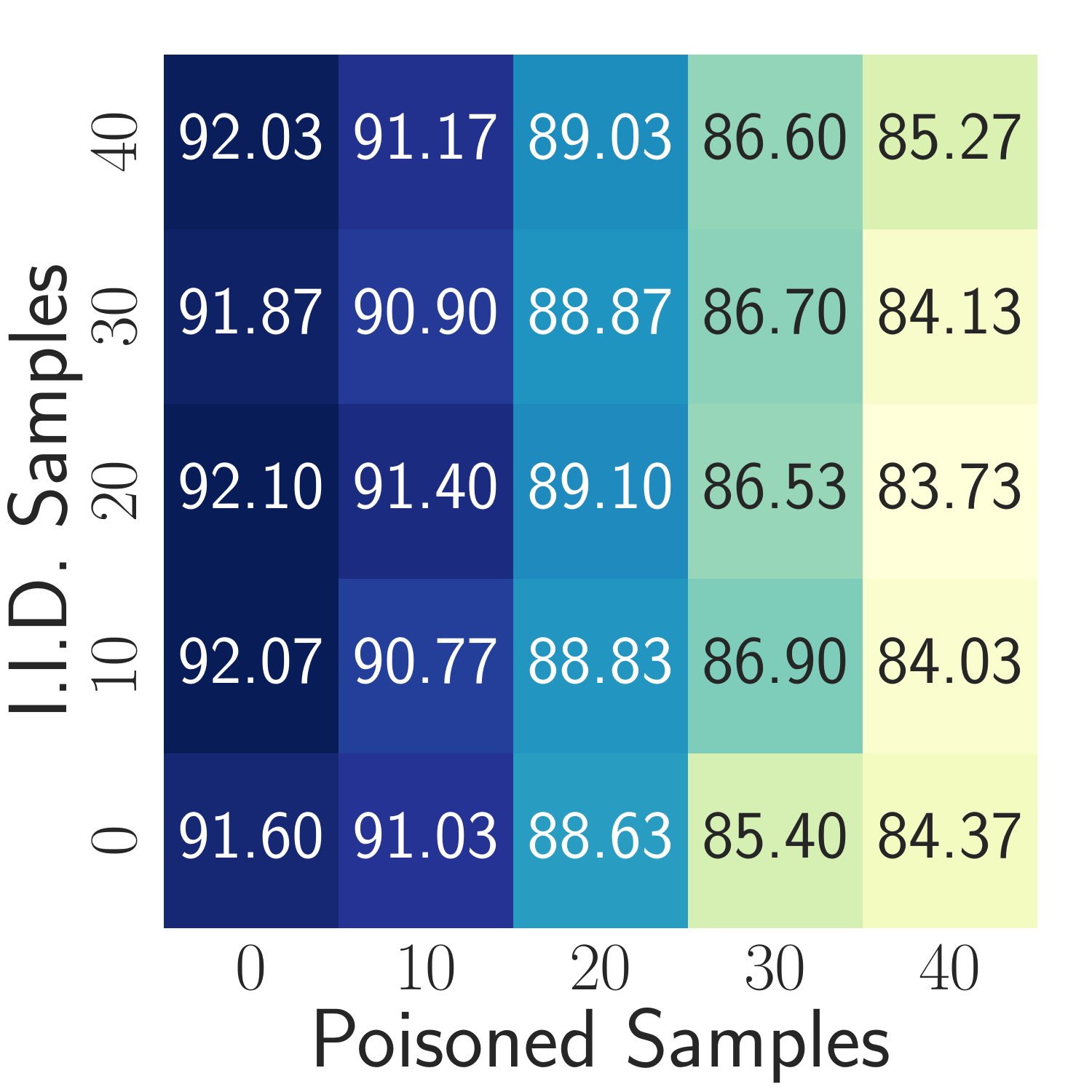}}
\hspace{6mm}
\subfloat[RPL \label{fig:clean_poison_ori_d}]{%
\includegraphics[width=0.15\linewidth]{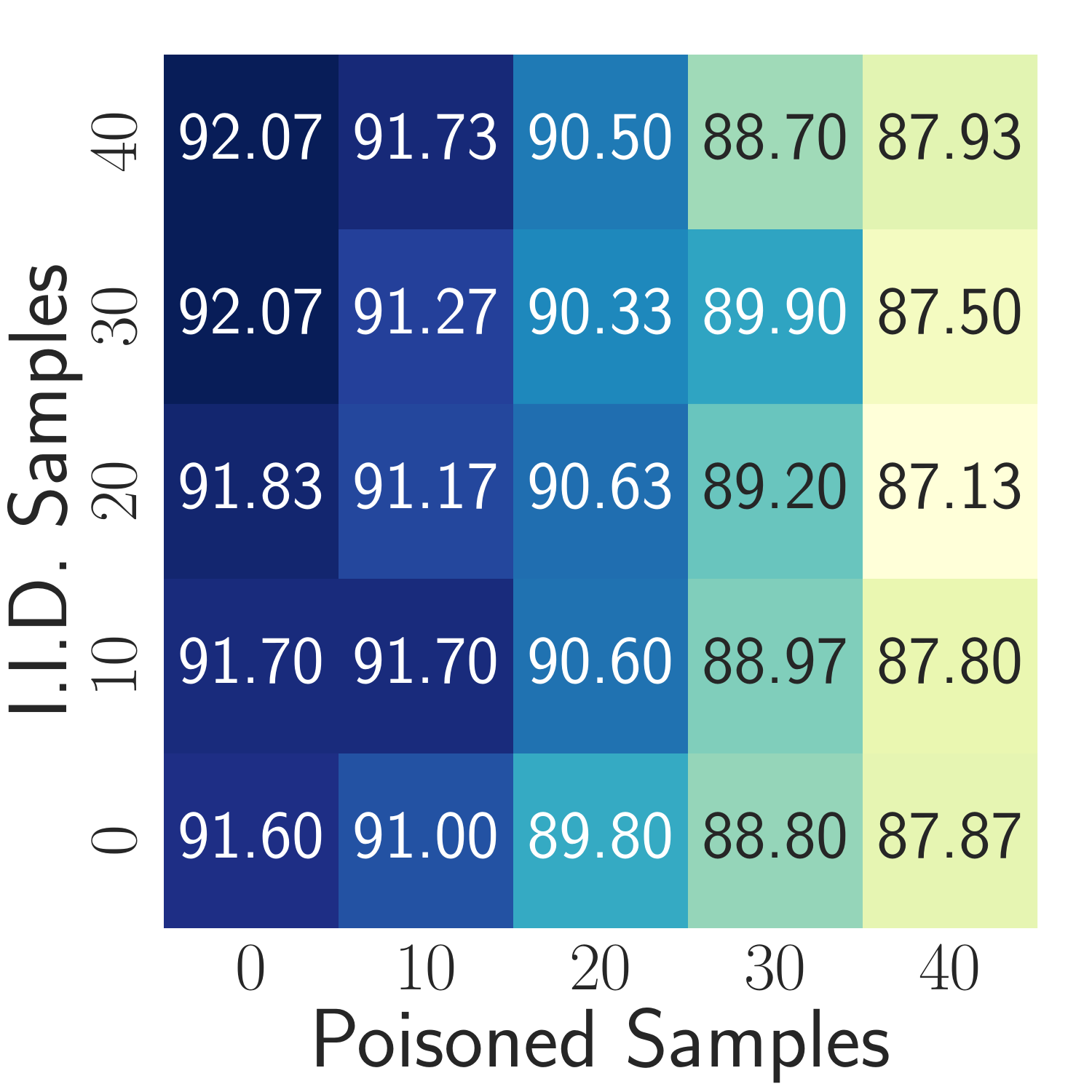}}
\caption{Warming-up before Poisoning. 
The target model is ResNet-18 trained on CIFAR-10. 
The y-axis and the x-axis represent the number of the i.i.d.\ samples and the poisoned samples, respectively. 
We fix the evaluation dataset to Ori.}
\label{fig:clean_poison_ori} 
\end{figure*}

\begin{figure*}[!t]
\centering
\subfloat[TTT \label{fig:poison_clean_ori_a}]{%
\includegraphics[width=0.15\linewidth]{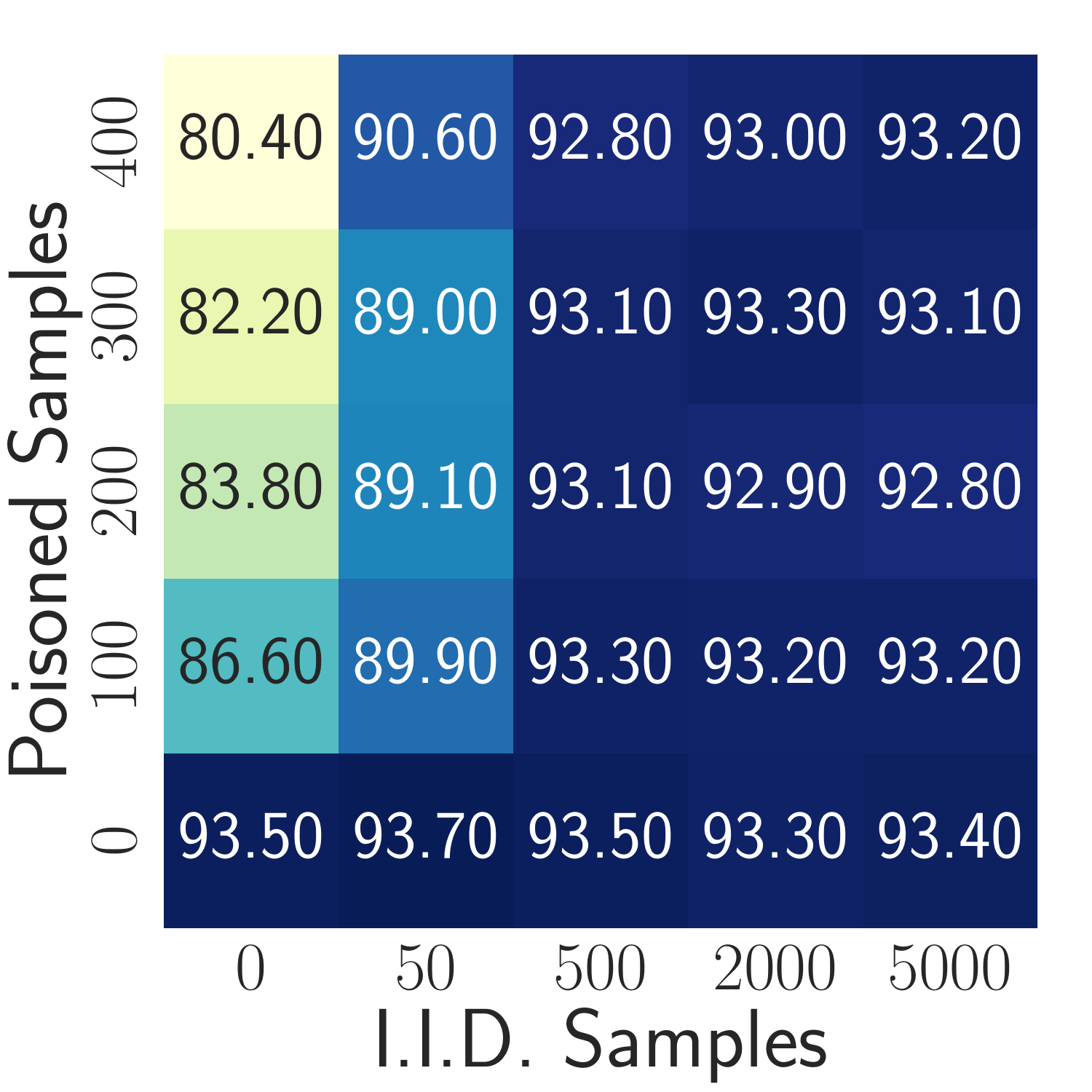}}
\hspace{6mm}
\subfloat[DUA \label{fig:poison_clean_ori_b}]{%
\includegraphics[width=0.15\linewidth]{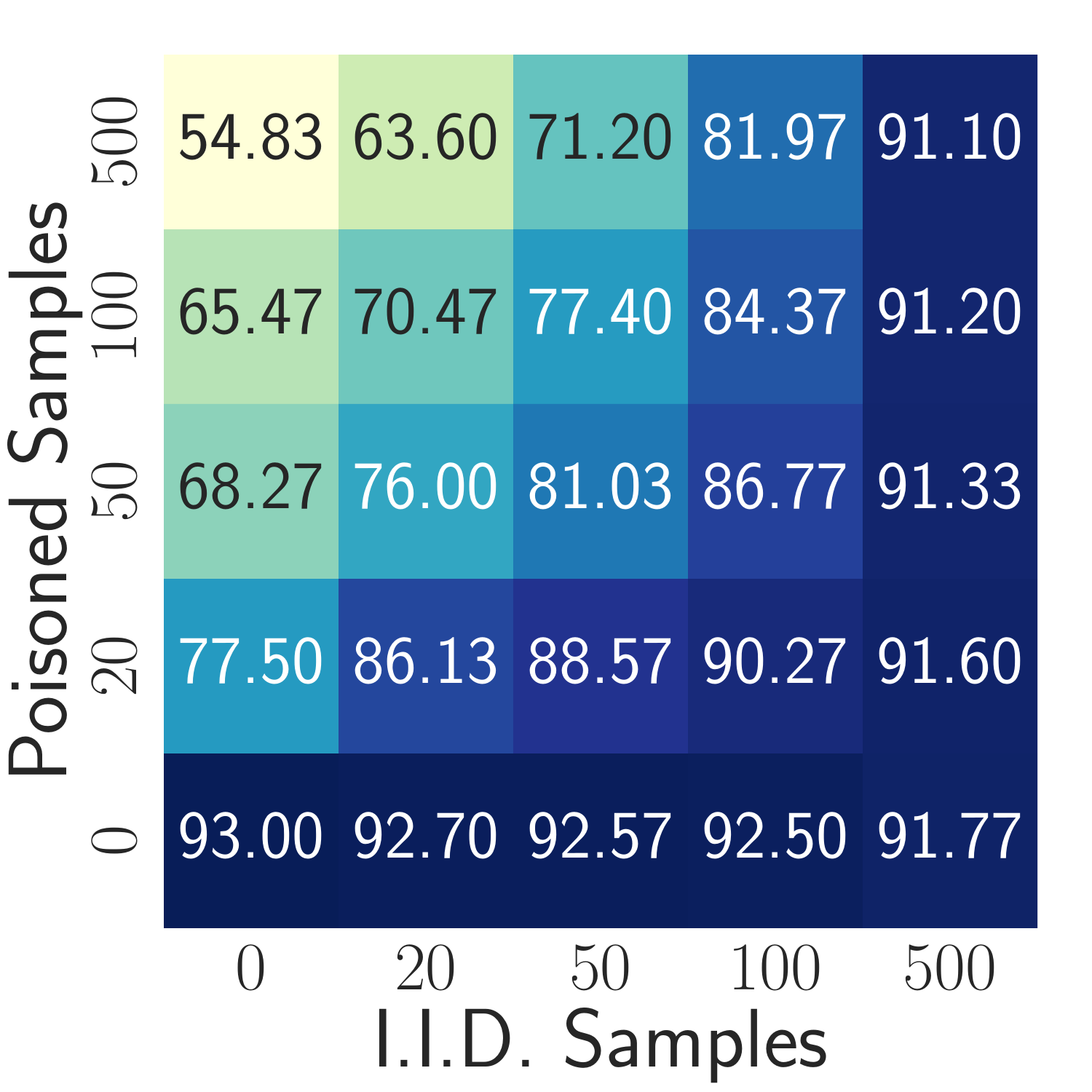}}
\hspace{6mm}
\subfloat[TENT\label{fig:poison_clean_ori_c}]{%
\includegraphics[width=0.15\linewidth]{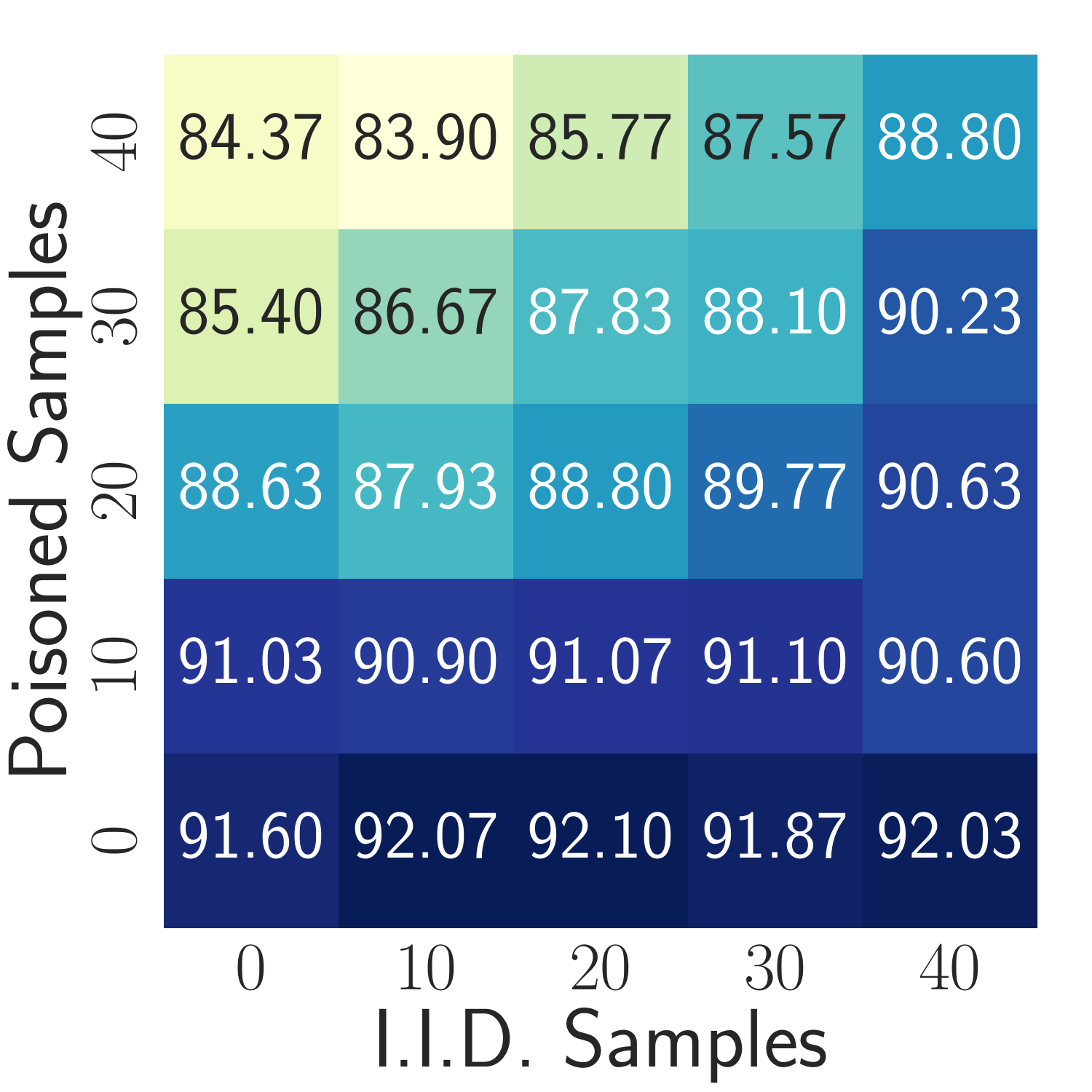}}
\hspace{6mm}
\subfloat[RPL \label{fig:poison_clean_ori_d}]{%
\includegraphics[width=0.15\linewidth]{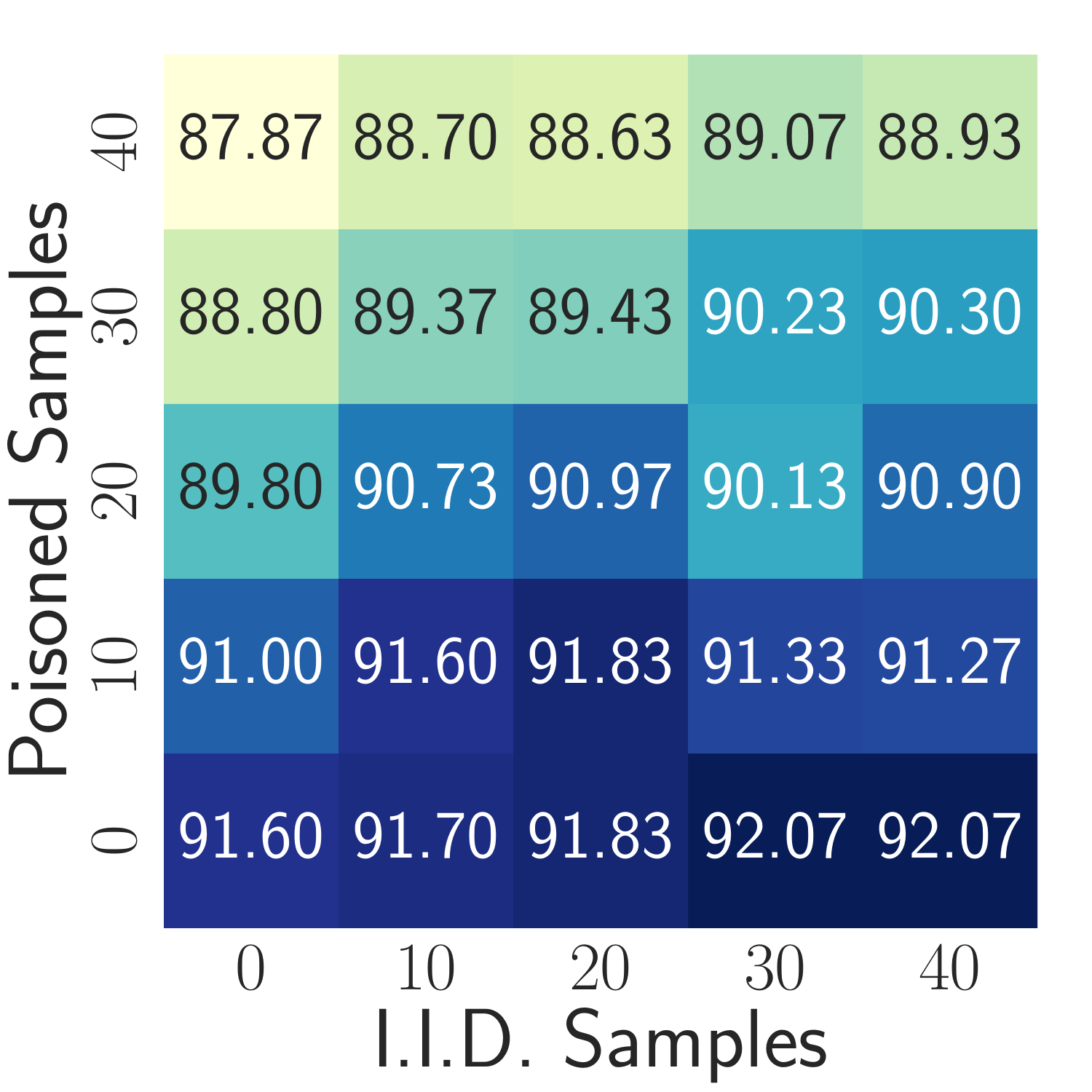}}
\caption{Warming-up after Poisoning. 
The target model is ResNet-18 trained on CIFAR-10. 
The y-axis and the x-axis represent the number of poisoned samples and the i.i.d.\ samples, respectively. 
We fix the evaluation dataset to Ori.}
\label{fig:poison_clean_ori} 
\end{figure*}

\begin{figure*}[!t]
\centering
\subfloat[AT \label{fig:defense_iid_a}]{%
\includegraphics[width=0.2\linewidth]{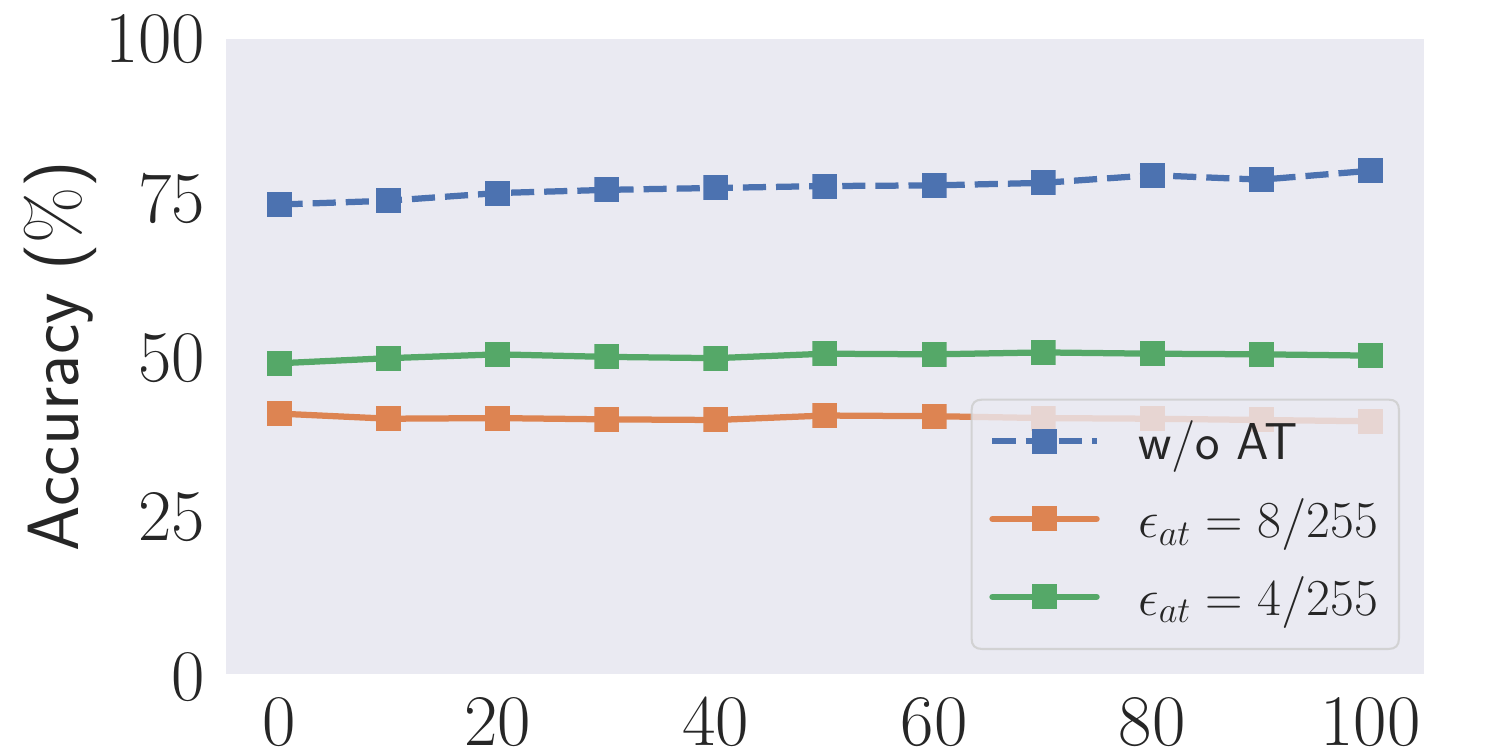}}
\hspace{2mm}
\subfloat[BDR \label{fig:defense_iid_b}]{%
\includegraphics[width=0.2\linewidth]{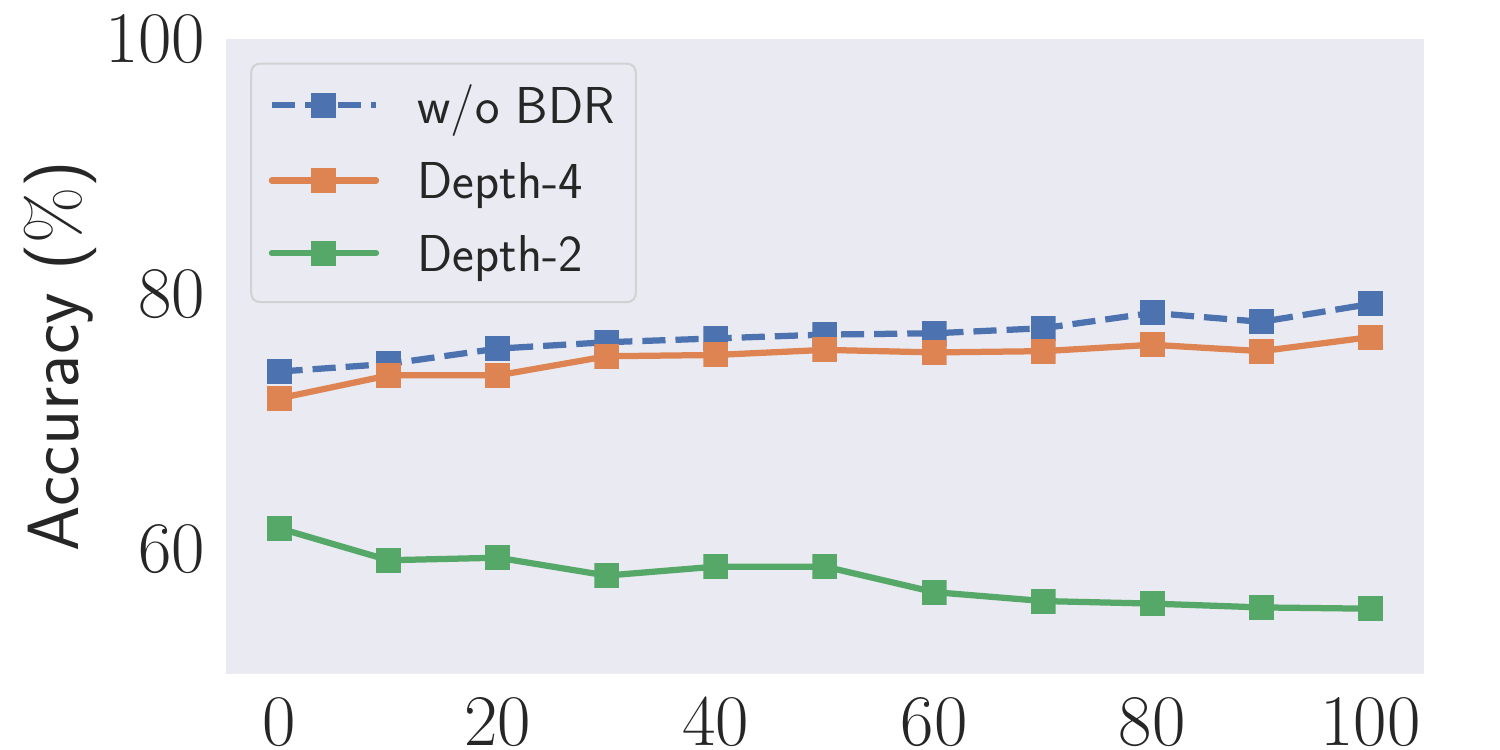}}
\hspace{2mm}
\subfloat[RRP\label{fig:defense_iid_c}]{%
\includegraphics[width=0.2\linewidth]{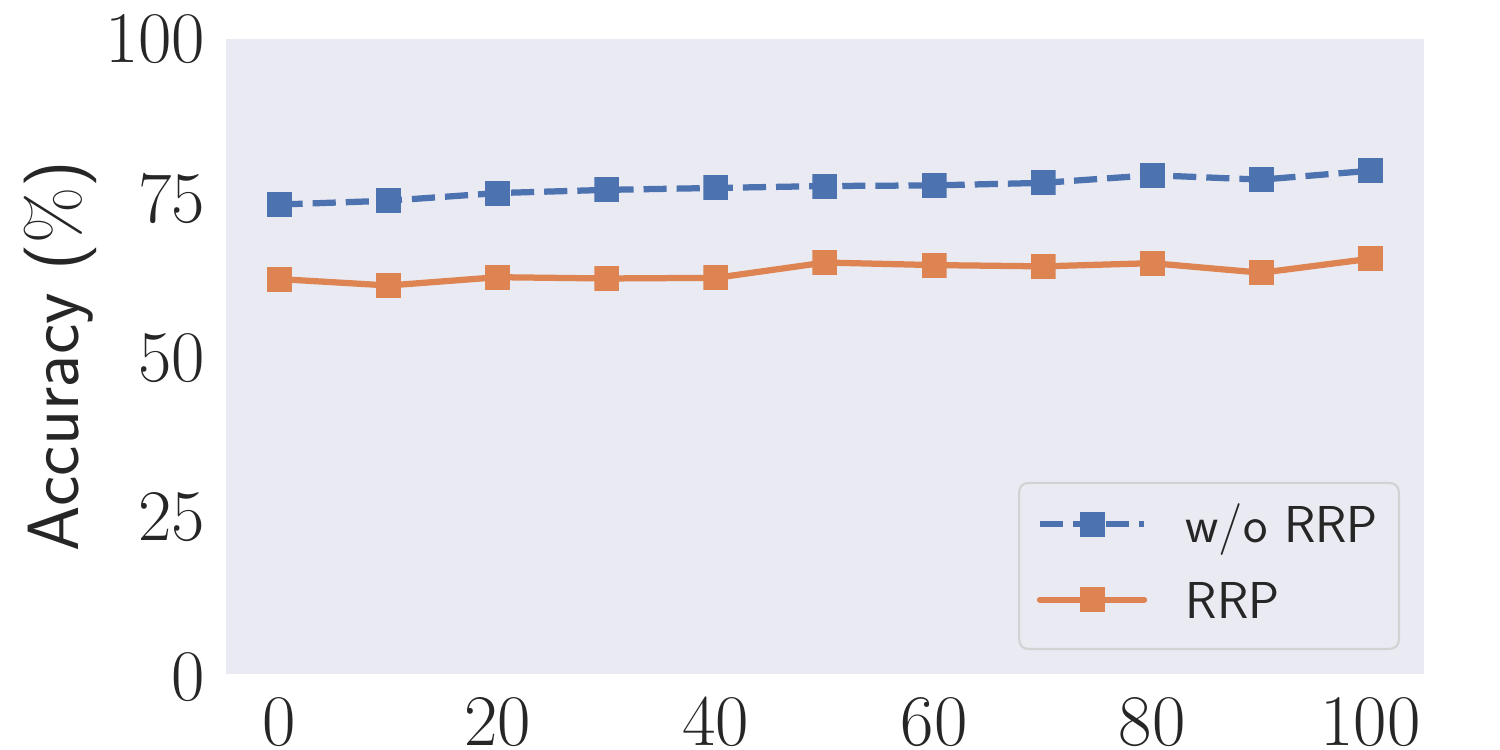}}
\hspace{2mm}
\subfloat[JC \label{fig:defense_iid_d}]{%
\includegraphics[width=0.2\linewidth]{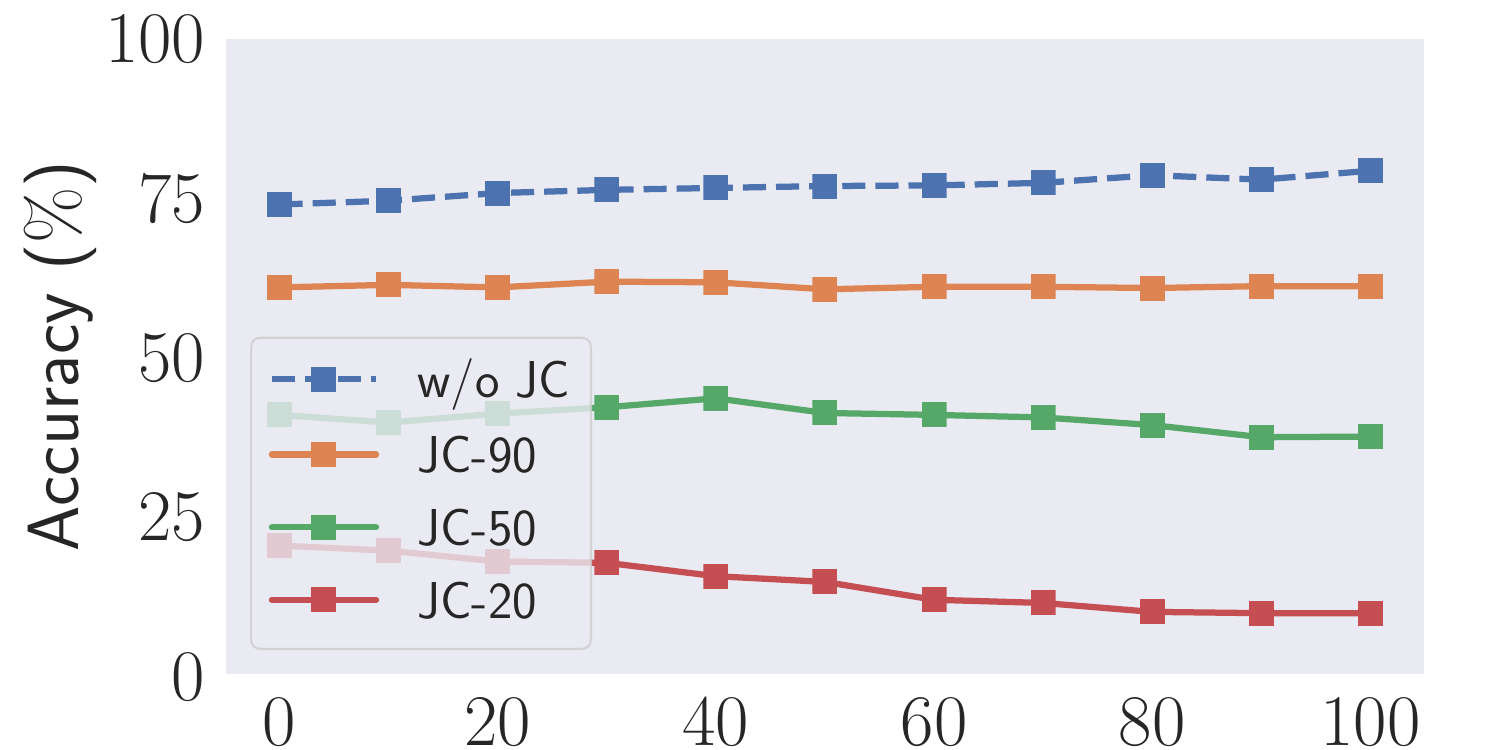}}
\caption{The impact of the four defense mechanisms on the benign samples. 
The x-axis represents the number of benign samples. 
We fix the target model to C10-Res18@Y4 and the evaluation dataset to Fog-5 of CIFAR-10-C.}
\label{fig:defense_iid} 
\end{figure*}

More experimental results are shown in~\autoref{fig::poison_eps32}, \autoref{fig::CIFAR-10_c}, \autoref{fig:non_iid}, \autoref{fig:utility_tta_res50_CIFAR-100}, \autoref{fig:clean_poison_ori}, \autoref{fig:poison_clean_ori}, and \autoref{fig:defense_iid}.

\begin{algorithm}[!t]
\caption{Adversarial training for TTT-models.}
\label{alg:at_ttt}
\SetKwInput{KwInput}{Input}                
\SetKwInput{KwOutput}{Output}             
\DontPrintSemicolon
  
\KwInput{Training dataset $\mathcal{D}_{train}$, initial network $f$, the perturbation budget $\epsilon_{at}$;}
\KwOutput{Robust TTT-model $f_{at}$;}

\SetKwFunction{FMain}{AT}
\SetKwProg{Fn}{Main function}{:}{\KwRet}

\Fn{\FMain($\mathcal{D}_{train}$, $f$, $\epsilon_{at}$)}{
\commentcolor{\%Transfer $f$ into a Y-structured network with a feature extractor $e(\cdot)$, a main task branch $\pi_m(\cdot)$, and an auxiliary task branch $\pi_s(\cdot)$.} \;
$e^{(0)},\pi^{(0)}_s,\pi^{(0)}_m \leftarrow f$; \;
\For {$t$ {\rm in} $[0,T_{max}]$}
{
\For {$(x_i,y_i)$ {\rm in} $\mathcal{D}_{train}$}
{
\commentcolor{\%Train the Y-structured network with the original training samples.} \;
$\mathcal{L}_1 = \mathcal{L}_m(x_i,y_i, e^{(t)}, \pi_m^{(t)})$; \;
$\mathcal{L}_2 = \mathcal{L}_s(x_i,y_i, e^{(t)}, \pi_s^{(t)})$; \;
\commentcolor{\%Train the Y-structured network with the adversarial examples.} \;
$f_{rot} = e^{(t)} \circ \pi^{(t)}_s;$\;
$x_i' = \texttt{PoiGen}(\text{TTT}, x_i,f_{rot}, \mathcal{L}_s, \epsilon_{at})$; \;
$\mathcal{L}_3 = \mathcal{L}_m(x'_i,y_i, e^{(t)}, \pi_m^{(t)})$; \;
$\mathcal{L}_4 = \mathcal{L}_s(x'_i, e^{(t)}, \pi_s^{(t)})$; \;
\commentcolor{\%Compute the final loss function for AT and update the model.} \;
$\mathcal{L}_{at} = \mathcal{L}_1 +  \mathcal{L}_2 +  \mathcal{L}_3 +  \mathcal{L}_4$; \;
$e^{(t+1)},\pi^{(t+1)}_s,\pi^{(t+1)}_m \leftarrow {\rm Optimizer}(e^{(t)},\pi^{(t)}_s,\pi^{(t)}_m, \mathcal{L}_{at})$; \;
}
}
\commentcolor{\%Generate the final Y-structured model $f_{at}$.} \;
$f_{at} \leftarrow e^{(T_{max})},\pi^{(T_{max})}_s,\pi^{(T_{max})}_m$; \;
\KwRet $f_{at}$.\;
}
\end{algorithm}

\end{document}